\documentclass{aa}
\usepackage{graphicx}
\usepackage{amssymb,amsmath}
\usepackage{latexsym}
\usepackage{mathtools}
\usepackage{epsfig}
\usepackage{natbib}
\usepackage{txfonts}
\usepackage{mathrsfs}
\begin{document}
   \title{Future destabilisation of Titan as a result of Saturn's tilting}
%
   \author{Melaine Saillenfest\inst{1}
          \and
          Giacomo Lari\inst{2}
          }
   \authorrunning{Saillenfest and Lari}
   \institute{IMCCE, Observatoire de Paris, PSL Research University, CNRS, Sorbonne Universit\'e, Universit\'e de Lille, 75014 Paris, France
              \email{melaine.saillenfest@obspm.fr}
              \and
              Department of Mathematics, University of Pisa, Largo Bruno Pontecorvo 5, 56127 Pisa, Italy
             }
   \date{Received 4 June 2021 / Accepted 6 July 2021}


  \abstract
  {As a result of Titan's migration and Saturn's probable capture in secular spin-orbit resonance, recent works show that Saturn's obliquity could be steadily increasing today and may reach large values in the next billions of years. Satellites around high-obliquity planets are known to be unstable in the vicinity of their Laplace radius, but the approximations used so far for Saturn's spin axis are invalidated in this regime.}
  {We aim to investigate the behaviour of a planet and its satellite when the satellite crosses its Laplace radius while the planet is locked in secular spin-orbit resonance.}
  {We expand on previous works and revisit the concept of Laplace surface. We use it to build an averaged analytical model that couples the planetary spin-axis and satellite dynamics.}
  {We show that the dynamics is organised around a critical point, S$_1$, at which the phase-space structure is singular, located at $90^\circ$ obliquity and near the Laplace radius. If the spin-axis precession rate of the planet is maintained fixed by a resonance while the satellite migrates outwards or inwards, then S$_1$ acts as an attractor towards which the system is forced to evolve. When it reaches the vicinity of S$_1$, the entire system breaks down, either because the planet is expelled from the secular spin-orbit resonance or because the satellite is ejected or collides into the planet.}
  {Provided that Titan's migration is not halted in the future, Titan and Saturn may reach instability between a few gigayears and several tens of gigayears from now, depending on Titan's migration rate. The evolution would destabilise Titan and drive Saturn towards an obliquity of $90^\circ$. Our findings may have important consequences for Uranus. They also provide a straightforward mechanism for producing transiting exoplanets with a face-on massive ring, a configuration that is often put forward to explain some super-puff exoplanets.}


  \maketitle

\section{Introduction}
A secular spin-orbit resonance occurs when the spin-axis precession rate of a planet becomes commensurate with a frequency (or a combination of frequencies) that appears in its orbital precession. Secular spin-orbit resonances were first studied individually and linked to Cassini's laws \citep{Colombo_1966,Peale_1969,Ward_1975,Henrard-Murigande_1987}. The effect on the spin-axis dynamics of a whole multi-harmonic orbital precession spectrum has also been investigated, and the overlap of several secular spin-orbit resonances has been identified as responsible for large chaotic regions in the inner Solar System \citep{Ward_1973,Ward_1982,Laskar-Robutel_1993,NerondeSurgy-Laskar_1997,Laskar-etal_2004}. More recently, higher-order resonances have been characterised in a systematic way, and their relation to the emergence of chaos has been assessed \citep{Li-Batygin_2014,Saillenfest-etal_2019a}. In fact, secular spin-orbit resonances are found to rule the long-term spin-axis dynamics of planets not only in the Solar System but also in extrasolar systems (see e.g. \citealp{Atobe-etal_2004,Deitrick-etal_2018,Shan-Li_2018,Millholland-Laughlin_2019}). 

As shown by \cite{Ward-Hamilton_2004}, Saturn is today very close to or inside a secular spin-orbit resonance with the nodal orbital precession mode of Neptune, noted $s_8$. The current large $26.7^\circ$ obliquity of Saturn probably results from this resonance \citep{Hamilton-Ward_2004}. It was first thought that the resonance trapping occurred more than four billion years ago during the late planetary migration \citep{Boue-etal_2009,Brasser-Lee_2015,Vokrouhlicky-Nesvorny_2015}. However, this would require Saturn's satellites to not have migrated much since this event, which contradicts the fast migration of the satellites measured by \cite{Lainey-etal_2020}. Instead, \cite{Saillenfest-etal_2021a} have shown that the migration of Saturn's satellites, and in particular of Titan, is likely responsible for the resonance encounter. The resonant interaction therefore began more recently than previously thought, perhaps about one billion years ago. Using Monte Carlo simulations, \cite{Saillenfest-etal_2021} show that in order to reproduce Saturn's current state, the most likely dynamical pathway is a gradual tilting starting from a few degrees before the resonance encounter. Since a near-zero primordial obliquity is also what is expected from planetary formation theories (see e.g. \citealp{Ward-Hamilton_2004}, \citealp{Rogoszinski-Hamilton_2020}, and references therein), and even though primordial non-zero obliquities are not totally excluded \citep{Millholland-Batygin_2019,Martin-Armitage_2021}, this scenario appears quite promising.

If Saturn did follow its expected pathway (and was not affected by an accidental major impact; see e.g. \citealp{Li-Lai_2020}), then Saturn should still be trapped inside the resonance today. As Titan continues migrating, Saturn should therefore continue to follow the drift of the resonance centre in the future. Because of this mechanism, Saturn may reach very large obliquity values in the next few billions of years \citep{Saillenfest-etal_2021}.

Yet, the final outcome of this dynamical mechanism remains unknown. The obliquity of a planet cannot increase forever, and there must exist some kind of dynamical barrier, either on the planet's or on the satellite's side, that would halt the tilting at some point. Even though this final outcome may not be directly relevant for Saturn and Titan because of the large timescales involved (see below), its generic nature makes it important even from the point of view of pure celestial mechanics as other planets and exoplanets may have been affected. Hence, we aim to characterise the full tilting mechanism in a general way, without any assumption about the distance of the satellite, and with a special focus on the case of Saturn and Titan.

Since Titan is still far from its Laplace radius today and may only reach it after billions of years of continuous orbital expansion (if not tens of billions of years), previous studies have been restricted to the close-in satellite regime, in which Titan's orbit lies in Saturn's equatorial plane (see e.g. \citealp{Goldreich_1966}). Out of this regime, regular satellites are known to oscillate around their local Laplace plane, which is inclined halfway between the equator and the orbital plane of the host planet \citep{Laplace_1805}. Most importantly, \cite{Tremaine-etal_2009} found that the satellite is unstable in the vicinity of its Laplace radius if the planet's obliquity is larger than about $69^\circ$ (for a circular satellite) or $71^\circ$ (for an eccentric satellite). According to \cite{Saillenfest-etal_2021}, Saturn's obliquity may exceed these thresholds in a few gigayears from now, depending on Titan's migration rate. Hence, if Titan happens to be located near its Laplace radius at this stage of the evolution, a non-trivial dynamics is expected for the system. The simulations of \cite{Tremaine-etal_2009} and \cite{Tamayo-etal_2013} revealed that, in some ranges of parameters, strong chaotic transitions in the satellite's eccentricity and inclination are possible. \cite{Tamayo-etal_2013} drew a parallel between the eccentricity increase of the satellite and the ZKL mechanism (for `von Zeipel-Lidov-Kozai'; see \citealp{Ito-Ohtsuka_2019}), in which large eccentricity oscillations occur while the satellite's argument of pericentre oscillates around a fixed value. Beyond some eccentricity threshold, the effect of planetary oblateness re-initiates the apsidal precession, with the result of averaging to zero the solar torque and stopping the eccentricity increase\footnote{The same mechanism was described by \cite{Saillenfest-etal_2019} as a protection mechanism for inner Oort cloud objects against the action of galactic tides.}. More recently, \cite{Speedie-etal_2020} performed an extensive numerical exploration of the stability of particles initially located near their local Laplace plane. Their study confirms the secular instabilities reported by \cite{Tremaine-etal_2009} and \cite{Tamayo-etal_2013}, and their fully unaveraged model allows for other kinds of instability to appear, driven by the evection and `ivection' resonances\footnote{The `ivection' resonance mentioned by \cite{Speedie-etal_2020} is order zero in eccentricity; it should not be confused with the mixed-type `eviction' resonance described by \cite{Touma-Wisdom_1998}, which is order two in eccentricity and can only be triggered once the eccentricity is high enough. See the preprint of \cite{Xu-Fabrycky_2019} for more details.}.

These previous results show that studying Saturn's tilting mechanism in a general way requires one to keep an eye on both the satellite's and planet's dynamics. Out of the close-in satellite regime, and a fortiori if the satellite becomes unstable, the model used by \cite{Saillenfest-etal_2021a,Saillenfest-etal_2021} for Saturn's spin-axis dynamics is invalidated. As a first step before developing a complete numerical model, our goal in this article is to establish a qualitative understanding of what happens to the planet and its satellite when the secular spin-orbit resonance leads them to their ultimate large-obliquity regime, where previous models fail.

In Sect.~\ref{sec:orb}, we revisit the concept of Laplace surface introduced by \cite{Tremaine-etal_2009}; we go further in the analytical characterisation of the equilibria and focus on the large-obliquity regime. In Sect.~\ref{sec:plspin}, we describe the influence of the satellite's dynamics on the spin-axis motion of the planet. We provide simplified formulas that allow the planet's obliquity evolution to be described as a function of the satellite's properties. In Sect.~\ref{sec:sattit}, we apply our findings to Saturn and Titan and explore their coupled dynamics as Titan migrates outwards. We conclude in Sect.~\ref{sec:ccl} and present some further applications of our results in other contexts.

\section{Orbital motion of the satellite}\label{sec:orb}
   
In order to investigate the way satellites interact with the spin axis of their host planet, we must get a clear understanding of their orbital dynamics as well. In this section, we first consider a massless satellite orbiting an oblate planet, which has itself a fixed orbit around the star (or an orbit that can be regarded as fixed over the interval of time considered).
   
   \subsection{Equations of motion}
   The Hamiltonian function describing the orbital motion of the massless satellite around the planet can be written $\mathcal{K} = \mathcal{K}_0 + \epsilon\mathcal{K}_1$, where $\mathcal{K}_0$ is the Keplerian part and $\epsilon\mathcal{K}_1$ gathers the orbital perturbations. The parameter $\epsilon\ll 1$ stresses that the orbital perturbations are small; neglecting $\mathcal{O}(\epsilon^2)$, the long-term behaviour of the satellite is described by the secular Hamiltonian $\mathcal{H}$, which can be written
   \begin{equation}\label{eq:Hfirst}
      \mathcal{H} = k_\mathrm{P}\mathcal{H}_\mathrm{P} + k_\odot\mathcal{H}_\odot\,,
   \end{equation}
   where $k_\mathrm{P}\mathcal{H}_\mathrm{P}$ comes from the planet's oblateness and $k_\odot\mathcal{H}_\odot$ comes from the star's gravitational attraction. The secular semi-major axis $a$ of the satellite is a constant of motion and a parameter of $\mathcal{H}$. Considering that $a$ is much larger than the planet's equatorial radius $R_\mathrm{eq}$ and much smaller than the star's semi-major axis $a_\odot$ in its orbit around the planet, both terms of Eq.~\eqref{eq:Hfirst} can be expanded in Legendre polynomials. As \cite{Tremaine-etal_2009}, we first limit the expansion to the quadrupolar approximation, which amounts to neglecting $(R_\mathrm{eq}/a)^3$ and $(a/a_\odot)^3$. This leads us to define the two fixed parameters of Eq.~\eqref{eq:Hfirst} as
   \begin{equation}\label{eq:kpks}
      k_\mathrm{P} = \frac{3}{4}\frac{\mu_\mathrm{P}}{a}J_2\frac{R_\mathrm{eq}^2}{a^2}
      \quad\text{and}\quad
      k_\odot = \frac{3}{8}\frac{\mu_\odot}{a_\odot(1-e_\odot^2)^{3/2}}\frac{ a^2}{a_\odot^2} \,,
   \end{equation}
   and the two parts of the Hamiltonian function as
   \begin{equation}\label{eq:HpHs}
      \left\{
      \begin{aligned}
         \mathcal{H}_\mathrm{P} &= \frac{1-3\cos^2I_\mathrm{Q}}{3(1-e^2)^{3/2}} \,,\\
         \mathcal{H}_\odot &= \sin^2I_\mathrm{C}\left(1 + \frac{3}{2}e^2 - \frac{5}{2}e^2\cos(2\omega_\mathrm{C})\right) - e^2 \,.
      \end{aligned}
      \right.
   \end{equation}
   In these expressions, $\mu_\mathrm{P}$ and $J_2$ are the gravitational parameter and the second zonal gravity coefficient of the host planet, and $\mu_\odot$, $a_\odot$, and $e_\odot$ are the gravitational parameter, the semi-major axis, and the eccentricity of the star, respectively. The usual orbital elements of the satellite are written $(e,I,\omega,\Omega)$, and we use the index Q for quantities measured with respect to the planet's equator, and the index C for quantities measured with respect to the planet's orbital plane (that we improperly call the `ecliptic'). Since Eq.~\eqref{eq:HpHs} is obtained through a multi-polar development only, it is valid for arbitrary eccentricity and inclination of the planet and of its satellite \citep{Laskar-Boue_2010}.
   
   The Hamiltonian function $\mathcal{H}_\odot$ in Eq.~\eqref{eq:HpHs} has been averaged over the star's orbital motion. As stressed by \cite{Tremaine-etal_2009}, this approximation requires not only that $a\ll a_\odot$, but even that $a\ll r_\mathrm{H}$, where $r_\mathrm{H}$ is the Hill radius of the planet. In particular, Eq.~\eqref{eq:HpHs} does not contain the evection resonance, which can have important effects for far-away satellites (see e.g. \citealp{Frouard-etal_2010,Speedie-etal_2020}). In Sect.~\ref{sec:sattit}, we verify that Eq.~\eqref{eq:HpHs} provides a fair approximation of Titan's orbital dynamics.
   
   The coordinates of the satellite measured with respect to the equator (Q) or to the ecliptic (C) are linked through the obliquity $\varepsilon$ of the planet via the relations given in Appendix~\ref{asec:CQ}. These relations can be used to express $\mathcal{H}$ in terms of the equator or ecliptic coordinates only. Noting $C=\cos\varepsilon$ and $S=\sin\varepsilon$, the Hamiltonian $\mathcal{H}_\mathrm{P}$ in Eq.~\eqref{eq:HpHs} can be written equivalently as
   \begin{equation}\label{eq:HpC}
      \begin{aligned}
         \mathcal{H}_\mathrm{P} = \frac{-1}{6(1-e^2)^{3/2}}&\bigg[
             (3C^2 - 1)(3\cos^2I_\mathrm{C} - 1) \\
          &+ 12CS\cos I_\mathrm{C}\sin I_\mathrm{C}\cos\delta_\mathrm{C} \\
          &+ 3S^2\sin^2I_\mathrm{C}\cos(2\delta_\mathrm{C})
         \bigg]\,,
      \end{aligned}
   \end{equation}
   where $\delta_\mathrm{C} \equiv \Omega_\mathrm{C}-\Omega_\mathrm{P}$ and $\Omega_\mathrm{P}$ is the ascending node of the planet's equator measured along the ecliptic. Likewise, the Hamiltonian $\mathcal{H}_\odot$ in Eq.~\eqref{eq:HpHs} can be written equivalently as
   \begin{equation}\label{eq:HsQ}
      \begin{aligned}
         \mathcal{H}_\odot = - \frac{1}{8}&\bigg[
           8e^2 + 2(3e^2 + 2)(2C^2\cos^2I_\mathrm{Q} + S^2\sin^2I_\mathrm{Q} - 2) \\
         &+ 8CS(3e^2 + 2)\cos I_\mathrm{Q}\sin I_\mathrm{Q}\cos\delta_\mathrm{Q} \\
         &+ 5S^2e^2(\cos I_\mathrm{Q} + 1)^2\cos(2\omega_\mathrm{Q} + 2\delta_\mathrm{Q}) \\
         &- 20CSe^2(\cos I_\mathrm{Q} + 1)\sin I_\mathrm{Q}\cos(2\omega_\mathrm{Q} + \delta_\mathrm{Q}) \\
         &+ 10(3C^2 - 1)e^2\sin^2I_\mathrm{Q}\cos(2\omega_\mathrm{Q}) \\
         &- 20CSe^2(\cos I_\mathrm{Q} - 1)\sin I_\mathrm{Q}\cos(2\omega_\mathrm{Q} - \delta_\mathrm{Q}) \\
         &+ 5S^2e^2(\cos I_\mathrm{Q} - 1)^2\cos(2\omega_\mathrm{Q} - 2\delta_\mathrm{Q}) \\
         &+ 2S^2(3e^2 + 2)\sin^2I_\mathrm{Q}\cos(2\delta_\mathrm{Q})
         \bigg]\,,
      \end{aligned}
   \end{equation}
   where $\delta_\mathrm{Q}\equiv\Omega_\mathrm{Q}-\Omega_\odot$ and $\Omega_\odot$ is the ascending node of the star measured along the equator of the planet.
   
   If the planet's axis of figure has a fixed orientation in space, then $\Omega_\mathrm{P}$ is a constant angle, and both the ecliptic and equatorial reference frames are inertial. This is equivalent to considering that the spin-axis precession of the planet is infinitely slow compared to the timescales relevant for the satellite. The validity of this hypothesis will be discussed in Sect.~\ref{sec:plspin}. For now, we consider that $\Omega_\mathrm{P}$ is constant and examine the dynamical system described by Eq.~\eqref{eq:HpHs}, expanding on the work of \cite{Tremaine-etal_2009}.
   
   First of all, the parameters $k_\mathrm{P}$ and $k_\odot$ in Eq.~\eqref{eq:kpks} make appear a characteristic length called `Laplace radius' defined by
   \begin{equation}\label{eq:rL}
      r_\mathrm{L}^5 = \frac{1}{2}\frac{k_\mathrm{P}}{k_\odot}a^5 = \frac{\mu_\mathrm{P}}{\mu_\odot}J_2R_\mathrm{eq}^2a_\odot^3(1-e_\odot^2)^{3/2}\,.
   \end{equation}
   We also introduce a critical radius $r_\mathrm{M}$, already used by \cite{Goldreich_1966}, that we define by
   \begin{equation}\label{eq:rM}
      r_\mathrm{M}^5 = 2\,r_\mathrm{L}^5 \,.
   \end{equation}
   As noticed by \cite{Tamayo-etal_2013}, it is more natural to use $r_\mathrm{M}$ as a reference radius than the conventional $r_\mathrm{L}$ of \cite{Tremaine-etal_2009}. The symbol M stands here for `midpoint' and the dynamical meaning of $r_\mathrm{M}$ will appear clear below\footnote{\cite{Cuk-etal_2016} and \cite{Speedie-etal_2020} go one step further and redefine $r_\mathrm{L}$ by adding the factor $2$ in Eq.~\eqref{eq:rL}. We rather prefer to introduce a different symbol, because using differing definitions for the classic `Laplace radius' is misleading when it comes to comparison with previous works.}. Using this definition, we can rewrite $\mathcal{H}$ as
   \begin{equation}
      \mathcal{H} = k_\mathrm{P}\left[\mathcal{H}_\mathrm{P} + \frac{a^5}{r_\mathrm{M}^5}\mathcal{H}_\odot\right] = k_\odot\left[\frac{r_\mathrm{M}^5}{a^5}\mathcal{H}_\mathrm{P} + \mathcal{H}_\odot\right] \,,
   \end{equation}
   where a change of timescale could be used to remove the leading constant factor. In order to investigate the dynamics of a slowly migrating satellite, it is more convenient to introduce a timescale that does not involve its semi-major axis. As shown below, the frequency $\kappa$, defined as
   \begin{equation}\label{eq:kappa}
      \kappa^2 = \frac{9}{4}\frac{\mu_\odot^2r_\mathrm{M}^3}{\mu_\mathrm{P}a_\odot^6(1-e_\odot^2)^3}\,,
   \end{equation}
   naturally appears in the dynamics, and it is therefore a good choice of characteristic timescale. We define the corresponding period as $\tau=2\pi/\kappa$. Hence, we can describe the full variety of trajectories of the satellite by the only two parameters $a/r_\mathrm{M}$ and~$\varepsilon$, and their evolution timescale is provided by the period $\tau$. Table~\ref{tab:param} lists these parameters for various satellites in the Solar System. We also include the case of distant trans-Neptunian objects perturbed by the galactic tides, which have an almost identical dynamics (see \citealp{Saillenfest-etal_2019}).
   
   \begin{table*}
      \caption{Parameters and dynamical timescales of some satellites and their host planets in the Solar System.}
      \label{tab:param}
      \vspace{-0.7cm}
      \begin{equation*}
         \begin{array}{rllllrll}
            \hline
            \text{satellite} & \phantom{0}r_\mathrm{M} & \phantom{0}a/r_\mathrm{M} & \phantom{0}\tau & \phantom{0}\eta & \text{planet} & \varepsilon & T \\
            \hline
            \hline
            \text{the Moon}\phantom{0} & \phantom{00}9.7~R_\mathrm{eq} & \phantom{0}6.23 & \phantom{000}139~\text{yrs} & \phantom{0}532 & \text{the Earth} & 23.4^\circ & 73\,700~\text{yrs} \\
           \text{Deimos}\phantom{0} & \phantom{0}13.0~R_\mathrm{eq} & \phantom{0}0.53& \phantom{000}260~\text{yrs} & \phantom{0}1.22\times 10^{-4} &  \text{Mars} & 25.2^\circ & 0.154~\text{Myrs} \\
          \text{Callisto}\phantom{0} & \phantom{0}36.3~R_\mathrm{eq} & \phantom{0}0.73 & \phantom{00}1269~\text{yrs} & \phantom{0}1.46 & \text{Jupiter} & \phantom{0}3.1^\circ & \phantom{0}1.40~\text{Myrs} \\
            \begin{array}{r}
               \text{Titan}\\
               \text{Iapetus}
            \end{array} &
            \begin{array}{l}
               41.7~R_\mathrm{eq}\\
               54.8~R_\mathrm{eq}
            \end{array} & 
            \begin{array}{l}
               0.49\\
               1.08
            \end{array} &
            \begin{array}{l}
               \phantom{0}4501~\text{yrs}\\
               \phantom{0}2982~\text{yrs}
            \end{array} &
            \left.
            \begin{array}{l}
               12.4\\
               6.45\times 10^{-2}
            \end{array}\right\} &
            \text{Saturn} & 26.7^\circ & \phantom{0}6.74~\text{Myrs} \\
         \text{Oberon}\phantom{0} & \phantom{0}65.0~R_\mathrm{eq} & \phantom{0}0.34 & \phantom{0}25\,674~\text{yrs} & \phantom{0}7.27 & \text{Uranus} & 97.9^\circ & \phantom{0.}165~\text{Myrs} \\
        \text{TNO}\phantom{0} &  \phantom{0}1038~\text{au} & \phantom{0}- &  \phantom{0}209~\text{Gyrs} & \phantom{0}0 & \text{Solar System} & 61.7^\circ & \infty \\
         \hline
         \end{array}
      \end{equation*}
      \vspace{-0.3cm}
      \tablefoot{See text for the definition of the parameters.
      The $J_2$ used in Eq.~\eqref{eq:rL} is enhanced by the contribution of inner satellites if there are any \citep{Tremaine-etal_2009}, thus the difference of $r_\mathrm{M}$ and $\tau$ between Titan and Iapetus.
      The parameter $\lambda$ appearing in Eq.~\eqref{eq:alpha} is not well known for the giant planets, so the values of $T$ given here are only approximate. We used $\lambda=0.25$, $0.23$, and $0.23$ for Jupiter, Saturn, and Uranus, respectively.
      The line `TNO' refers to the motion of a trans-Neptunian object perturbed by the quadrupolar perturbations from the planets and from the galactic tides \citep{Saillenfest-etal_2019}; in this case, $\varepsilon$ is the tilt of the galactic plane with respect to the invariable plane of the Solar System \citep{Murray_1989,Souami-Souchay_2012}, and the other parameters are $r_\mathrm{M}^5=(3/2)\sum_i\mu_ia_i^2/\mathcal{G}_3$ and $\kappa^2=r_\mathrm{M}^3\mathcal{G}_3^2/\mu_\odot$, where $\mu_i$ and $a_i$ are the gravitational parameters and semi-major axes of the planets of the Solar System, and $\mathcal{G}_3$ is a constant incompassing the mass distribution within the Galaxy \citep{Fouchard_2004}.}
   \end{table*}
   
   In order to study the orbital dynamics of the satellite, it is more suitable to use a set of coordinates that are not singular for circular and/or zero-inclination orbits, as the usual rectangular coordinates
   \begin{equation}
      \begin{aligned}
         k &= e\cos(\omega_\mathrm{Q}+\delta_\mathrm{Q})\,, \\
         q &= \sin\frac{I_\mathrm{Q}}{2}\cos(\delta_\mathrm{Q})\,,
      \end{aligned}
      \quad
      \begin{aligned}
         h &= e\sin(\omega_\mathrm{Q}+\delta_\mathrm{Q})\,, \\
         p &= \sin\frac{I_\mathrm{Q}}{2}\sin(\delta_\mathrm{Q})\,.
      \end{aligned}
   \end{equation}
   Alternatively, one can use a vectorial formulation as \cite{Tremaine-etal_2009}.
   
   \subsection{The Laplace states}\label{ssec:Lap}
   By writing down the equations of motion in a non-singular set of coordinates, we see that the condition $e=0$ is an equilibrium point for the satellite whatever its other orbital elements. Moreover, linear stability analysis shows that the eccentricity and inclination degrees of freedom are decoupled in the vicinity of $e=0$. Assuming that the satellite's eccentricity is zero (for instance, if it has been damped at the time of its formation in a circumplanetary disc), we can therefore study the evolution of its inclination degree of freedom in a decoupled way.
   
\begin{figure}
      \includegraphics[width=0.9\columnwidth]{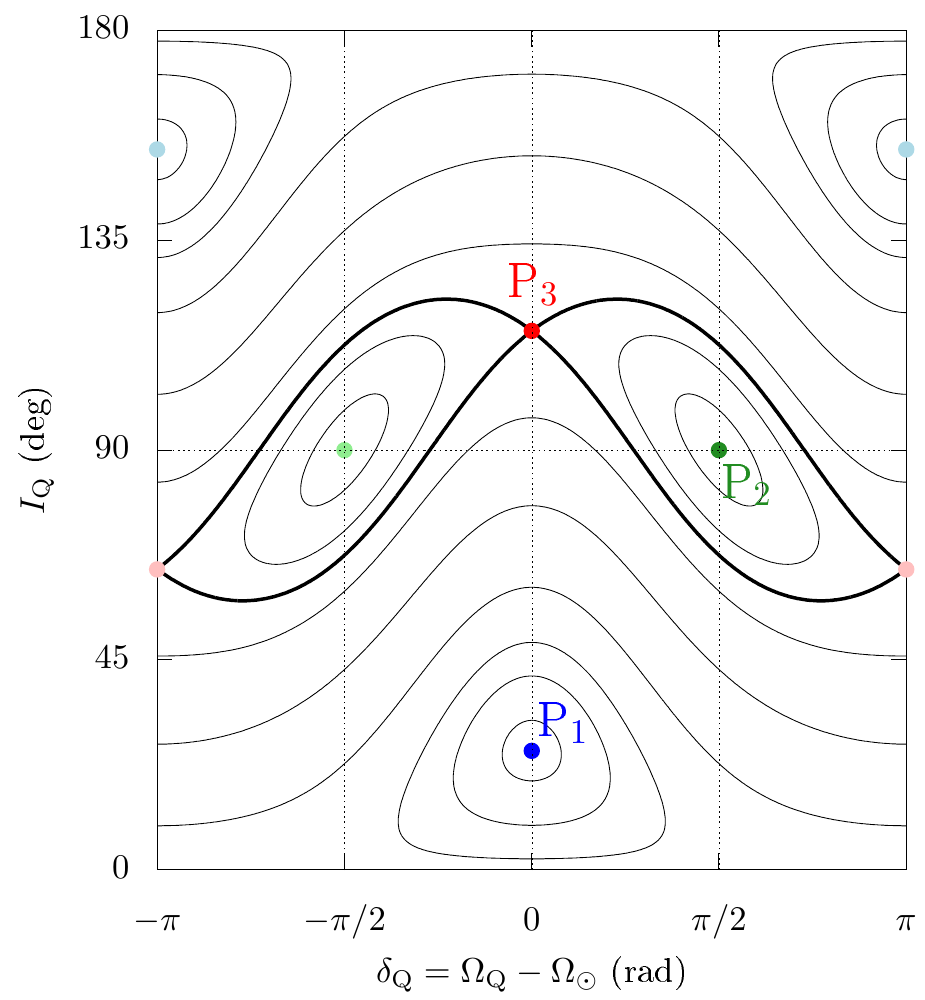}
      \caption{Level curves of the Hamiltonian function $\mathcal{H}$ for a circular orbit. The parameters are $a/r_\mathrm{M}=1.1$ and $\varepsilon=40^\circ$. The separatrix is shown by a thicker black curve. The coloured dots represent the three kinds of equilibrium points (`Laplace states'), labelled as in the text. A dark colour is used for points P$_1$ and P$_3$ lying at $\delta_\mathrm{Q}=0$ and for P$_2$ lying at $\delta_\mathrm{Q}=\pi/2$. A light colour is used for the symmetric equilibrium point that corresponds to the same Laplace state with reversed orbital motion. Figure~\ref{fig:phase3D}a shows the same phase portrait plotted on the sphere.}
      \label{fig:phase}
   \end{figure}

\begin{figure*}
      \includegraphics[width=0.33\textwidth]{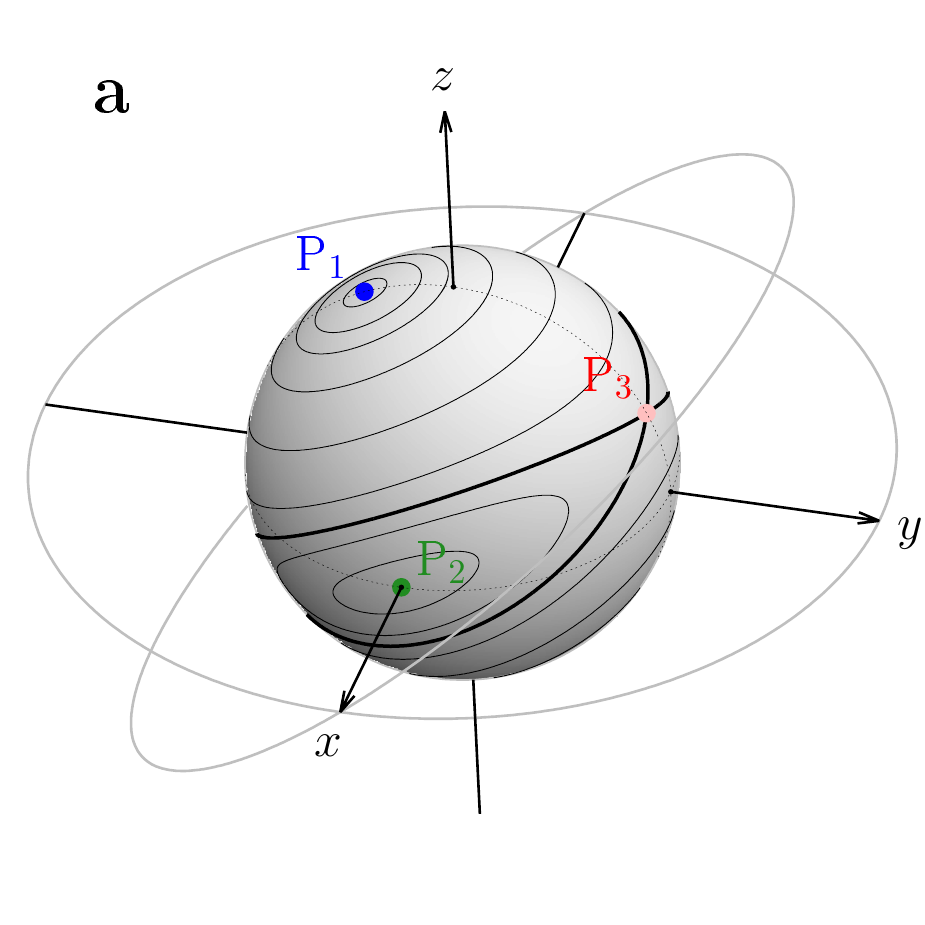}
      \includegraphics[width=0.33\textwidth]{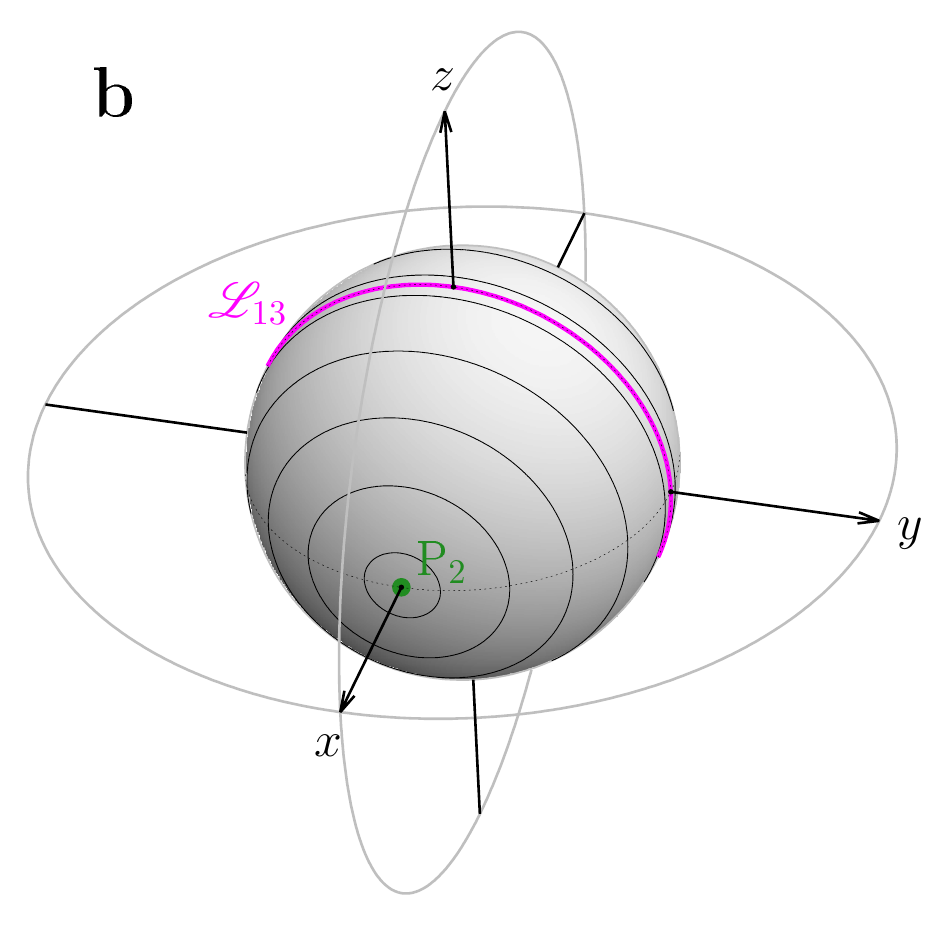}
      \includegraphics[width=0.33\textwidth]{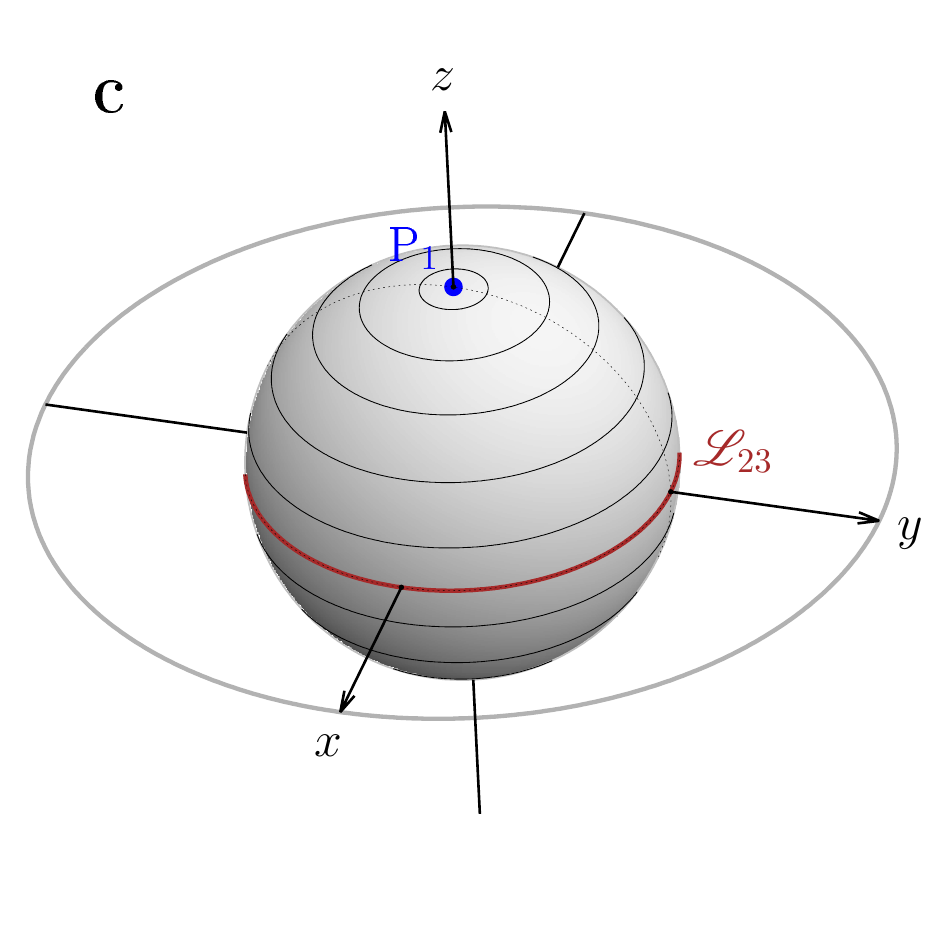}
      \caption{Level curves of the Hamiltonian function $\mathcal{H}$ with $e=0$ plotted on the sphere. The $z$-axis is along the spin axis of the planet. The $x$-axis is along the intersection of the equatorial and ecliptic planes (i.e. the line joining both equinoxes of the planet) and directed towards the ascending node of the star. The equator $xy$ plane and the ecliptic plane are highlighted by the two outer grey circles. A point on the sphere represents the tip of the orbital angular momentum of the satellite, which has coordinates $(x,y,z)=(\sin I_\mathrm{Q}\sin\delta_\mathrm{Q},-\sin I_\mathrm{Q}\cos\delta_\mathrm{Q},\cos I_\mathrm{Q})$. The colour code is the same as in Fig.~\ref{fig:phase}. Panel~\textbf{a}: same parameters as Fig.~\ref{fig:phase}. Panel~\textbf{b}: parameter region S$_1$, defined by $a/r_\mathrm{M}=1$ and $\varepsilon=90^\circ$. The magenta curve $\mathscr{L}_{13}$ is made of an infinity of stable equilibria resulting from the merging of P$_1$ with the separatrix emerging from P$_3$. Panel~\textbf{c}: parameter region S$_2$, defined by $a/r_\mathrm{M}>0$ and  $\varepsilon=0^\circ$ (or $180^\circ$). The brown curve $\mathscr{L}_{23}$ is made of an infinity of stable equilibria resulting from the merging of P$_2$ with the separatrix emerging from P$_3$.}
      \label{fig:phase3D}
   \end{figure*}

   Figure~\ref{fig:phase} shows examples of trajectories for the satellites obtained by plotting the level curves of $\mathcal{H}$ for $e=0$. The dynamics of the satellite is described by the direction of its orbital angular momentum; since the dynamics actually lie on a sphere, any planar representation of the trajectories has coordinate singularities. In Fig.~\ref{fig:phase3D}a, we show the same phase portrait as Fig.~\ref{fig:phase} on the sphere. The system being secular, it is independent of whether the orbits and spins are prograde or retrograde. This is traduced by the invariance of the phase space to the transformations $(\delta_\mathrm{Q},I_\mathrm{Q})\rightarrow(\pi+\delta_\mathrm{Q},\pi-I_\mathrm{Q})$ and $(\varepsilon,I_\mathrm{Q})\rightarrow(\pi-\varepsilon,\pi-I_\mathrm{Q})$. Three kinds of equilibrium points can be seen, which we label P$_1$, P$_2$, and P$_3$. In the work of \cite{Tremaine-etal_2009}, the points P$_1$ and P$_3$ are called `circular coplanar Laplace equilibria', and the point P$_2$ is called `circular orthogonal Laplace equilibrium'. This denomination clearly reflects the geometry of these configurations. For the sake of succinctness, we call them `Laplace states' $1$, $2$, and~$3$ (in reference to the famous `Cassini states' described below).
   
\begin{figure}[h!]
      \includegraphics[width=0.95\columnwidth]{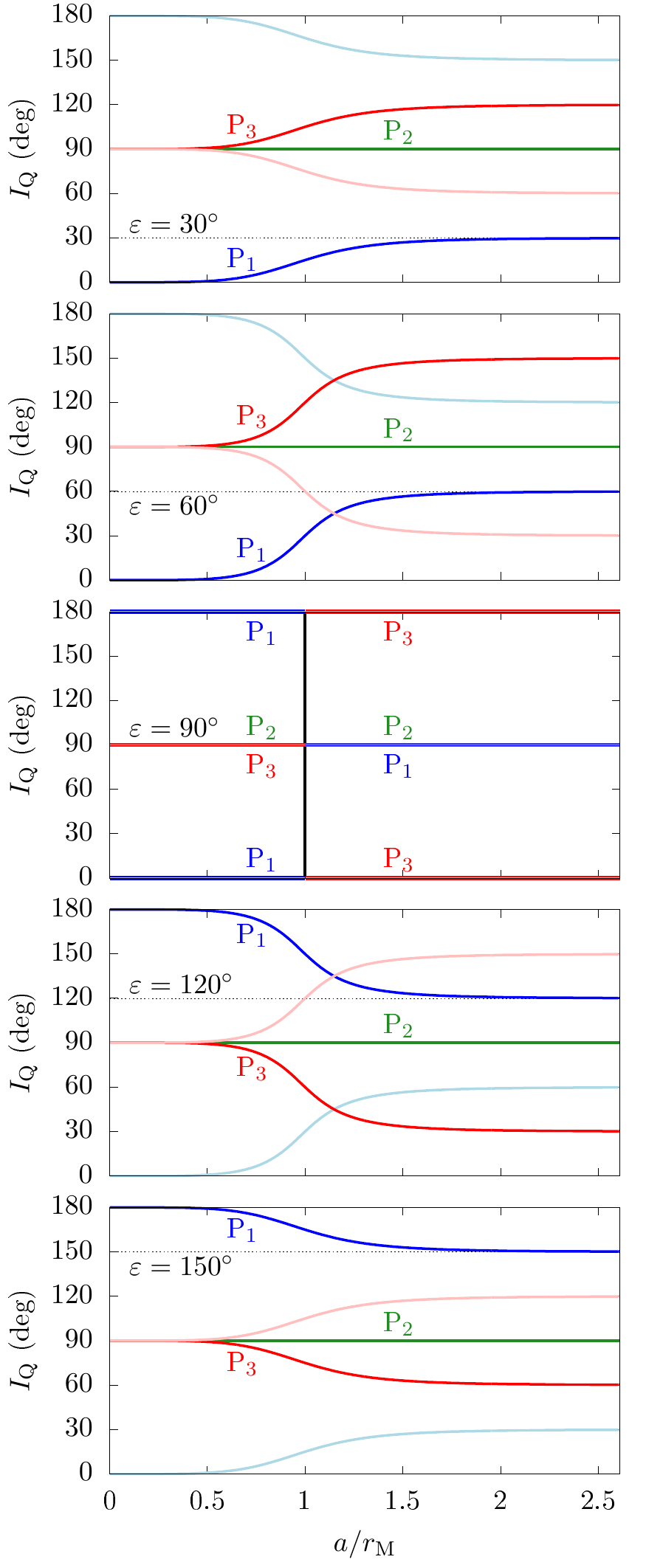}
      \caption{Location of the Laplace states as a function of $a/r_\mathrm{M}$. Each panel features a fixed value of $\varepsilon$ (see labels). The level $I_\mathrm{Q}=\varepsilon$ is shown by a horizontal dotted line. The colour code is the same as in Fig.~\ref{fig:phase}. The black vertical line in the central panel shows the degenerate equilibrium circle produced by the merging of P$_1$ and P$_3$.}
      \label{fig:geomcutob}
   \end{figure}

\begin{figure}
      \includegraphics[width=0.95\columnwidth]{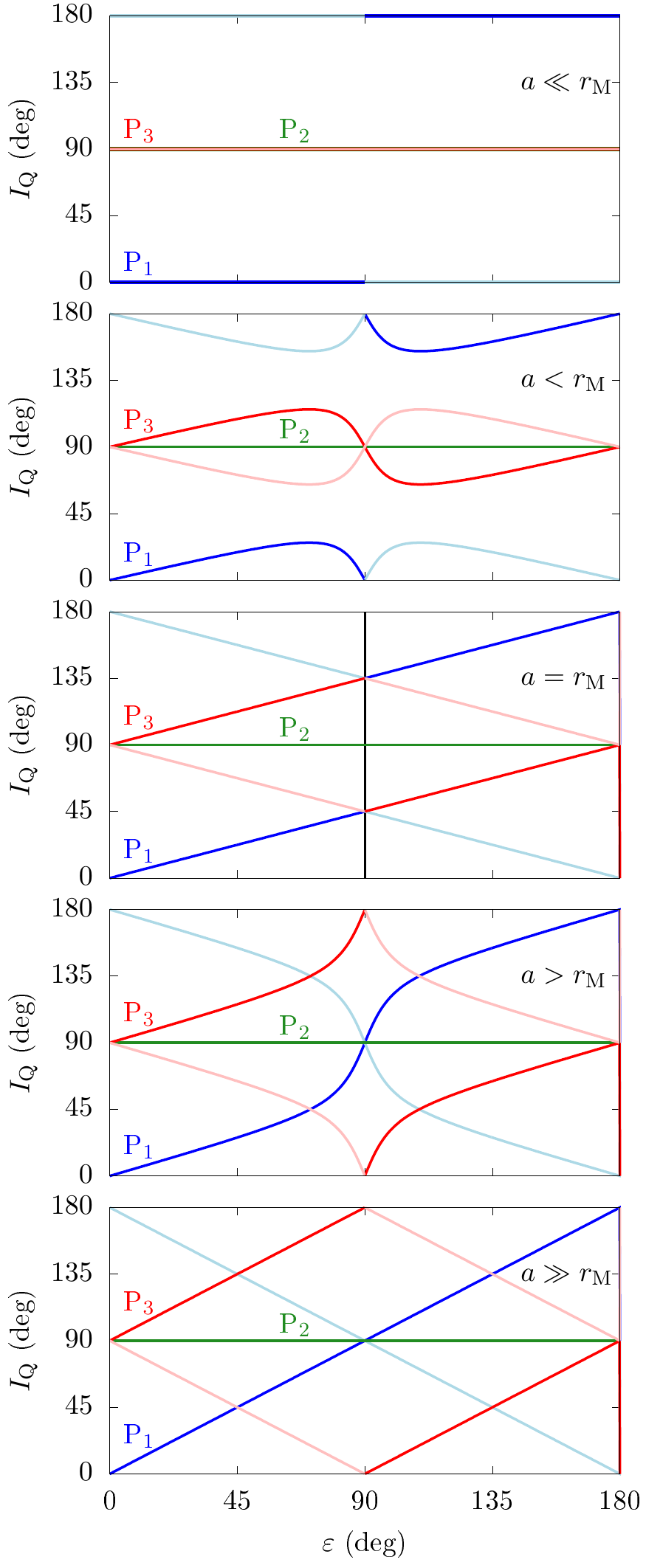}
      \caption{Location of the Laplace states as a function of $\varepsilon$. Each panel features a fixed value of $a/r_\mathrm{M}$ (from top to bottom: $0.01$, $0.95$, $1$, $1.05$, and $20$, respectively). Colours have the same meaning as in Fig.~\ref{fig:geomcutob}.}
      \label{fig:geomcuta}
   \end{figure}

   The geometry of the phase portraits for any value of the parameters can be described by the location of the equilibrium points and the shape of the separatrix. The respective locations of the Laplace states when varying the parameters are illustrated in Figs.~\ref{fig:geomcutob} and~\ref{fig:geomcuta}. Because of the symmetries mentioned above, each equilibrium point has a twin obtained by the transformation $(\delta_\mathrm{Q},I_\mathrm{Q})\rightarrow(\pi+\delta_\mathrm{Q},\pi-I_\mathrm{Q})$ that corresponds to the same Laplace state with reversed orbital motion.
   
   In the space of parameters, there is a critical point, that we call S$_1$, defined by
   \begin{equation}
      \mathrm{S}_1 = \Big\{a/r_\mathrm{M}=1,\:\varepsilon=90^\circ\Big\}\,.
   \end{equation}
   At point S$_1$, the Laplace states P$_1$ and P$_3$ and the separatrix degenerate into an equilibrium circle $\mathscr{L}_{13}$ spanning all values of inclination (see Fig.~\ref{fig:phase3D}b). All points of this circle are stable equilibrium configurations in which the linearised problem has zero eigenfrequency for inclination variations (we note it $\xi_{13}^2=0$). As shown by Figs.~\ref{fig:geomcutob} and~\ref{fig:geomcuta}, going through point S$_1$ by smoothly changing the parameters inverts the locations of P$_1$ and P$_3$. We stress that in Fig.~\ref{fig:geomcuta} the apparent jumps of P$_1$ (for $a<r_\mathrm{M}$) and P$_3$ (for $a>r_\mathrm{M}$) are only coordinate singularities in which P$_1$ or P$_3$ smoothly pass through the pole of the sphere (see Fig.~\ref{fig:phase3D}). On the contrary, the jump observed at point S$_1$ ($a=r_\mathrm{M}$) is a real singularity.
   
\begin{figure}
      \centering
      \includegraphics[width=\columnwidth]{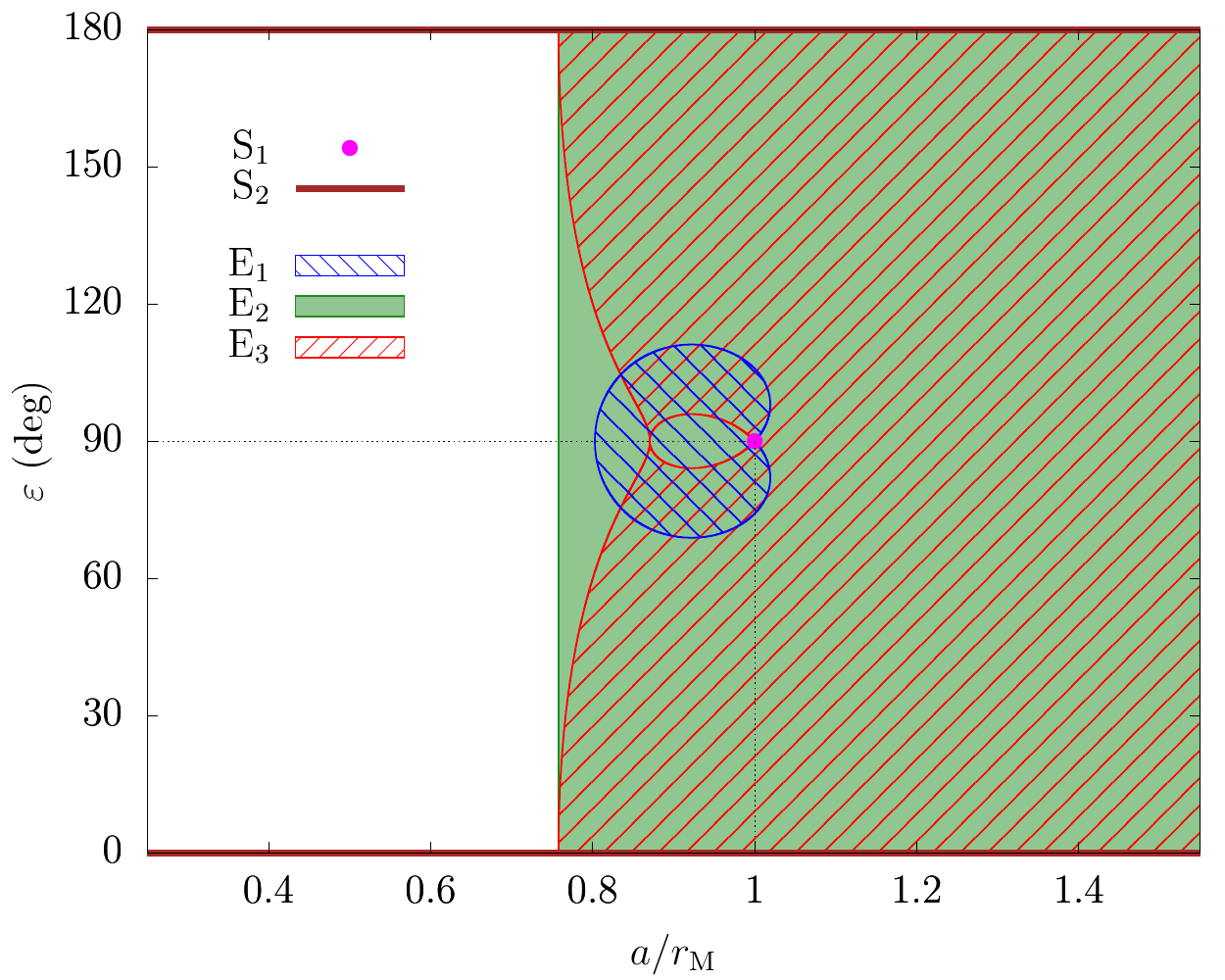}
      \caption{Summary of the regions of the parameter space governing the dynamics of a massless satellite on a nearly circular orbit. The phase space features three equilibrium points (`Laplace states'), among which P$_1$ and P$_2$ are stable to inclination variations and P$_3$ is unstable. At point S$_1$ of the parameter space, P$_1$ and P$_3$ degenerate into the equilibrium circle $\mathscr{L}_{13}$ that is stable to inclination variations. In region S$_2$ of the parameter space, P$_2$ and P$_3$ degenerate into the equilibrium circle $\mathscr{L}_{23}$ that is stable to inclination variations. In regions E$_1$, E$_2$, and E$_3$, respectively, P$_1$, P$_2$, and P$_3$ are unstable to eccentricity variations. In region $\mathrm{S}_2\cup\mathrm{E}_2$, $\mathscr{L}_{23}$ is unstable to eccentricity variations. At point S$_1$, $\mathscr{L}_{13}$ is stable to eccentricity variations provided that the equatorial inclination of the satellite verifies $\cos^2I_\mathrm{Q}\leqslant 1/5$.}
      \label{fig:recap}
   \end{figure}

   Another singularity occurs for a null or $180^\circ$ obliquity. We call S$_2$ the corresponding region of the parameter space, defined by
   \begin{equation}
      \mathrm{S}_2 = \Big\{a/r_\mathrm{M}>0,\:\varepsilon=0^\circ\;\text{or}\;180^\circ\Big\}\,.
   \end{equation}
   In region S$_2$, the Laplace states P$_2$ and P$_3$ and the separatrix degenerate into an equilibrium circle $\mathscr{L}_{23}$ spanning all values of $\delta_\mathrm{Q}$ (see Fig.~\ref{fig:phase3D}c). All points of this circle are stable equilibrium configurations in which the linearised problem has zero eigenfrequency for inclination variations  (we note it $\xi_{23}^2=0$). The regions S$_1$ and S$_2$ of the parameter space can be visualised in Fig.~\ref{fig:recap}.
   
   Apart from regions S$_1$ (in which P$_1$ becomes singular) and S$_2$ (in which P$_2$ becomes singular), the phase space keeps the same topology whatever the parameters $a/r_\mathrm{M}>0$ and $\varepsilon$. This means that the Laplace states are smoothly transported by a continuous change of parameters, and they keep their stability nature against inclination variations. On Fig.~\ref{fig:phase3D}a, such a continuous change of parameter would simply produce the rotation of the sphere around the $x$-axis and the narrowing or widening of the black separatrix. More precisely, Fig.~\ref{fig:geomcuta} shows that for $a<r_\mathrm{M}$, varying the obliquity produces an oscillation of P$_1$ around the pole and P$_3$ remains near $90^\circ$; for $a>r_\mathrm{M}$, on the contrary, varying the obliquity makes P$_1$ and P$_3$ roll all over the sphere. The opposite behaviour would be obtained by representing the ecliptic inclination $I_\mathrm{C}$ instead of $I_\mathrm{Q}$. The location and stability nature of the Laplace states play a fundamental role in the combined dynamics of the satellite's orbit and the planet's spin axis. For this reason, we review here their basic properties and go deeper than previous works in their analytic characterisation.
   
\begin{figure}
      \includegraphics[width=\columnwidth]{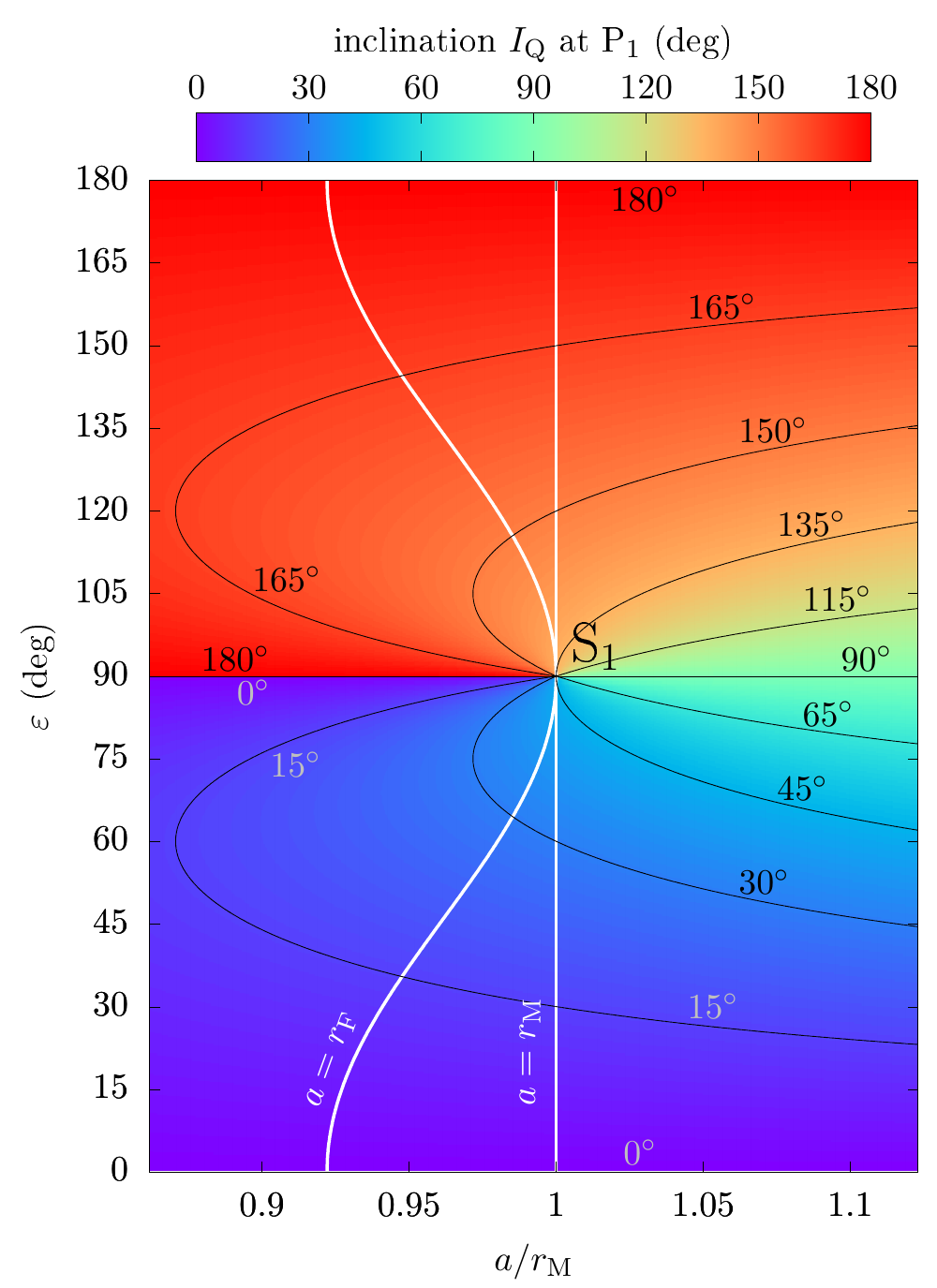}
      \caption{Inclination of the Laplace state P$_1$ as a function of the parameters (see Eq.~\ref{eq:ILap13}). Some level curves are labelled in black. The critical radii and $r_\mathrm{M}$ and $r_\mathrm{F}$ are represented in white. At $a=r_\mathrm{M}$, P$_1$ lies exactly halfway between the equator and the ecliptic (i.e. $I_\mathrm{Q}=\varepsilon/2$ for a prograde spin). The abrupt transition of $I_\mathrm{Q}$ from $0^\circ$ to $180^\circ$ at $\varepsilon=90^\circ$ is a coordinate singularity. The S$_1$ point is a real singularity (see text).}
      \label{fig:P1detail}
   \end{figure}

   P$_1$ is stable to inclination variations. As illustrated by Fig.~\ref{fig:geomcutob}, it corresponds to an orbit lying on the equator plane for close-in satellites ($a\ll r_\mathrm{M}$), and on the ecliptic plane for far-away satellites ($a\gg r_\mathrm{M}$). In between, P$_1$ corresponds to an intermediate tilt between the equator and the ecliptic. As a result of eccentricity and inclination damping, P$_1$ is expected to be the birth place of regular satellites formed in a circumplanetary disc. P$_1$ is therefore particularly important in satellite dynamics studies; for this reason, it is called `classical Laplace equilibrium' by \cite{Tremaine-etal_2009}. For $\delta_\mathrm{Q}=0$ (dark blue colour in the figures), the inclination of P$_1$ is given by one of the two solutions of the equation\footnote{There seems to be a typographical error in Eqs.~(22) and (23) of \cite{Tremaine-etal_2009}: for both equations, the first equality is correct but not the second one. We give the correct expression in Eq.~\eqref{eq:ILap}.}
   \begin{equation}\label{eq:ILap}
      \tan(2I_\mathrm{Q}) = \frac{\sin(2\varepsilon)}{\cos(2\varepsilon) + r_\mathrm{M}^5/a^5}\,,
   \end{equation}
   the second solution being the inclination of P$_3$. We note them $I_{\mathrm{Q}1}$ and $I_{\mathrm{Q}3}$. Their closed form expressions can be written
   \begin{equation}\label{eq:ILap13}
      \begin{aligned}
         I_{\mathrm{Q}1} &= \frac{\pi}{2} + \frac{1}{2}\mathrm{atan2}\big[-\sin(2\varepsilon),-u-\cos(2\varepsilon)\big]\,, \\
         I_{\mathrm{Q}3} &= \frac{\pi}{2} + \frac{1}{2}\mathrm{atan2}\big[\sin(2\varepsilon),u+\cos(2\varepsilon)\big]\,, \\
      \end{aligned}
   \end{equation}
   where $u\equiv r_\mathrm{M}^5/a^5$. At $a=r_\mathrm{M}$, the Laplace state P$_1$ lies exactly halfway between the equator and the ecliptic planes (i.e. $I_{\mathrm{Q}1}=\varepsilon/2$ for $\varepsilon<90^\circ$). This is why we use the index M, for `midpoint', introduced in Eq.~\eqref{eq:rM}. Interestingly, this midpoint does not depend on the value of the obliquity $\varepsilon$, but only on the distance of the satellite. The curve described by Eq.~\eqref{eq:ILap} and illustrated in Fig.~\ref{fig:geomcutob}, however, is not exactly symmetric with respect to $a=r_\mathrm{M}$. Its inflexion point F (for `flex') is reached at radius $r_\mathrm{F}(\varepsilon)$, defined by
   \begin{equation}\label{eq:rF}
      r_\mathrm{F}^5 = \frac{\sqrt{\cos^2(2\varepsilon) + 24} - \cos(2\varepsilon)}{6}\,r_\mathrm{M}^5
   \end{equation}
   and illustrated in Fig.~\ref{fig:P1detail}. The distance between $r_\mathrm{F}$ and $r_\mathrm{M}$ is a way to quantify the asymmetry of $I_{\mathrm{Q}1}$ as a function of $a$. We stress that all level curves in Fig.~\ref{fig:P1detail} converge at S$_1$. Through a smooth variation of parameters, the satellite can therefore reach the singular point $\mathrm{S}_1$ from any orbital inclination between $0^\circ$ and $180^\circ$. This property has important consequences for the spin-axis dynamics of the host planet, as we discuss in Sect.~\ref{sec:plspin}. We also show that $r_\mathrm{F}$ divides the close-in and far-away satellite regimes considered in previous works.
   
\begin{figure}
      \includegraphics[width=\columnwidth]{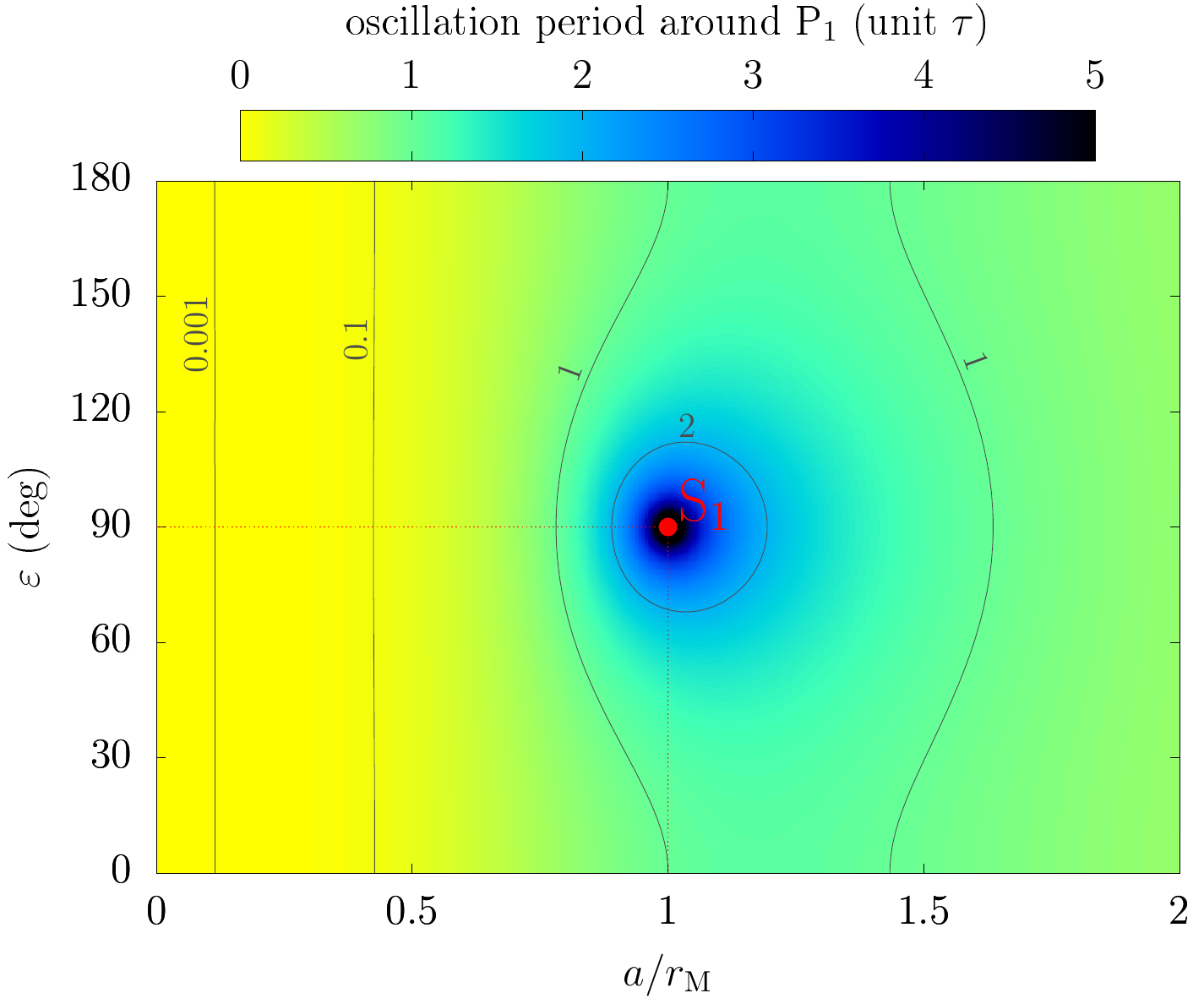}
      \caption{Period of small inclination oscillations around the Laplace state P$_1$ as a function of the parameters. The period is given by Eq.~\eqref{eq:nu1} and represented here in unit of the characteristic period $\tau=2\pi/\kappa$. Some level curves are highlighted and labelled in black. The singular point S$_1$ is indicated in red.}
      \label{fig:libT1}
   \end{figure}

   \cite{Tremaine-etal_2009} give a compact expression for the frequency of small-amplitude oscillations around P$_1$, which can be written as
   \begin{equation}\label{eq:nu1}
      \xi_1^2 = -\frac{\kappa^2}{4}\left(\frac{a}{r_\mathrm{M}}\right)^3\frac{\cos\varepsilon\sin^2\varepsilon}{\cos I_{\mathrm{Q}1}\sin^2I_{\mathrm{Q}1}}\cos(\varepsilon-I_{\mathrm{Q}1})\,,
   \end{equation}
   where $I_{\mathrm{Q}1}$ is the equatorial inclination at P$_1$ given in Eq.~\eqref{eq:ILap13}. A negative value of $\xi_1^2$ means that the equilibrium point is stable. As expected, $\xi_1^2$ is negative all over the parameter space. For a zero-obliquity planet, Eq.~\eqref{eq:nu1} simplifies to
   \begin{equation}\label{eq:nu10}
      \xi_1^2\Big|_{\varepsilon=0} = -\frac{\kappa^2}{4}\left(\frac{r_\mathrm{M}^5}{a^5} + 1\right)^2\frac{a^3}{r_\mathrm{M}^3}\,.
   \end{equation}
   It shows that the timescale parameter $\kappa$ defined in Eq.~\eqref{eq:kappa} is the oscillation frequency around P$_1$ for $\varepsilon=0$ and at a radius $a=r_\mathrm{M}$. See Appendix~\ref{asec:circeq} for the limit value of $\xi_1$ in the regions of parameter space where Eq.~\eqref{eq:nu1} looks undefined. As a summary, Fig.~\ref{fig:libT1} shows the libration period around P$_1$ in the whole parameter space. The stability properties of $\mathrm{P}_2$ and $\mathrm{P}_3$ are not crucial for the dynamics of a regular satellite, but they can play a role if the satellite becomes unstable during its orbital migration (see Sect.~\ref{sec:sattit}). For this reason, a brief description of $\mathrm{P}_2$ and $\mathrm{P}_3$ is provided in Appendix~\ref{asec:circeq}.
   
   From P$_3$ emerges the separatrix that divides the regions of oscillations around P$_1$ and around P$_2$. Noting $u=r_\mathrm{M}^5/a^5$, the extent of the separatrix can be expressed as
   \begin{equation}\label{eq:widthP2}
      \cos^2I_\mathrm{Q} = \frac{1+u-\sqrt{1+u^2+2u\cos(2\varepsilon)}}{2u}\,,
   \end{equation}
   in which the two solutions are the minimum and maximum value of $I_\mathrm{Q}$ along the separatrix  (see Fig.~\ref{fig:phase}). From Eq.~\eqref{eq:widthP2}, we deduce that the width of the island surrounding P$_2$ is zero in parameter region S$_2$ (as expected from Fig.~\ref{fig:phase3D}), and that it increases for growing $a$ and for decreasing $\cos^2\varepsilon$. This has important consequences for the emergence of chaos discussed in Sect.~\ref{sec:sattit}. At the singular point $\mathrm{S}_1$, the island covers the whole sphere.
   
   Apart from the singular regions S$_1$ and S$_2$, the continuous behaviour of the Laplace states all over the parameter space is crucial for the long-term satellite dynamics, because if some physical mechanism induces a slow change of parameters (e.g. if the satellite migrates, or if the planet's spin axis is gradually tilted), then the satellite would adiabatically follow the equilibrium point around which it oscillates, while conserving the phase space area $J$ spanned by its trajectory. If the system never transits through point S$_1$ (for oscillations around P$_1$) or S$_2$ (for oscillations around P$_2$), then this adiabatic drift can go on as long as the phase space area delimited by the separatrix is wide enough to contain $J$. This last condition is always verified if $J=0$, that is, if the satellite lies exactly on a stable Laplace state.
   
   \subsection{Stability to variations in eccentricity}\label{ssec:eccstab}
   Up to now, we assumed that the eccentricity $e$ of the satellite is zero, which is an equilibrium point. Since the linearised system in the vicinity of $e=0$ produces a decoupling between eccentricity and inclination variations, the previous analysis is valid up to order $\mathcal{O}(e)$ and it neglects $\mathcal{O}(e^2)$. Since the condition $e=0$ for a real satellite is never exactly verified, we must consider the stability of the Laplace states to eccentricity variations, that is, we must determine whether a small non-zero offset of eccentricity remains small or grows big over time. As before, we focus on the Laplace state $\mathrm{P}_1$; the description of P$_2$, P$_3$, and the degenerate circles $\mathscr{L}_{13}$ and  $\mathscr{L}_{23}$ are provided in Appendix~\ref{asec:circeq}.
   
\begin{figure}
      \centering
      \includegraphics[width=0.9\columnwidth]{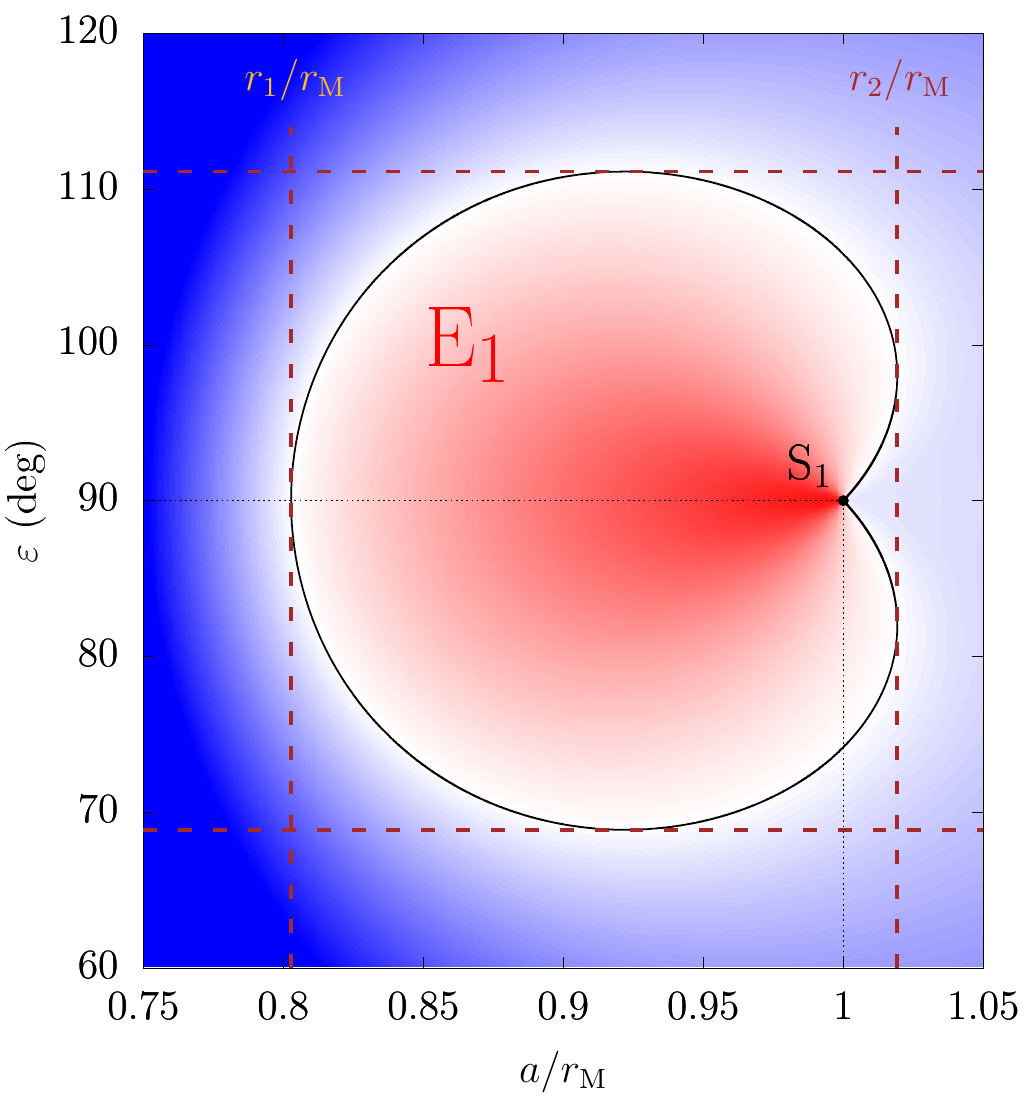}
      \caption{Region E$_1$ of the parameter space where the Laplace state P$_1$ is unstable to eccentricity variations. The background colour shows the normalised value of $\mu_1^2$, with blue for negative values (stable) and red for positive values (unstable). The border of E$_1$ is highlighted with a black contour, obtained using the closed-form expression in Eqs.~\eqref{eq:E1a} and~\eqref{eq:E1b}. Region E$_1$ is entirely contained inside the dashed brown lines, whose left and right limits $r_1$ and $r_2$ are given in Eq.~\eqref{eq:r1r2}, and whose bottom and top limits are given by Eq.~\eqref{eq:emin}. The Laplace state P$_1$ is singular at point S$_1$, in which $\mu_1^2$ is discontinuous.}
      \label{fig:E1detail}
   \end{figure}

   The eigenvalues of the eccentricity linear sub-system inform us about their stability against eccentricity growth. \cite{Tremaine-etal_2009} give a compact expression for the frequency of small-amplitude eccentricity oscillations around P$_1$. It can be written
   \begin{equation}\label{eq:mu1}
      \begin{aligned}
         \mu_1^2 = -\frac{\kappa^2\left(a/r_\mathrm{M}\right)^3}{512\sin^2(2I_{\mathrm{Q}1})}\Bigg[
      -106 + 24\cos(2I_{\mathrm{Q}1}) + 146\cos(4I_{\mathrm{Q}1})& \\
      - 100\cos(6I_{\mathrm{Q}1}-2\varepsilon) - 24\cos(2I_{\mathrm{Q}1}-4\varepsilon)& \\
      + 224\cos(2I_{\mathrm{Q}1}-2\varepsilon) - 54\cos(4I_{\mathrm{Q}1}-4\varepsilon)& \\
      - 8\cos(2\varepsilon) - 11\cos(4\varepsilon) - 124\cos(2I_{\mathrm{Q}1}+2\varepsilon)& \\
      + 25\cos(8I_{\mathrm{Q}1}-4\varepsilon) + 8\cos(4I_{\mathrm{Q}1}-2\varepsilon)
      \Bigg]\,,&
      \end{aligned}
   \end{equation}
   where $I_{\mathrm{Q}1}$ is the equatorial inclination at P$_1$ given by Eq.~\eqref{eq:ILap13}. See Appendix~\ref{asec:circeq} for the limit value of $\mu_1$ in the regions of parameter space where Eq.~\eqref{eq:mu1} looks undefined. A negative value of $\mu_1^2$ means that the equilibrium point is stable to eccentricity variations. As noted by \cite{Tremaine-etal_2009}, P$_1$ is stable in all the parameter space except in a small closed region resembling a cardioid. We call E$_1$ this region of the parameter space; it can be visualised in Figs.~\ref{fig:recap} and~\ref{fig:E1detail}.
   
   The boundary of E$_1$ is given by the roots of $\mu_1^2$, which have a closed-form analytical expression. We first define two critical radii $r_1$ and $r_2$ as
   \begin{equation}\label{eq:r1r2}
      r_1^5 = \frac{1}{3}r_\mathrm{M}^5
      \quad\text{and}\quad
      r_2^5 = \frac{10\sqrt{22}-4}{39}r_\mathrm{M}^5 \,.
   \end{equation}
   As shown in Fig.~\ref{fig:E1detail}, the radii $r_1$ and $r_2$ mark the leftmost and rightmost limits of $\mathrm{E}_1$, and the boundary of $\mathrm{E}_1$ has a cusp at the singular point S$_1$. Noting $u=r_\mathrm{M}^5/a^5$, the boundary of E$_1$ can be expressed piecewise as
   \begin{equation}\label{eq:E1a}
      \begin{aligned}
         \cos^2\varepsilon &= \frac{-32u^2+113u-72 + (4-u)\sqrt{56u^2-8u-39}}{242u}
      \end{aligned}
   \end{equation}
   for $r_1 \leqslant a\leqslant r_2$, and
   \begin{equation}\label{eq:E1b}
      \begin{aligned}
         \cos^2\varepsilon &= \frac{-32u^2+113u-72 - (4-u)\sqrt{56u^2-8u-39}}{242u}
      \end{aligned}
   \end{equation}
   for $r_\mathrm{M} \leqslant a\leqslant r_2$. Equation~\eqref{eq:E1b} corresponds to the cusp portion of the curve, and the two portions meet at $a=r_2$ (see Fig.~\ref{fig:E1detail}). The obliquity $\varepsilon\approx 68.875^\circ$ quoted by \cite{Tremaine-etal_2009} as the minimum value where P$_1$ can be unstable is reached at $a^5/r_\mathrm{M}^5 = 2/3$. It has actually the following closed-form:
   \begin{equation}\label{eq:emin}
      \cos^2\varepsilon = \frac{51+25\sqrt{3}}{726}\,.
   \end{equation}
   Interestingly, $\mu_1^2$ does not go to zero at the singular point $\mathrm{S}_1$, but is discontinuous (see Appendix~\ref{asec:circeq}). Moreover, the value of $\mu_1^2$ at $\varepsilon=90^\circ$ and $a\rightarrow r_\mathrm{M}^-$ is the largest (positive) value that $\mu_1^2$ can ever reach in the whole parameter space: it is therefore the most unstable location of P$_1$ to eccentricity variations. This explains the numerical results of \cite{Tamayo-etal_2013}, who note that for Uranus, whose obliquity is not far from $90^\circ$, the radius $r_\mathrm{M}$ is the approximate location at which the eccentricity grows most rapidly. This also explains why they find that the instability is more violent if the satellite reaches E$_1$ while migrating inwards rather than outwards (see Fig.~\ref{fig:E1detail}).
   
   The stability properties of $\mathrm{P}_2$ and $\mathrm{P}_3$ to eccentricity variations are given in Appendix~\ref{asec:circeq}. We show that they are unstable in the regions $\mathrm{E}_2$ and $\mathrm{E}_3$, respectively, illustrated in Fig.~\ref{fig:recap}. We note that E$_2$ entirely contains the region E$_1$; hence, in region E$_1$ all Laplace states are unstable to at least eccentricity or inclination variations.
   
   \subsection{Eccentric Laplace states}\label{ssec:eccLap}
   Along the boundaries of the regions E$_1$ and E$_2$, where the Laplace states P$_1$ and P$_2$ become unstable to eccentricity variations, \cite{Tremaine-etal_2009} show that they both bifurcate into equilibrium configurations with an eccentric orbit. We call these configurations P$_1'$ and P$_2'$. Likewise, we show in Appendix~\ref{asec:ecceq} that along the two boundaries of the E$_3$ region (the V-shaped boundary for $a<r_\mathrm{L}$ and the drop-like boundary for $a>r_\mathrm{L}$; see Fig.~\ref{fig:recap} and Appendix~\ref{asec:circeq}), the Laplace state P$_3$ bifurcates into two eccentric equilibria that we call P$_3'$ and P$_3''$, respectively.
   
   In the space of parameters, there exist stable regions for all of these eccentric equilibria. Therefore, satellites reaching the unstable regions E$_1$, E$_2$, or E$_3$ via a smooth parameter change are not bound to destabilise; they can instead bifurcate to a stable eccentric configuration. The properties of the eccentric equilibria are recalled in Appendix~\ref{asec:ecceq}; we provide formulas that can be used to easily compute their locations as a function of the parameters. In our case, we are mostly interested in the equilibrium P$_1'$, because it bifurcates from the classic Laplace state P$_1$ in which regular satellites are formed.
   
   At equilibrium P$_1'$, the orbital angles of the satellite are $\omega_\mathrm{Q}=\pi/2\mod\pi$ and $\delta_\mathrm{Q}=0$ (or $\delta_\mathrm{Q}=\pi$ for the twin equilibrium with reversed orbital motion). The equatorial inclination of the satellite at P$_1'$ can be written as
   \begin{equation}\label{eq:IQ1p}
   I_{\mathrm{Q}1}' = \frac{\pi}{2} + \frac{1}{2}\mathrm{atan2}\big[-\sin(2\varepsilon),-v-\cos(2\varepsilon)\big]\,,
   \end{equation}
   where we define $v$ as
   \begin{equation}\label{eq:uprime}
   v = \frac{r_\mathrm{M}^5}{a^5}\frac{1}{(1-e^2)^{3/2}(1 + 4e^2)}\,,
   \end{equation}
   in which $e$ is the satellite's eccentricity at equilibrium. We note that the inclination $I_{\mathrm{Q}1}'$ in Eq.~\eqref{eq:IQ1p} has the same form as $I_{\mathrm{Q}1}$ in Eq.~\eqref{eq:ILap13}, but where $u=r_\mathrm{M}^5/a^5$ is replaced by $v$. The behaviour of $I_{\mathrm{Q}1}'$ as a function of the parameters is therefore very similar to $I_{\mathrm{Q}1}$ except that the non-zero eccentricity of the satellite acts like a modified orbital distance. For $e=0$, the definitions of $I_{\mathrm{Q}1}'$ and $I_{\mathrm{Q}1}$ coincide. As shown in Appendix~\ref{asec:ecceq}, the eccentricity at equilibrium can be computed in the general case as a three-dimensional surface with an explicit parametric representation.
   
\begin{figure}
      \centering
      \includegraphics[width=\columnwidth]{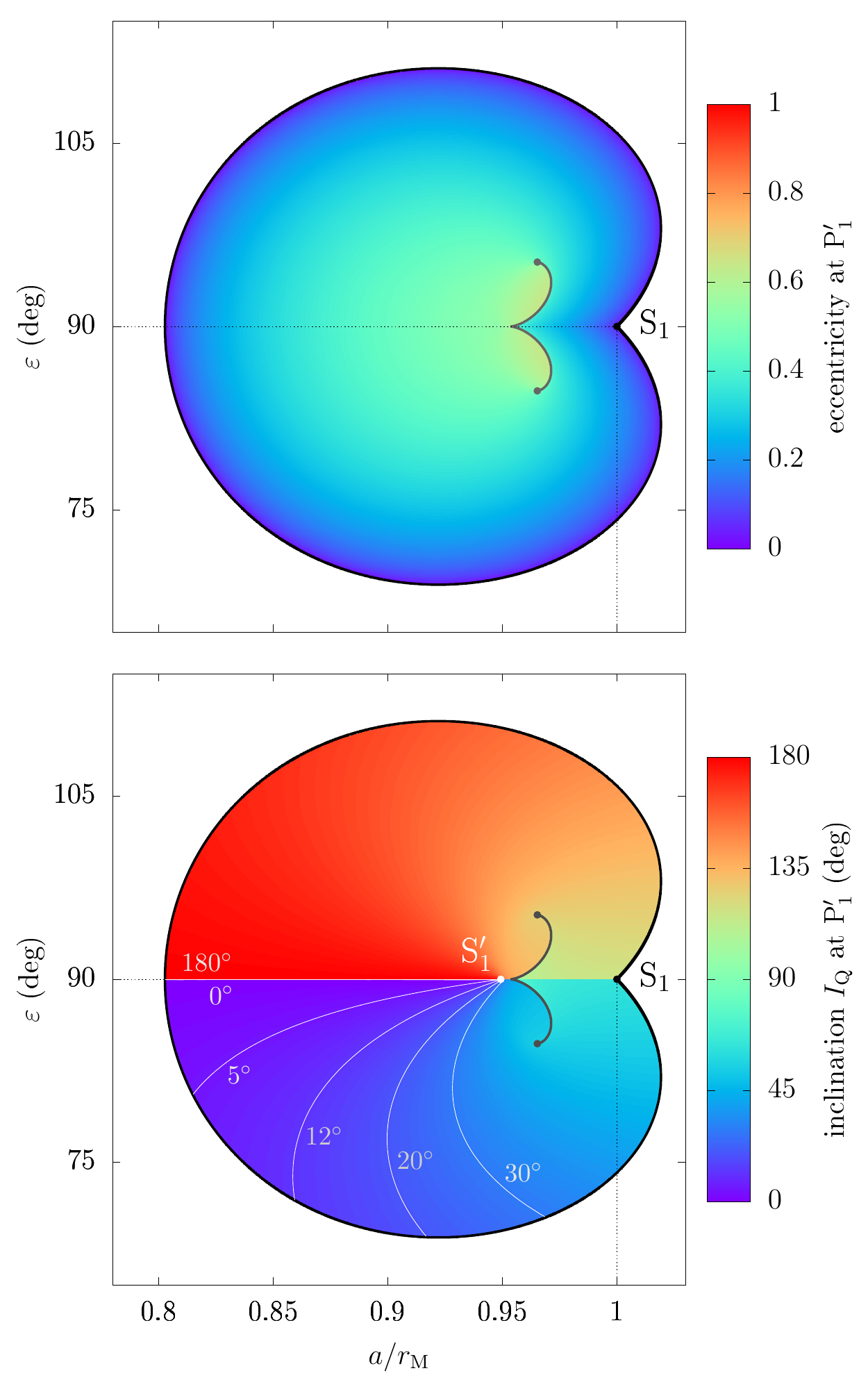}
      \caption{Eccentricity and inclination of the satellite at the eccentric equilibrium P$_1'$. The three-dimensional surface of equilibrium has been cut along the grey line, as shown in Fig.~\ref{fig:P1prime3D}. In the bottom panel, some level curves are plotted in white (see labels). Noting $u=r_\mathrm{M}^5/a^5$, the extremity of the grey cutting line have coordinates $u=27\sqrt{2}/32$ and $\cos^2\varepsilon=(5-2\sqrt{6})/12$ at the top and bottom points, and $u=25\sqrt{30}/108$ and $\varepsilon=90^\circ$ at the middle.}
      \label{fig:P1prime}
   \end{figure}

\begin{figure}
      \centering
      \includegraphics[width=\columnwidth]{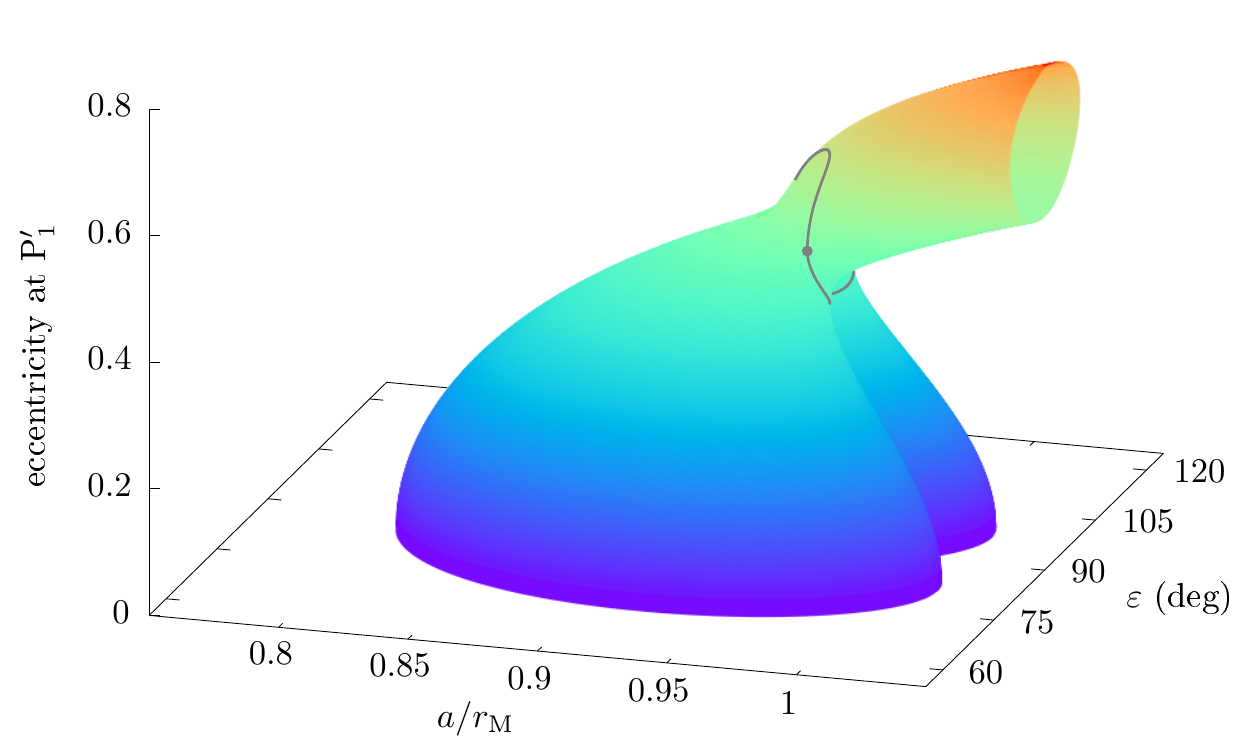}
      \caption{Eccentricity at the eccentric equilibrium P$_1'$ seen as a three-dimensional surface. The colour is the same as in Fig.~\ref{fig:P1prime}, top panel. In Fig.~\ref{fig:P1prime}, the top tube-like portion of the surface has been cut off along the grey line. As detailed in Appendix~\ref{asec:ecceq}, the top portion extends up to $a\rightarrow\infty$, where it tends to $e=1$. Sections of this surface can be seen in Fig.~6 of \cite{Tremaine-etal_2009}.}
      \label{fig:P1prime3D}
   \end{figure}

   The eccentricity and inclination of the satellite at equilibrium P$_1'$ are shown in Fig.~\ref{fig:P1prime} as a function of the parameters. We recognise the cardioid-like boundary of the E$_1$ region. Since Fig.~\ref{fig:P1prime} is the projection of a complex three-dimensional surface, a portion of this surface has been cut off for the purpose of the figure. The removed portion of the surface connects to the grey line near the centre of the figure (see the colour discontinuity), and it can be visualised in Fig.~\ref{fig:P1prime3D}. Along the cutting line, the right portion of the three-dimensional surface turns round to higher semi-major axes again.
   
   Along the three-dimensional curve $\mathbb{S}_1$ defined by
   \begin{equation}\label{eq:Ssing3D}
   \mathbb{S}_1 = \left\{\frac{a^5}{r_\mathrm{M}^5}=\frac{1}{(1-e^2)^{3/2}(1+4e^2)},\:\varepsilon=90^\circ\right\}\,,
   \end{equation}
   the inclination $I_{\mathrm{Q}1}'$ of the satellite given in Eq.~\eqref{eq:IQ1p} is undefined. This is a real singularity, where P$_1'$ does not exist. Indeed, the curve $\mathbb{S}_1$ is the eccentric continuation of the singular point S$_1$, at which P$_1$ and P$_3$ are degenerate. In Fig.~\ref{fig:P1prime}, the curve $\mathbb{S}_1$ is visible between the points labelled S$_1$ and S$_1'$. Along this line, the orbital inclination $I_{\mathrm{Q}1}'$ has two different limits (different from $0^\circ$ and $180^\circ$) according to whether the system tends to $\varepsilon=90^\circ$ from below or from above. As shown in Appendix~\ref{asec:ecceq}, the point S$_1'$ is the location where $\mathbb{S}_1$ pierces the three-dimensional surface of equilibrium. Noting $u=r_\mathrm{M}^5/a^5$, the point S$_1'$ has coordinates $u=75\sqrt{35}/343$ and $\varepsilon=90^\circ$. By comparing Figs.~\ref{fig:P1prime} and \ref{fig:P1detail}, we see that S$_1'$ can be seen as the eccentric counterpart of S$_1$, where inclination level curves converge. This property will be important in Sect.~\ref{sec:plspin}.
   
\begin{figure}
      \centering
      \includegraphics[width=\columnwidth]{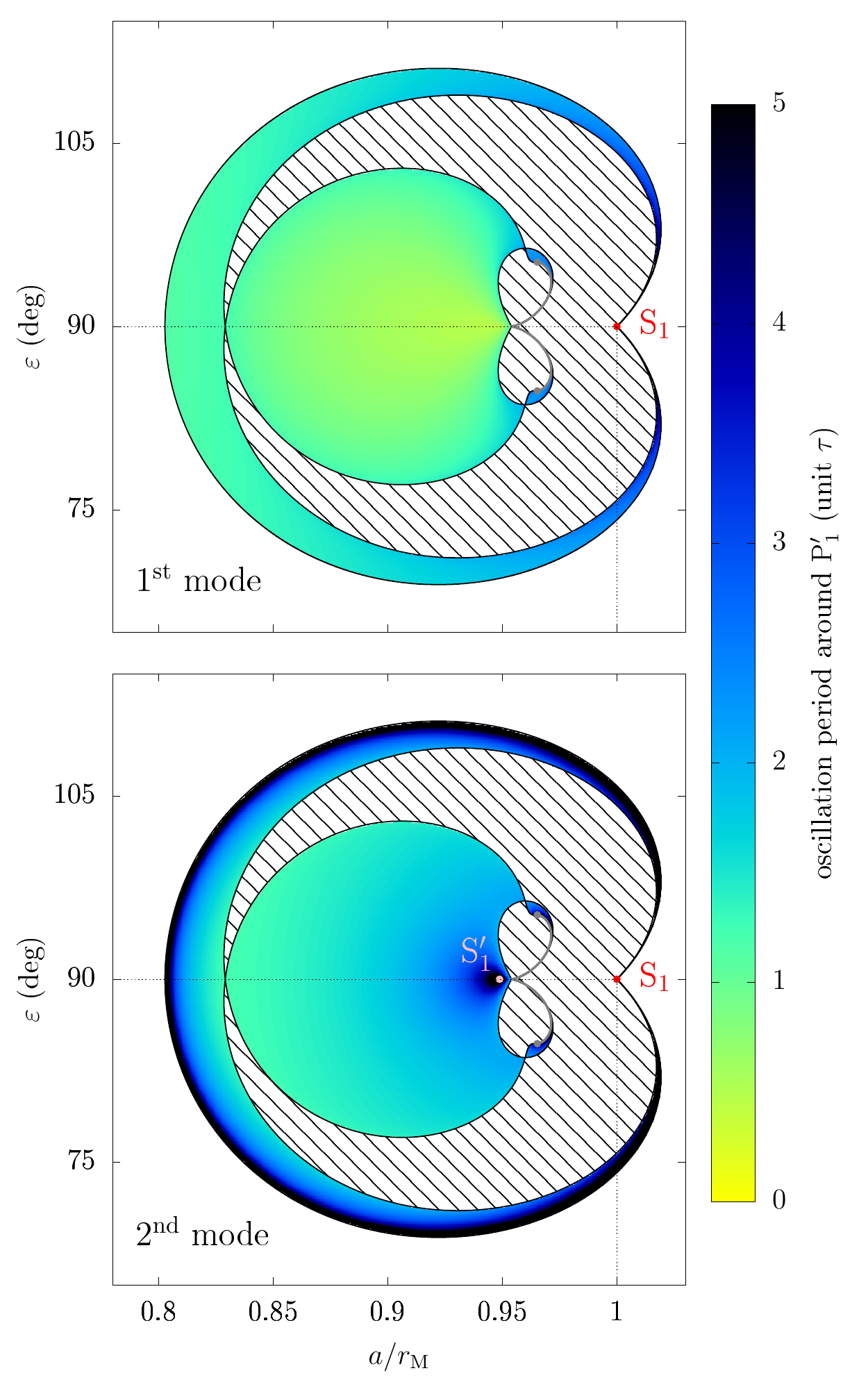}
      \caption{Period of small oscillations around P$_1'$ as a function of the parameters. There are two oscillation modes, shown in the two panels. In the hatched area, the equilibrium is unstable. The singular points S$_1$ and S$_1'$ are labelled, and the cutting line of the three-dimensional surface is shown by a grey line (same as Fig.~\ref{fig:P1prime}). Three-dimensional figures are provided in Appendix~\ref{asec:ecceq}, where the geometry of the stable regions is clearer.}
      \label{fig:P1primeStab}
   \end{figure}

   Contrary to the circular case, the eccentricity and inclination degrees of freedom are fully coupled at P$_1'$. Therefore, in the vicinity of P$_1'$, the eccentricity and inclination of the satellite both vary according to two distinct eigenfrequencies (plus their opposite). The periods of these two oscillation modes in the stable regions are shown in Fig.~\ref{fig:P1primeStab}. By comparing with Fig.~\ref{fig:libT1}, we see that the oscillation timescale near P$_1'$ has the same order of magnitude as in the circular case. Moreover, we note that one frequency tends to zero at point S$_1'$, in the same way as $\xi_1^2$ tends to zero at S$_1$. Along the boundary of the E$_1$ region, the two eigenfrequencies tend to the oscillation frequencies $\xi_1$ and $\mu_1$ around P$_1$ given at Eqs.~\eqref{eq:nu1} and \eqref{eq:mu1}, confirming that the eccentric equilibrium P$_1'$ bifurcates from the circular equilibrium P$_1$.
   
   As stressed by \cite{Tremaine-etal_2009}, the eccentric equilibrium P$_1'$ is stable near its bifurcation from P$_1$ and in the central region of Fig.~\ref{fig:P1primeStab}. These properties will be important for the future evolution of Titan described in Sect.~\ref{sec:sattit}. On the top tube-like portion of the equilibrium surface (not shown in Fig.~\ref{fig:P1primeStab}), we show in Appendix~\ref{asec:ecceq} that P$_1'$ is mostly unstable, even though a small stable region exists at very high eccentricities.
   
   Before concluding this section, we stress that the linear instability of a Laplace state does not necessarily mean that the satellite's trajectory is chaotic, and it gives no information about the amount of eccentricity and inclination increase suffered by the satellite. Interestingly, the simulations of \cite{Tremaine-etal_2009}, \cite{Tamayo-etal_2013}, and \cite{Speedie-etal_2020} reveal more chaos than expected in the E$_1$ region, even where the eccentric equilibrium P$_1'$ should theoretically be stable. In the case of trans-Neptunian objects perturbed by the galactic tides (see Table~\ref{tab:param}), \cite{Saillenfest-etal_2019} find that at $a\approx r_\mathrm{M}$ the phase space is covered by chaos, allowing for transitions between circular and quasi-parabolic orbits. The emergence of violent chaos in the orbit of Titan is confirmed numerically in Sect.~\ref{sec:sattit}. But before speaking of chaos, we must first understand the mechanism through which Titan is brought into the unstable region. In the next section, we see that it results from an interplay between the dynamics of Titan's orbit and Saturn's spin axis.
   
\section{Spin-axis dynamics of the planet}\label{sec:plspin}

In the previous section, the spin axis of the host planet was assumed to be fixed in an inertial frame. Actually, because of the torque applied by the star and the satellite on its equatorial bulge, the spin axis of the planet is made to slowly precess over time. In this section, we aim to get a qualitative understanding of the effect of the satellite on the spin-axis motion of its host planet, with an eye on the case where the planet is locked in a secular spin-orbit resonance, that is, where additional perturbations maintain the planet's spin-axis precession frequency to a fixed value.

A self-consistent model for the dynamics of a satellite and the spin axis of its host planet has been derived by \cite{Boue-Laskar_2006}: under the assumption that the satellite's argument of pericentre stably circulates, they obtained a full analytical characterisation of the averaged dynamics, which was proven to be integrable. However, this model does not hold if the system is affected by additional perturbations. In particular, mutual interactions between planets result in their nodal and apsidal precession motions (see e.g. \citealp{Murray-Dermott_1999}), whose multiple modes and harmonics are responsible for the secular spin-orbit resonances. Besides, the assumptions of \cite{Boue-Laskar_2006} cannot apply if the system reaches the region $\mathrm{E}_1$, as the satellite's pericentre can become stationary near the equilibrium $\mathrm{P}_1'$ (see Sect.~\ref{sec:orb}). Consequently, the model of \cite{Boue-Laskar_2006} will serve us as a reference for the `instantaneous' value of the secular spin-axis precession rate of the planet, but it cannot be used (as such) to describe the dynamics inside a secular spin-orbit resonance.

In this section, we first recall the properties of secular spin-orbit resonances (Sect.~\ref{ssec:ssores}), and then we study the effect of a satellite on a resonantly locked planet (Sect.~\ref{ssec:spinandsat}).

\subsection{Secular spin-orbit resonance}\label{ssec:ssores}

   In the approximation of rigid rotation, the secular spin-axis dynamics of an oblate planet is ruled by the Hamiltonian function
   \begin{equation}\label{eq:Hamrot}
      \mathcal{M} = \mathcal{M}_\odot(X) + \mathcal{M}_\mathrm{P}(X,\psi,t)\,,
   \end{equation}
   where the conjugate canonical coordinates used here are $X=\cos\varepsilon$ (cosine of obliquity) and $-\psi$ (minus the precession angle). The first part comes from the torque exerted by the star on the equatorial bulge of the planet at quadrupolar order. It can be written 
   \begin{equation}\label{eq:Modot}
      \mathcal{M}_\odot = -\frac{\alpha}{2}\frac{X^2}{\big(1-e_\odot^2\big)^{3/2}}\,,
   \end{equation}
   where the parameter $\alpha$ is called the `precession constant'. In the absence of satellite, the expression of the precession constant is given for instance by \cite{NerondeSurgy-Laskar_1997} as
   \begin{equation}\label{eq:alpha}
      \alpha = \frac{3}{2}\frac{\mu_\odot}{\omega a_\odot^3}\frac{J_2}{\lambda}\,,
   \end{equation}
   where $\omega$ is the spin rate of the planet and $\lambda$ is its normalised polar moment of inertia. The parameters $J_2$ and $\lambda$ are related to the moments of inertia $A$, $B$, and $C$ of the planet through
   \begin{equation}\label{eq:J2lb}
      J_2 = \frac{2C-A-B}{2MR_\mathrm{eq}^2}
      \quad\text{and}\quad
      \lambda = \frac{C}{MR_\mathrm{eq}^2}\,,
   \end{equation}
   where $M$ is the mass of the planet. The second part of the Hamiltonian function stems from the motion of the planet's orbital plane, produced for instance through mutual perturbations with other planets. It can be written
   \begin{equation}
      \mathcal{M}_\mathrm{P} = - \sqrt{1-X^2}\bigg(\mathcal{A}(t)\sin\psi + \mathcal{B}(t)\cos\psi\bigg) + 2X\mathcal{C}(t)\,,
   \end{equation}
   where $\mathcal{A}$, $\mathcal{B}$, and $\mathcal{C}$ are explicit functions of time whose expressions in terms of the planet's classical orbital elements are given for instance by \cite{Laskar-Robutel_1993} and \cite{NerondeSurgy-Laskar_1997}. If the planet's orbit were fixed, then $\mathcal{M}_\mathrm{P}$ would be identically zero.
   
   We assume that the orbit of the planet is long-term stable, so that its secular orbital motion can be expressed (at least locally) in convergent quasi-periodic series. Truncating the series describing its orbital inclination motion to $N$ terms, it can be expressed as
   \begin{equation}\label{eq:zeta}
      \zeta = \sin\frac{I}{2}\exp(i\Omega) = \sum_{k=1}^NS_k\exp(i\phi_k)\,,
   \end{equation}
   where $S_k$ is a positive real constant and $\phi_k$ evolves linearly over time with frequency $\nu_k$, that is,
   \begin{equation}\label{eq:nuk}
     \phi_k = \nu_k\,t + \phi_k^{(0)}\,,
   \end{equation}
   for any $k=1,2...N$. In Eq.~\eqref{eq:zeta}, $I$ and $\Omega$ are the orbital inclination and the longitude of ascending node of the planet measured in an inertial reference frame, not to be confused with the satellite's orbital elements used in Sect.~\ref{sec:orb}. As shown by \cite{Saillenfest-etal_2019a}, the Hamiltonian function $\mathcal{M}_\mathrm{P}$ is proportional to the amplitudes $S_k$ of the quasi-periodic decomposition. Therefore, if the planet is not much inclined with respect to the invariable plane of the system (i.e. $S_k\ll 1$ for $\nu_k\neq 0$), which is what we expect in a long-term stable planetary system, then the Hamiltonian $\mathcal{M}_\mathrm{P}$ in Eq.~\eqref{eq:Hamrot} can be considered as a perturbation to the unperturbed Hamiltonian $\mathcal{M}_\odot$. In this setting, the long-term spin-axis dynamics of the planet is shaped by resonances between the unperturbed spin-axis precession frequency and the forcing frequencies appearing in Eq.~\eqref{eq:nuk}. In the Solar System, the orbital precession motions of the terrestrial planets contain numerous large-amplitude harmonics, creating a collection of wide secular spin-orbit resonances which overlap with each other and create wide chaotic zones \citep{Laskar-Robutel_1993}. The orbital precession motions of the giant planets, on the contrary, are composed of many fewer strong harmonics, so that large secular spin-orbit resonances are rare and isolated from each other. Depending on their spin-axis precession frequency, the giant planets of the Solar System can therefore be captured into an isolated resonance and oscillate stably within its separatrix (see e.g. \citealp{Ward-Hamilton_2004,Ward-Canup_2006,Saillenfest-etal_2020,Saillenfest-etal_2021}).
   
   Using a perturbative approach, \cite{Saillenfest-etal_2019a} have described the properties of all resonances up to order three in the amplitudes $\{S_k\}$. The largest resonances are those of order $1$, for which the resonance angle is $\sigma=\psi+\phi_j$, where $j$ is a given index in the orbital series in Eq.~\eqref{eq:zeta}. Second-order resonances involve two terms in the series, and third-order resonances involve three. Eccentricity-driven resonances only appear at order three and beyond. In any case, the resonance angle is a linear combination involving $\psi$ and one or several $\phi_k$. If the planet is trapped inside one of those resonances, then the resonance angle oscillates around a fixed value, which means that the spin-axis precession frequency $\dot{\psi}$ is forced to remain approximatively constant, equal to a combination of frequencies $\nu_k$.
   
   In the vicinity of a first-order secular spin-orbit resonance, the Hamiltonian function reduces to the well-known `Colombos's top Hamiltonian' \citep{Colombo_1966,Henrard-Murigande_1987}. This Hamiltonian can be written
   \begin{equation}\label{eq:Fres}
      \mathcal{F} = -\frac{1}{2}X^2 + \gamma X + \beta\sqrt{1-X^2}\cos\sigma\,,
   \end{equation}
   where the conjugate coordinates are $X=\cos\varepsilon$ and $-\sigma=-\psi-\phi_j$ (i.e. minus the resonance angle). Neglecting terms of order four and higher in the amplitudes $\{S_k\}$, the parameters $\gamma$ and $\beta$ in Eq.~\eqref{eq:Fres} are
   \begin{equation}\label{eq:gambet}
      \begin{aligned}
         \gamma &= \frac{-1}{p}\left(\nu_j - 2\sum_{k=1}^N\nu_kS_k^2\right)\,, \\
         \beta &= \frac{-S_j}{p}\left(2\nu_j + \nu_jS_j^2 - 2\sum_{k=1}^N\nu_kS_k^2\right)\,, \\
      \end{aligned}
   \end{equation}
   where $p = \alpha (1-e_\odot^2)^{-3/2}$ is the characteristic spin-axis precession frequency of the planet. Contrary to \cite{Saillenfest-etal_2019a}, we do not expand the eccentricity variations of the planet in quasi-periodic series: since eccentricity variations only appear at third order in the amplitudes, they are not important for our present qualitative description of the dynamics. Written as in Eq.~\eqref{eq:gambet}, eccentricity variations simply produce slight fluctuations in the resonance parameters $\gamma$ and $\beta$.
   
\begin{figure}
      \includegraphics[width=0.9\columnwidth]{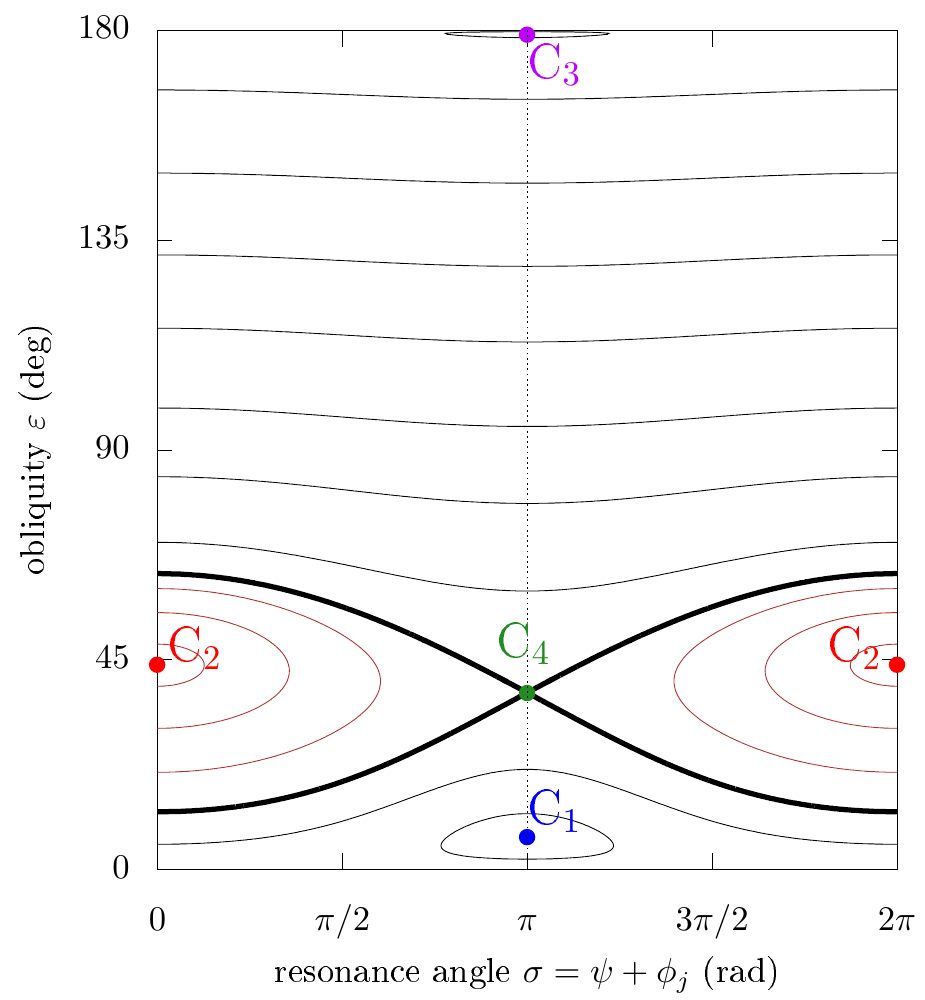}
      \caption{Spin-axis dynamics in the vicinity of a first-order secular spin-orbit resonance. Some level curves of the Hamiltonian function $\mathcal{F}$ in Eq.~\eqref{eq:Fres} are represented in black or in red, according to whether they are outside or inside the resonance, respectively. The separatrix is shown by a thick black curve. The constant parameters are $\gamma=0.75$ and $\beta=0.03$. The equilibrium points (`Cassini states') are represented by coloured dots. Figure~\ref{fig:phaseC3D}b shows the same phase portrait plotted on the sphere.}
      \label{fig:phaseC}
   \end{figure}

   As defined in Eq.~\eqref{eq:gambet}, for small amplitudes $S_k$, the parameters $\gamma$ and $\beta$ are both positive if $\nu_j<0$. This corresponds to a prograde resonance, for which the resonance centre is located at an obliquity $\varepsilon\leqslant 90^\circ$. An example is presented in Fig.~\ref{fig:phaseC}. On the contrary, retrograde resonances are obtained for $\nu_j > 0$, for which $\gamma$ and $\beta$ are negative. Due to symmetries, changing the sign of $\gamma$ is equivalent to replacing $X$ by $-X$, and changing the sign of $\beta$ is equivalent to replacing $\sigma$ by $\sigma+\pi$. Following \cite{Peale_1969}, the equilibrium points are usually called `Cassini states', numbered from 1 to 4, as labelled in Fig.~\ref{fig:phaseC}.
   
\begin{figure*}
      \includegraphics[width=0.33\textwidth]{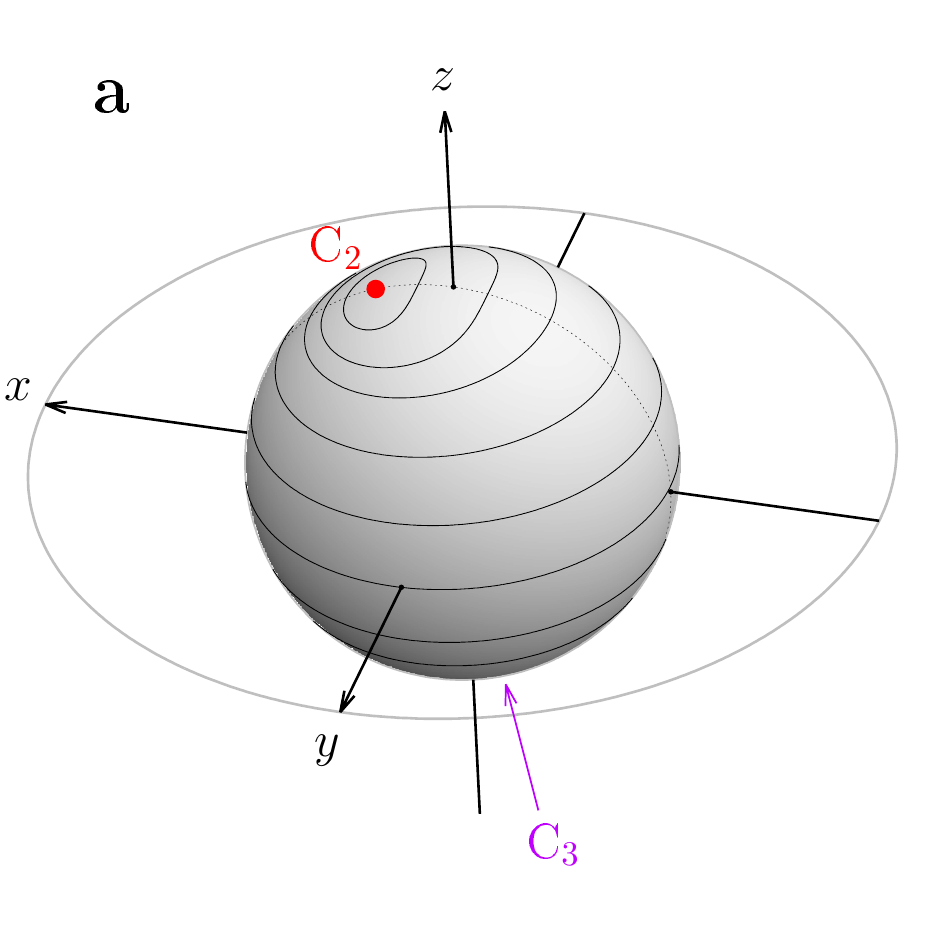}
      \includegraphics[width=0.33\textwidth]{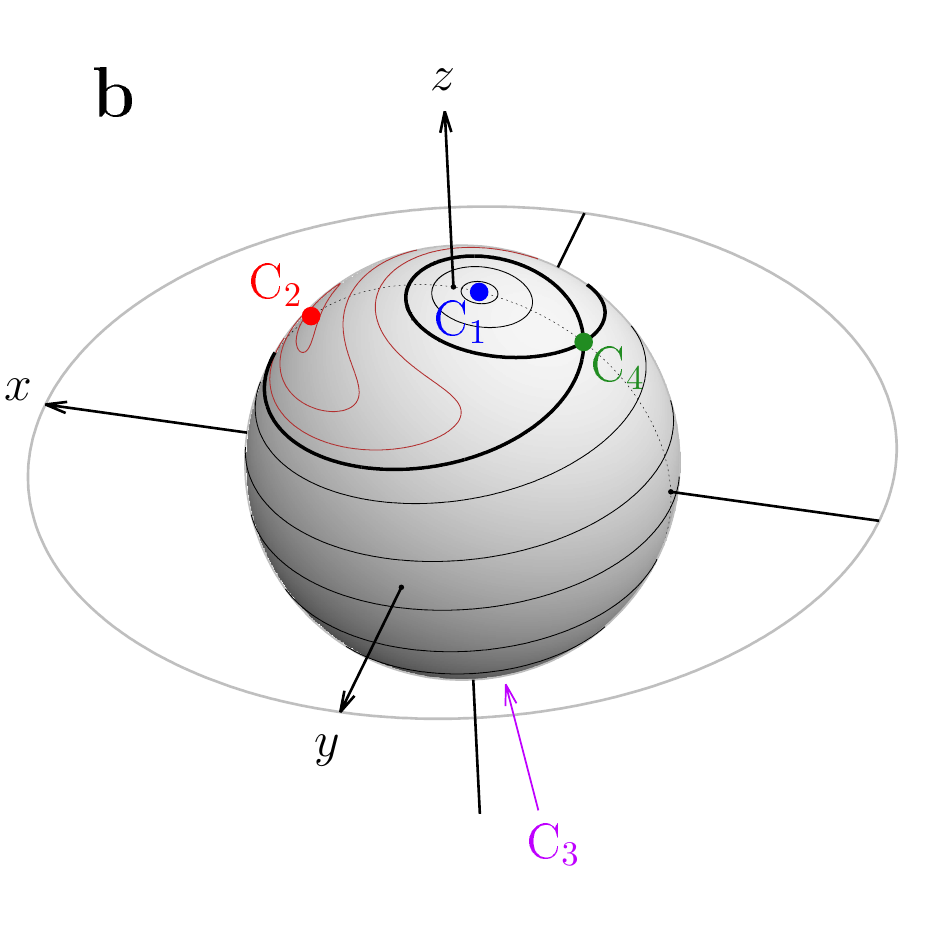}
      \includegraphics[width=0.33\textwidth]{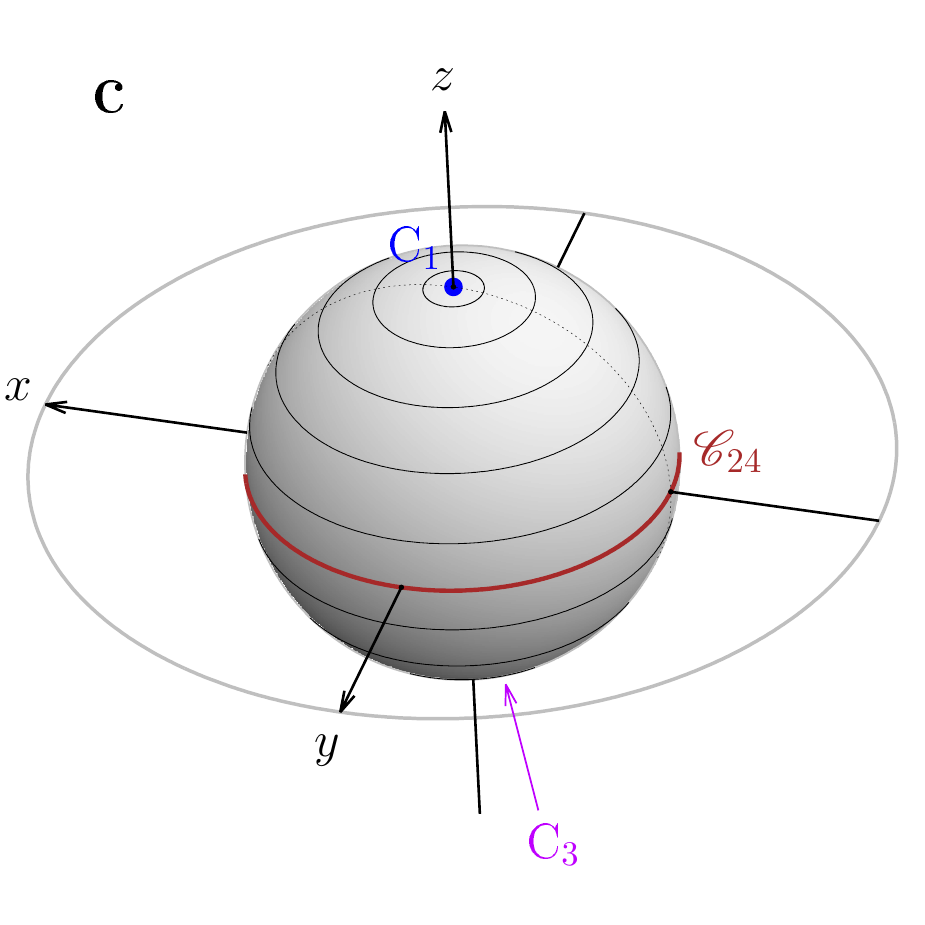}
      \caption{Level curves of the Hamiltonian $\mathcal{F}$ plotted on the sphere. The planet's orbit lies in the $xy$-plane. The obliquity $\varepsilon$ is the tilt from the $z$-axis, and the resonance angle $\sigma$ is the polar angle measured in the $xy$-plane. The colour code is the same as in Fig.~\ref{fig:phaseC}. Panel \textbf{a}: geometry for $\gamma^{2/3}+\beta^{2/3}>1$. Panel \textbf{b}: geometry for $\gamma^{2/3}+\beta^{2/3}<1$ (same parameters as in Fig.~\ref{fig:phaseC}). Panel \textbf{c}: geometry for $\gamma$ and $\beta$ tending to zero.}
      \label{fig:phaseC3D}
   \end{figure*}

   \cite{Henrard-Murigande_1987} showed that the phase space has two different topologies according to whether $\gamma^{2/3}+\beta^{2/3}$ is smaller or larger than $1$. If $\gamma^{2/3}+\beta^{2/3}>1$, then there is no resonance (i.e. no separatrix) and only the Cassini states C$_2$ and C$_3$ are present (see Fig.~\ref{fig:phaseC3D}a). If $\gamma^{2/3}+\beta^{2/3}<1$, then all four Cassini states are present, and C$_2$ becomes the resonance centre (see Fig.~\ref{fig:phaseC3D}b). As noted by \cite{Saillenfest-etal_2019a}, in some parameter region, the resonance contains the north pole and/or the south pole of the sphere. We stress that the resonance can be quite large even for a moderate amplitude $S_j$.
   
   As detailed below, the effect of satellites on the spin-axis motion can be modelled by replacing $\alpha$ in Eq.~\eqref{eq:Modot} by an effective precession constant $\alpha'$, whose value depends on the distance of the satellites. Therefore, we are interested in how the geometry of the phase space is changed when modifying the parameter $\alpha$ appearing in Eq.~\eqref{eq:gambet} via $p$. Analytical formulas for the locations of the equilibrium points and the separatrix are provided by \cite{Saillenfest-etal_2019a} or \cite{Haponiak-etal_2020}; we recall here their behaviour when varying the parameter $\alpha$. Due to symmetries, we only describe the case of a prograde resonance, for which $\gamma$ and $\beta$ are positive.
   
\begin{figure*}
   	  \centering
      \includegraphics[width=0.9\textwidth]{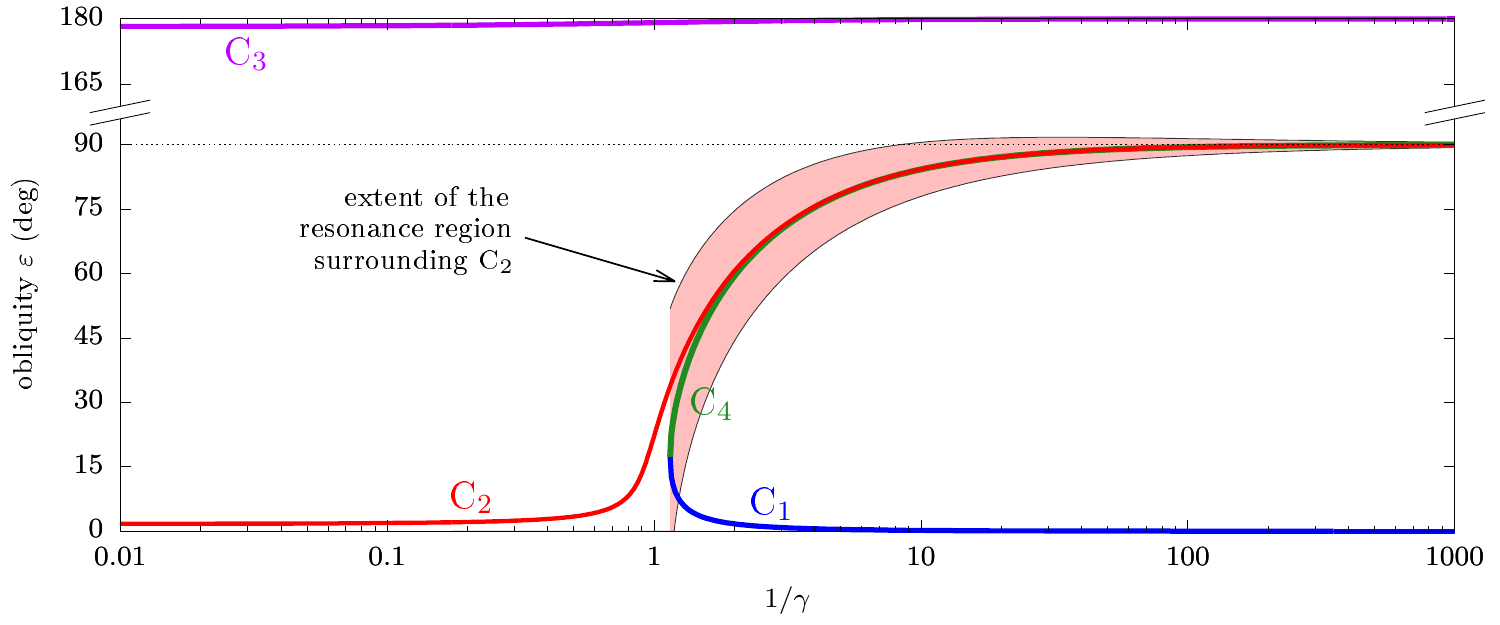}
      \caption{Bifurcation diagram of the system as a function of $1/\gamma$, which is proportional to $\alpha$. The second parameter used in this example is $\beta=0.03\gamma$. The Cassini states are labelled with the same colour code as previous figures.}
      \label{fig:Cassinivaralpha}
   \end{figure*}

   Figure~\ref{fig:Cassinivaralpha} shows the generic behaviour of the system with respect to the parameter $1/\gamma$, which is proportional to $\alpha$. For $\alpha\rightarrow 0$, both $\gamma$ and $\beta$ tend to infinity, so only Cassini states 2 and 3 are present (see Fig.~\ref{fig:phaseC3D}a). Their asymptotic locations are
   \begin{equation}
      \lim\limits_{\alpha\rightarrow 0}\cos\varepsilon_2 = - \lim\limits_{\alpha\rightarrow 0}\cos\varepsilon_3 = \frac{1}{\sqrt{1 + \beta^2/\gamma^2}} \,,
   \end{equation}
   which, for a small amplitude S$_j$, represents just a small offset from $\varepsilon_2 = 0^\circ$ and $\varepsilon_3 = 180^\circ$ (see Fig.~\ref{fig:Cassinivaralpha}). In other words, for $\alpha\rightarrow 0$ the resonance is infinitely far away and it only contributes through a residual shift of the equilibrium points at the north and south poles of the sphere. Therefore, the obliquity is almost constant and $\sigma$ circulates.
   
   If we increase $\alpha$ above zero, the Cassini state C$_2$ is tilted away from the north pole of the sphere (see Figs.~\ref{fig:phaseC3D}a and \ref{fig:Cassinivaralpha}); then, when $\gamma^{2/3}+\beta^{2/3}$ becomes smaller than $1$, the Cassini states C$_1$ and C$_4$ appear together with the separatrix. As shown in previous articles (see in particular Figs.~B.1 and B.2 of \citealp{Saillenfest-etal_2020}), for increasing $\alpha$ the resonance first contains the north pole of the sphere, and then the separatrix crosses the north pole and moves down to larger obliquities. This transition is visible in Fig.~\ref{fig:Cassinivaralpha} as the very narrow interval of $1/\gamma$ in which the pink region touches $\varepsilon=0^\circ$.
   
   For $\alpha\rightarrow\infty$, the parameters $\gamma$ and $\beta$ both tend to zero and the location of the Cassini states tend to
   \begin{equation}
      \begin{aligned}
         &\lim\limits_{\alpha\rightarrow \infty}\varepsilon_1 = 0^\circ\,,
         \quad\quad
         &\lim\limits_{\alpha\rightarrow \infty}\varepsilon_2 = 90^\circ\,, \\
         &\lim\limits_{\alpha\rightarrow \infty}\varepsilon_3 = 180^\circ\,,
         \quad\quad
         &\lim\limits_{\alpha\rightarrow \infty}\varepsilon_4 = 90^\circ\,.
      \end{aligned}
   \end{equation}
   Moreover, for $\alpha\rightarrow\infty$ the separatrix enclosing C$_2$ becomes vanishingly narrow and it merges with the Cassini states C$_2$ and C$_4$, producing the degenerate equilibrium circle $\mathscr{C}_{24}$ shown in Fig.~\ref{fig:phaseC3D}c. For large but finite values of $\alpha$, we note that the resonance width goes beyond $\varepsilon=90^\circ$ (there is no topological boundary at $\varepsilon=90^\circ$ for non-zero libration amplitudes).
   
   Hence, as a summary, if we increase the parameter $\alpha$ from $0$ to $\infty$, the Cassini state C$_2$ gradually passes from $\varepsilon_2\approx 0^\circ$ (without separatrix) to $\varepsilon_2=90^\circ$ (inside the resonance separatrix). These properties will be important below.
   
\subsection{Effect of a satellite}\label{ssec:spinandsat}

   The orbital angular momentum of the planet usually greatly exceeds its rotational angular momentum. Therefore, the planet's orbit remains almost unaffected by the spin-axis precession motion. In Sect.~\ref{ssec:ssores}, this property allowed us to treat the orbital variations as a forcing term in the spin-axis dynamics. Now, if the planet has a satellite that lies in its local Laplace plane (i.e. if it is in the Laplace state P$_1$ described in Sect.~\ref{ssec:Lap}), then the planet's spin axis and the satellite's orbit rigidly precess as a whole about the planet's orbital angular momentum \citep{Boue-Laskar_2006}. In Eq.~\eqref{eq:HpC}, this precession would be traduced by a drift of $\Omega_\mathrm{P}$ over time. 

   We define the characteristic spin-axis precession timescale of the planet as $T=2\pi/p$, where $p=\alpha (1-e_\odot^2)^{-3/2}$ has been introduced in Sect.~\ref{ssec:ssores}. Without satellite, the free spin-axis precession frequency of the planet is simply
   \begin{equation}
      \Omega_0 = p\cos\varepsilon\,,
   \end{equation}
   as obtained from Eq.~\eqref{eq:Modot}. The characteristic timescale $T$ is given in Table~\ref{tab:param} for some planets of the Solar System. The value of $T$ can be compared to the characteristic timescale $\tau$ of the satellite's orbital dynamics. We see that $\tau\ll T$ for all satellites listed in Table~\ref{tab:param}. This large separation of timescales justifies the approximation made by \cite{Tremaine-etal_2009} and used in Sect.~\ref{sec:orb} to consider that the planet's equatorial reference frame is inertial despite its precession motion. For a satellite oscillating about the Laplace state P$_1$, this timescale condition may be violated at the border of region E$_1$ or in the extreme vicinity of the singular point S$_1$, that is, where one of its orbital oscillation frequencies tends to zero (see Fig.~\ref{fig:libT1}). The behaviour of satellites in this critical regime will be investigated numerically in Sect.~\ref{sec:sattit}.

   Substantially massive satellites contribute to the spin-axis precession frequency of their host planet. If the satellites oscillate about the Laplace state P$_1$ with $\tau\ll T$, \cite{French-etal_1993} give an elegant expression for their contribution: one must simply replace $J_2$ and $\lambda$ in Eq.~\eqref{eq:alpha} by the effective values
   \begin{equation}\label{eq:J2prime}
      \begin{aligned}
         J_2' &= J_2 + \frac{1}{2}\sum_k\frac{m_k}{M}\frac{a_k^2}{R_\mathrm{eq}^2}\frac{\sin(2\varepsilon-2L_k)}{\sin(2\varepsilon)}\,, \\
         \lambda' &= \lambda + \sum_k\frac{m_k}{M}\frac{a_k^2}{R_\mathrm{eq}^2}\frac{n_k}{\omega}\frac{\sin|\varepsilon-L_k|}{\sin\varepsilon}\,,
      \end{aligned}
   \end{equation}
   where $m_k$, $a_k$ and $n_k$ are the mass, the semi-major axis, and the mean motion of the $k$th satellite, and $L_k$ is the inclination of the Laplace plane of the $k$th satellite with respect to the planet's equator\footnote{\cite{Ward-Hamilton_2004} give another expression for $J_2'$ and $\lambda'$ in their Eqs.~(2) and (3). When the results are compared to the self-consistent theory of \cite{Boue-Laskar_2006}, the expression of \cite{Ward-Hamilton_2004} appears to be erroneous. We suspect that Eq.~(2) of \cite{Ward-Hamilton_2004} actually contains a typographical error. This likely error seems to have been propagated in Eq.~(44) of \cite{Millholland-Laughlin_2019}.}. Using these expressions, the free spin-axis precession frequency $\Omega_0$ of the planet is still obtained from Eq.~\eqref{eq:Modot}, but where $\alpha$ is replaced by an effective precession constant $\alpha'$. We stress that Eq.~\eqref{eq:J2prime} is valid whatever the distance of the satellite, and not only in the close-in regime considered previously by \cite{Saillenfest-etal_2020} and \cite{Saillenfest-etal_2021}. It only requires that the satellites oscillate around the Laplace state P$_1$, which is what we expect for any regular satellite (see e.g. \citealp{Tremaine-etal_2009}), unless it reaches the high-obliquity unstable region E$_1$ (see Sect.~\ref{ssec:eccstab}).

   For small satellites, the value of $L_k$ can be directly taken from Eq.~\eqref{eq:ILap13}. In this case, the model is not self-consistent, because the satellite is considered to be massless when dealing with the orbital dynamics (Sect.~\ref{sec:orb}) but massive when computing its long-term influence on the planet's spin axis. Yet, because $\tau\ll T$ (see Table~\ref{tab:param}), this approximation results to be very accurate for satellites having a small mass ratio $m_k/M$. For Titan, whose mass is about $10^{-4}$ of Saturn's, the inclination $L_k$ and the precession rate $\Omega_0$ obtained through Eqs.~\eqref{eq:ILap13} and \eqref{eq:J2prime} are very close to those obtained using the self-consistent (but quite complicated) model of \cite{Boue-Laskar_2006}, as detailed in Appendix~\ref{asec:BL06}. The approximation is less good for very massive satellites like the Moon ($m_k/M\approx 0.01$), but we can check that the qualitative picture described below remains valid, which means that our analysis captures the essence of the dynamics even for large satellite-to-planet mass ratios.

   We note that Eq.~\eqref{eq:J2prime} looks undefined for $\varepsilon=0$ or $90^\circ$. However, computing $L_k$ using Eq.~\eqref{eq:ILap13}, the contribution of satellites around a zero-obliquity planet simplifies to
   \begin{equation}
      \begin{aligned}
         J_2'\Big|_{\varepsilon=0} &= J_2 + \frac{1}{2}\sum_k\frac{m_k}{M}\frac{r_\mathrm{M}^2}{R_\mathrm{eq}^2}\left(\frac{a_k^2/r_\mathrm{M}^2}{1 + a_k^5/r_\mathrm{M}^5}\right)\,, \\
         \lambda'\Big|_{\varepsilon=0} &= \lambda + \sum_k\frac{m_k}{M}\frac{r_\mathrm{M}^2}{R_\mathrm{eq}^2}\frac{n_k}{\omega}\left(\frac{a_k^2/r_\mathrm{M}^2}{1 + a_k^5/r_\mathrm{M}^5}\right)\,.
      \end{aligned}
   \end{equation}
   Likewise, for $\varepsilon=90^\circ$ the parameter $J_2'$ becomes
   \begin{equation}
      J_2'\Big|_{\varepsilon=90^\circ} = J_2 + \frac{1}{2}\sum_k\frac{m_k}{M}\frac{r_\mathrm{M}^2}{R_\mathrm{eq}^2}\left|\frac{a_k^2/r_\mathrm{M}^2}{1 - a_k^5/r_\mathrm{M}^5}\right|\,.
   \end{equation}
   However, since $\Omega_0$ is factored by $\cos\varepsilon$, the spin-axis precession frequency of the planet is in any case zero for $\varepsilon=90^\circ$, except at $a=r_\mathrm{M}$, where the Laplace state P$_1$ of the satellites is undefined (see Sect.~\ref{sec:orb}).

   In order to keep the discussion as general as possible before focussing on particular bodies, one more approximation can be made. Indeed, even for a fast spinning planet like Saturn, the oblateness coefficient $J_2$ appearing in Eq.~\eqref{eq:alpha} is a small parameter compared to $\lambda$ (whose order of magnitude is slightly less than unity). As a result, the relative increase in $J_2$ produced by satellites in Eq.~\eqref{eq:J2prime} is much larger than the relative increase in $\lambda$. This discrepancy is amplified by the factor $n_k/\omega$ appearing in Eq.~\eqref{eq:J2prime}, which further reduces the satellites' contribution to $\lambda'$. Therefore, as a first approximation, the contribution of satellites to $\lambda'$ is negligible compared to their contribution to $J_2'$. Assuming that $\lambda'\approx\lambda$, and considering that the planet has a single main satellite, the free precession frequency of the planet's spin axis simplifies to
   \begin{equation}\label{eq:psidotsimp}
      \Omega_0 = p\left(\cos\varepsilon + \eta\frac{a^2}{r_\mathrm{M}^2}\frac{\sin(2\varepsilon-2L)}{2\sin(\varepsilon)}\right)\,,
   \end{equation}
   where we have introduced the `mass parameter' $\eta$ of the satellite, defined as
   \begin{equation}\label{eq:betamass}
      \eta = \frac{1}{2}\frac{m}{M}\frac{r_\mathrm{M}^2}{J_2R_\mathrm{eq}^2} \,.
   \end{equation}
   The mass parameters of some satellites in the Solar System are given in Table~\ref{tab:param}. We stress that even a low-mass satellite can have a large mass parameter $\eta$. For instance, Titan has a mass of $m/M\approx 10^{-4}$ but a mass parameter $\eta\approx 10$, and Titan greatly affects Saturn's spin-axis dynamics (see below). Therefore, contrary to what one could think a priori (see e.g. \citealp{Li-Batygin_2014}), a large mass ratio $m/M$ is not necessarily required to substantially alter a planet's obliquity. Using Eq.~\eqref{eq:psidotsimp}, the spin-axis precession rate of the planet normalised by $p$ is only a function of $\varepsilon$, $\eta$, and $a/r_\mathrm{M}$. We can therefore study its behaviour in a very generic way, as we did for the satellite's dynamics in Sect.~\ref{sec:orb}.
   
\begin{figure}
      \centering
      \includegraphics[width=\columnwidth]{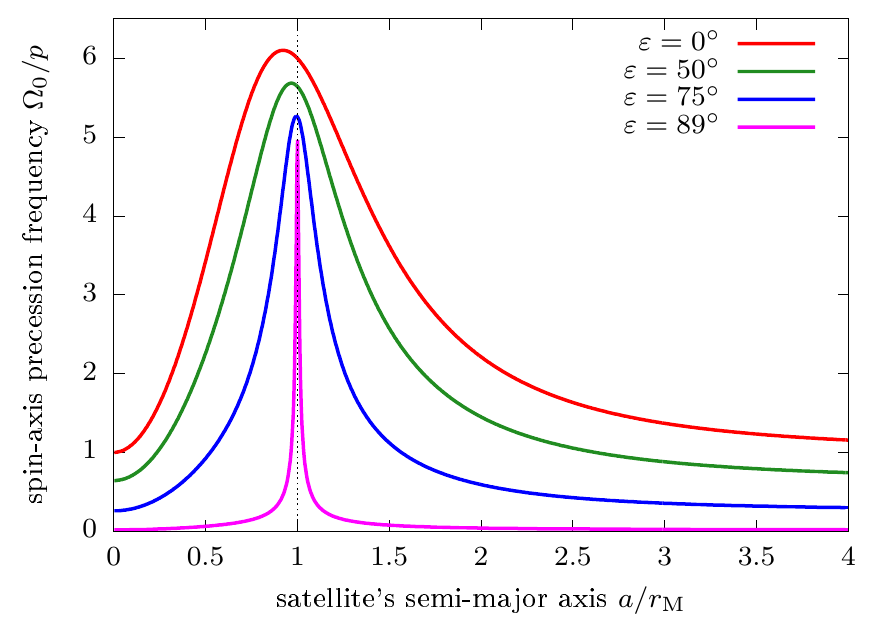}
      \caption{Spin-axis precession frequency of a planet orbited by a single satellite. The frequency is given by the simplified expression in Eq.~\eqref{eq:psidotsimp} and represented here in unit of the characteristic frequency $p$, for different obliquity values (see labels). In this example, the mass parameter of the satellite is set to $\eta=10$, which is close to Titan's value (see Table~\ref{tab:param}).}
      \label{fig:psidot}
   \end{figure}

   Figure~\ref{fig:psidot} shows the spin-axis precession frequency of the planet as a function of the distance of its satellite. We retrieve the classical curve shown in Fig.~5 of \cite{Boue-Laskar_2006}, with the close-in and far-away satellite regimes. For $a\rightarrow 0$ and $a\rightarrow\infty$, the precession frequency tends to the value $p\cos\varepsilon$ obtained without satellite. In between, the satellite enhances the precession frequency of the planet by an increment that is proportional to its mass parameter $\eta$ (see Eq.~\ref{eq:psidotsimp}). For a given obliquity, the maximum value of $\Omega_0$ divides the close-in and far-away satellite regimes, characterised by the well-known power laws in $a^2$ and $a^{-3}$, respectively. The magnitude of $\Omega_0$ differs when varying the satellite's mass parameter $\eta$, but the location of its maximum as a function of $a$ is independent of $\eta$. As shown in Fig.~\ref{fig:psidot}, the maximum of $\Omega_0$ is located somewhat below the midpoint radius $a=r_\mathrm{M}$. By analysing Eq.~\eqref{eq:psidotsimp}, we find that the maximum of the curve is located at $a=r_\mathrm{F}$, defined in Eq.~\eqref{eq:rF}. We recall that $r_\mathrm{F}$ is the inflexion point of the satellite's inclination, illustrated in Fig.~\ref{fig:P1detail}. We see here that the spin-axis precession frequency of the planet is intimately linked to the properties of the Laplace state P$_1$ of its satellite. We further analyse this relation below.

\begin{figure}
      \centering
      \includegraphics[width=\columnwidth]{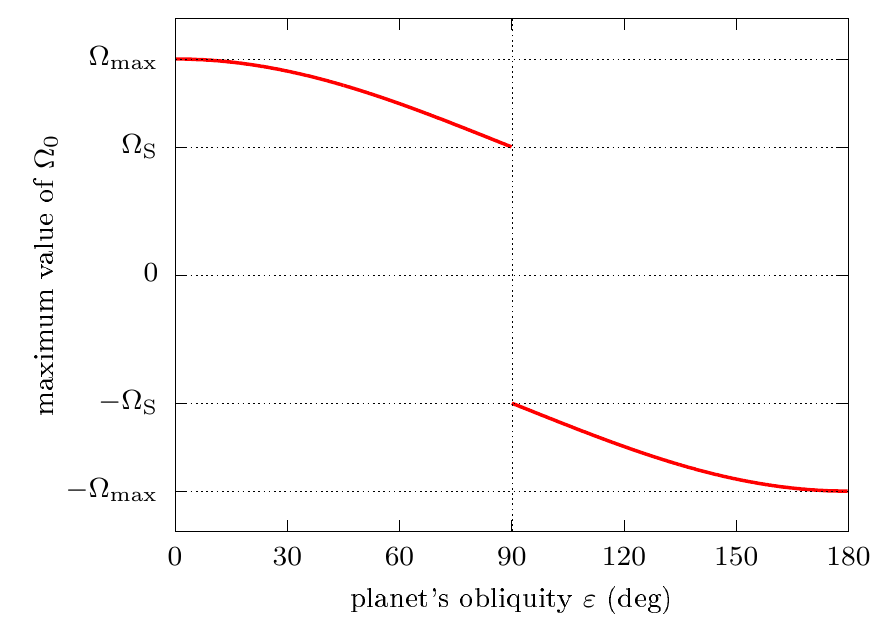}
      \caption{Maximum spin-axis precession frequency of the planet as a function of its obliquity. The curve is obtained by injecting $a=r_\mathrm{F}$ from Eq.~\eqref{eq:rF} in Eq.~\eqref{eq:psidotsimp}. The extreme values $\Omega_\mathrm{max}$ and $\Omega_\mathrm{S}$ are given in Eqs.~\eqref{eq:psidotmax} and \eqref{eq:psipm}. For $\varepsilon=90^\circ$, the value at the maximum is undefined.}
      \label{fig:psidotmax}
   \end{figure}

   Figure~\ref{fig:psidotmax} shows the value of $\Omega_0$ at its maximum (i.e. at $a=r_\mathrm{F}$) as a function of the planet's obliquity. The global maximum of $\Omega_0$ is reached at $\varepsilon=0^\circ$, for which $a^5/r_\mathrm{M}^5=2/3$. It has value
   \begin{equation}\label{eq:psidotmax}
      \Omega_\mathrm{max} = p\left(1 + \frac{108^{1/5}}{5}\eta\right)\,.
   \end{equation}
   For an obliquity $\varepsilon\rightarrow 90^\circ$, the maximum of $\Omega_0$ is reached at $a=r_\mathrm{M}$, that is, at the singular point S$_1$ of the satellite's dynamics (see Sect.~\ref{sec:orb}). Figure~\ref{fig:psidotmax} shows that the maximum value of $\Omega_0$ has two different limits at point S$_1$ according to whether the obliquity tends to $90^\circ$ from above or from below. These limits are $\Omega_0=\pm\,\Omega_\mathrm{S}$, where
   \begin{equation}\label{eq:psipm}
      \Omega_\mathrm{S} = \frac{1}{2}p\eta\,.
   \end{equation}
   We note that these limits are non-zero and well defined. This is far from obvious when modelling the satellites by an effective precession constant $\alpha'$, as one would never guess that $\alpha'\cos\varepsilon$ tends to a non-zero finite value when $\cos\varepsilon$ tends to zero ($\alpha'$ actually goes to infinity). We deduce that when the system approaches the singular point S$_1$, the notion of `precession constant' loses its meaning. In this regime, classical figures drawn in the plane $(\varepsilon,\alpha)$, like those used by \cite{Saillenfest-etal_2021} are inappropriate.

\begin{figure}[h!]
      \centering
      \includegraphics[width=\columnwidth]{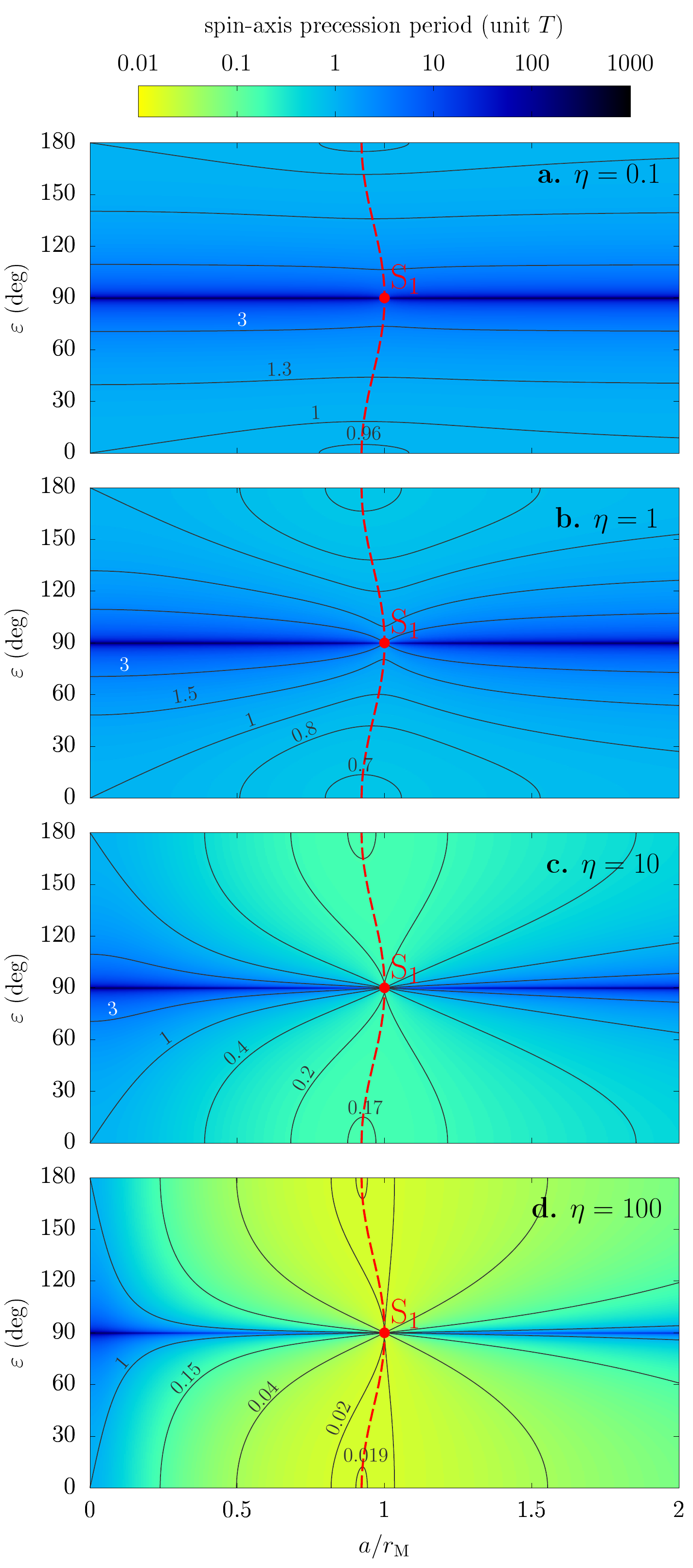}
      \caption{Spin-axis precession period of a planet orbited by a single satellite. The period is given by Eq.~\eqref{eq:psidotsimp} and represented here in unit of the characteristic period $T=2\pi/p$. For $\varepsilon >90^\circ$, the spin-axis precession frequency is negative. Each panel corresponds to a given value of the satellite's mass parameter $\eta$ (see Eq.~\ref{eq:betamass}), as labelled in the top right corner. Some level curves are highlighted in black. The singular point S$_1$ and the ridge line $a=r_\mathrm{F}$ are indicated in red.}
      \label{fig:precrate}
   \end{figure}

   We are interested in the level curves of $\Omega_0$ as a function of the parameters. Indeed, if the planet is trapped in a secular spin-orbit resonance during the migration of its satellite, its precession frequency $\dot{\psi}\approx\Omega_0$ is forced to oscillate around a constant value while the parameters $a$ and $\varepsilon$ vary (see Sect.~\ref{ssec:ssores}). This is likely the case for Saturn and Titan (see \citealp{Saillenfest-etal_2021}), and it will probably be the case for Jupiter and its satellites in the future (see \citealp{Saillenfest-etal_2020}). Figure~\ref{fig:precrate} shows the spin-axis precession period of the planet in the parameter space for different values of the satellite's mass parameter $\eta$. We choose here to plot the precession period $|2\pi/\Omega_0|$, instead of the frequency $\Omega_0$ for easier comparison with the satellite's oscillation period shown in Fig.~\ref{fig:libT1}. Values of $\tau$, $\eta$ and $T$ for real bodies can be found in Table~\ref{tab:param}. As shown in Appendix~\ref{asec:BL06}, Fig.~\ref{fig:precrate} presents a very good agreement with the precession period obtained in self-consistent models.

   For a very small mass parameter $\eta\ll 1$ (e.g. for Deimos), the effect of the satellite is negligible and the level curves of $\Omega_0$ would appear perfectly horizontal in Fig.~\ref{fig:precrate}. Therefore, even if the planet is trapped in a secular spin-orbit resonance, its mean obliquity would remain unaltered over the migration of its satellite. For a substantial value of $\eta$, on the contrary, Fig.~\ref{fig:precrate} shows that the level curves of $\Omega_0$ lose their horizontal shape and rearrange around the ridge line $a=r_\mathrm{F}$, where the satellite's contribution is maximum. Indeed, since the factor in front of $\eta$ in Eq.~\eqref{eq:psidotmax} is close to $1/2$, the relative difference between $\Omega_\mathrm{max}$ and $\Omega_\mathrm{S}$ is very small for a large value of $\eta$, namely
   \begin{equation}
      \lim\limits_{\eta\rightarrow\infty}\frac{\Omega_\mathrm{max}-\Omega_\mathrm{S}}{\Omega_\mathrm{max}} = 1 - 5\frac{72^{1/5}}{12} \approx 0.02\,.
   \end{equation}
   Therefore, a large value of $\eta$ implies that $\Omega_0$ is approximatively constant along the ridge line $a=r_\mathrm{F}$ (i.e. the two curves in Fig.~\ref{fig:psidotmax} are nearly horizontal).    Consequently, the ridge line $a=r_\mathrm{F}$ creates a barrier that cannot be crossed by the level curves of $\Omega_0$: instead, the level curves must go around the ridge line, and they converge at the singular point S$_1$ (see Fig.~\ref{fig:precrate}). More precisely, all level curves with frequency values $|\Omega_0|\leqslant\Omega_\mathrm{S}$ converge to S$_1$. For a large mass parameter $\eta$, this condition is verified for almost all level curves of $\Omega_0$. This property is related to the orbital plane of the satellite, which can reach S$_1$ from any inclination (see Fig.~\ref{fig:P1detail}).
   
   Figure~\ref{fig:precrate} shows that for a large enough value of $\eta$, numerous level curves of $\Omega_0$ connect the singular region S$_2$ (i.e. $\varepsilon=0$) to the singular point S$_1$. Therefore, if the planet is trapped in a secular spin-orbit resonance, the migration of its satellite through $a=r_\mathrm{M}$ forces its obliquity to increase all the way from $0^\circ$ to $90^\circ$. We see that such an extreme obliquity increase can take place over a very short migration range for the satellite (e.g. in Panel d if its distance decreases from $a\approx 1.05r_\mathrm{M}$ to $a=r_\mathrm{M}$). The theoretical limit for the obliquity reached through the mechanism described by \cite{Saillenfest-etal_2021} is therefore $90^\circ$, or even more if the resonance is large and its width extends beyond $\varepsilon=90^\circ$ (see Sect.~\ref{ssec:ssores}). If ever the system manages to go through the singular point S$_1$ in some way, one could even imagine a scenario where the planet then picks a retrograde resonance (see e.g. \citealp{Kreyche-etal_2020}) and goes on tilting up to $180^\circ$.

   The level curves of $\Omega_0$ going from $\varepsilon=0^\circ$ to $\varepsilon=90^\circ$ verify $p\leqslant\Omega_0\leqslant\Omega_\mathrm{S}$.
   Therefore, the minimum mass parameter allowing the planet to tilt all the way from $0^\circ$ to $90^\circ$ through the migration of its satellite is obtained by putting $\Omega_\mathrm{S}=p$ in Eq.~\eqref{eq:psipm}, which gives $\eta=2$. In other words, $\eta=2$ is the minimum mass parameter for which the level curve $\Omega_0=p$, which starts at $(a,\varepsilon)=(0,0^\circ)$, connects to $(a,\varepsilon)=(r_\mathrm{M},90^\circ)$: see the level curve labelled `1' in Fig.~\ref{fig:precrate}. In Table~\ref{tab:param}, the condition $\eta\geqslant 2$ is verified for the Moon, Titan, and Oberon, but not for Deimos, Callisto, and Iapetus.

   In practice, since the singular point S$_1$ is surrounded by the unstable region E$_1$ (see Sect.~\ref{ssec:eccstab}), we expect the satellite's orbit to become unstable before actually reaching S$_1$. Moreover, the width of the secular spin-orbit resonance of the planet decreases as the obliquity increases (see Fig.~\ref{fig:Cassinivaralpha}), and since many level curves converge to S$_1$, other secular spin-orbit resonances (if any) would necessarily overlap at some point and create a chaotic region. For these two reasons, we expect the planet to be released out of resonance before actually reaching $\varepsilon=90^\circ$. This double destabilisation of the planet and of its satellite is investigated in the next section.

\section{Saturn and Titan}\label{sec:sattit}

\subsection{Overview of the dynamics}\label{ssec:overview}

The Hamiltonian function in Eq.~\eqref{eq:Hamrot} explicitly depends on the orbit of the planet and on its temporal variations. In order to explore the long-term dynamics of Saturn's spin axis, we need an orbital solution that is valid over billions of years. As in previous studies, we use the secular solution of \cite{Laskar_1990} expanded in quasi-periodic series, that is, under the form given in Eq.~\eqref{eq:zeta}. The ten largest terms of the $\zeta$ series of Saturn are shown in Table~\ref{tab:zetashort}, ordered by decreasing amplitude.

\begin{table}
   \caption{Largest terms of Saturn's inclination series in the J2000 ecliptic and equinox reference frame.}
   \label{tab:zetashort}
   \vspace{-0.7cm}
   \begin{equation*}
      \begin{array}{rcrrr}
      \hline
      \hline
      k & \text{identification} & \nu_k\ (''\,\text{yr}^{-1}) & S_k\times 10^8 & \phi_k^{(0)}\ (^\text{o}) \\
      \hline   
        1 &          s_5 &   0.00000 & 1377395 & 107.59 \\
        2 &          s_6 & -26.33023 &  785009 & 127.29 \\
        3 &          s_8 &  -0.69189 &   55969 &  23.96 \\
        4 &          s_7 &  -3.00557 &   39101 & 140.33 \\
        5 &  g_5-g_6+s_7 & -26.97744 &    5889 &  43.05 \\
        6 &     2g_6-s_6 &  82.77163 &    3417 & 128.95 \\
        7 &  g_5+g_6-s_6 &  58.80017 &    2003 & 212.90 \\
        8 &     2g_5-s_6 &  34.82788 &    1583 & 294.12 \\
        9 &          s_1 &  -5.61755 &    1373 & 168.70 \\
       10 &          s_4 & -17.74818 &    1269 & 123.28 \\
      \hline
      \end{array}
   \end{equation*}
   \vspace{-0.5cm}
   \tablefoot{Adapted from \cite{Laskar_1990}. Each frequency $\nu_k$ is a combination of the fundamental frequencies $s_i$ and $g_i$ of the planets of the Solar System. See Appendix~A of \cite{Saillenfest-etal_2021} for the full quasi-periodic decomposition with amplitudes down to $10^{-8}$.}
\end{table}

As explained in Sect.~\ref{sec:plspin}, a satellite is able to tilt a planet from $\varepsilon = 0^\circ$ to $90^\circ$ if its mass parameter $\eta$ is larger than $2$. This condition is met for Titan, which has $\eta\approx 12.4$. Furthermore, the maximum spin-axis precession frequency of the planet is reached along $a=r_\mathrm{F}$, and the global maximum $\Omega_\mathrm{max}$ is given by Eq.~\eqref{eq:psidotmax}. For Saturn and Titan (see Table~\ref{tab:param}), we obtain $\Omega_\mathrm{max}\approx 1.41''\,$yr$^{-1}$. Therefore, all frequencies in Saturn's orbital decomposition whose magnitude exceeds this value are unreachable by Saturn, whatever the distance of Titan. This only leaves a handful of possible first-order secular spin-orbit resonances, even when considering the full quasi-periodic solution of \cite{Laskar_1990}. The largest resonance is with $\nu_3=s_8$ (see Table~\ref{tab:zetashort}). Other resonances can be identified in Table~A.2 of \cite{Saillenfest-etal_2021}: there are two prograde resonances ($\nu_{19}$ and $\nu_{51}$) and two retrograde resonances ($\nu_{28}$ and $\nu_{44}$). As revealed by their high index in the orbital series, these four resonances are very small. This explains why no relevant second- or higher-order resonance can possibly affect Saturn. Then, we know that all precession frequencies $\Omega_0$ verifying $p\leqslant\Omega_0\leqslant\Omega_\mathrm{S}$ have a level curve that connects $\varepsilon = 0^\circ$ to $90^\circ$ (or $\varepsilon = 180^\circ$ to $90^\circ$ for a retrograde spin). For Saturn and Titan, we have $p\approx0.19''\,$yr$^{-1}$ and $\Omega_\mathrm{S}\approx 1.20''\,$yr$^{-1}$, so this condition is met by all five resonances mentioned above (even though $\nu_{51}$ is right at the limit, since $|\nu_{51}|\approx \Omega_\mathrm{S}$).

\begin{figure*}
   \centering
   \includegraphics[width=\textwidth]{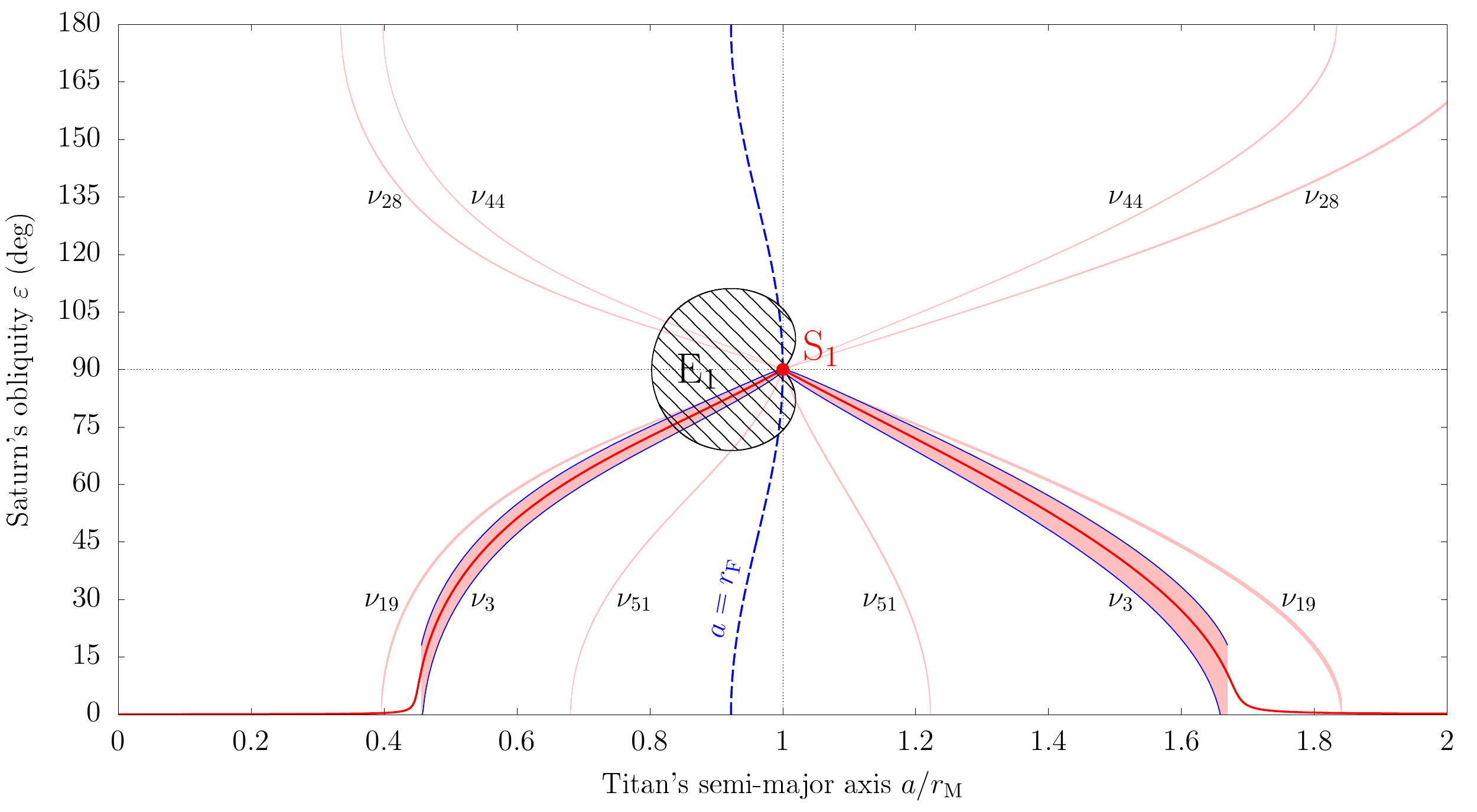}
   \caption{Location and width of every first-order secular spin-orbit resonance reachable by Saturn over the migration of Titan, with amplitudes down to $10^{-8}$. Each resonant angle is of the form $\sigma_j = \psi+\phi_j$ where $\phi_j$ has frequency $\nu_j$ labelled on the graph according to its index in the orbital series (see Table~\ref{tab:zetashort}). For a given value of Titan's semi-major axis, the interval of obliquity enclosed by the separatrix is shown in pink. The centre of the resonance (i.e. Cassini state 2) with $\phi_3$ is shown by a red line and the upper and lower separatrices are highlighted in blue. The ridge line $a=r_\mathrm{F}$ dividing the close-in and far-away satellite regimes is represented by a dashed blue line. The region E$_1$ where the satellite's Laplace state P$_1$ is unstable is shown in black. The singular point S$_1$ is shown by a red dot.}
   \label{fig:SaturnTitanRes}
\end{figure*}

Figure~\ref{fig:SaturnTitanRes} shows the location and width of the first-order secular spin-orbit resonances reachable by Saturn as a function of the distance of Titan. The full effect of Titan on Saturn's spin-axis is taken into account, including its contribution in $\lambda'$ (see Eq.~\ref{eq:J2prime}). For each resonance, the location of the Cassini states and the separatrix are obtained using the exact analytical formulas of \cite{Saillenfest-etal_2019a}; however, since Titan's contribution to the precession constant $\alpha$ itself depends on the obliquity $\varepsilon$ (see Eq.~\ref{eq:J2prime}), the equations become implicit and must be solved numerically (e.g. with the bisection method). As expected, all five resonances converge at the singular point S$_1$. If the planet is trapped in a secular spin-orbit resonance during the migration of its satellite, we see that it can behave very differently according to whether the satellite migrates outwards or inwards. For an outward migration, the system goes straight across the unstable region E$_1$ before reaching the singularity S$_1$; therefore, the satellite is expected to become gradually eccentric and eventually destabilise (see Sect.~\ref{ssec:eccLap} and \citealp{Tremaine-etal_2009}). On the contrary, for an inward migration, the system can go very close to S$_1$ before being brutally destabilised at the singularity.

Figure~\ref{fig:SaturnTitanRes} can be compared to Figs.~1 and 17 of \cite{Saillenfest-etal_2021}, drawn in terms of Saturn's effective precession constant (the vertical and horizontal axes are inverted). Figure~17 of \cite{Saillenfest-etal_2021} is a good example of how using a formalism with the precession constant $\alpha$ can be misleading when the satellite gets close to its Laplace radius. Indeed, because of the frequency cut at $\Omega_\mathrm{max}\approx 1.41''\,$yr$^{-1}$, Saturn would be unable to reach $\nu_{14}$ and $\nu_{15}$ for any distance of Titan, even though the trajectory in Fig.~17 of \cite{Saillenfest-etal_2021} appears at the same height as $\nu_{14}$ and $\nu_{15}$ on the graph. Moreover, as shown in Sect.~\ref{ssec:spinandsat}, $\alpha$ can tend to infinity while the spin-axis precession frequency remains finite, which is quite counter-intuitive. In order to prevent misinterpretations, we advocate avoiding using $\alpha$ as parameter when the satellite approaches $a=r_\mathrm{M}$, and using the general formula in Eq.~\eqref{eq:J2prime} or the model of \cite{Boue-Laskar_2006} for the spin-axis precession.

\begin{figure*}
   \centering
   \includegraphics[width=0.8\textwidth]{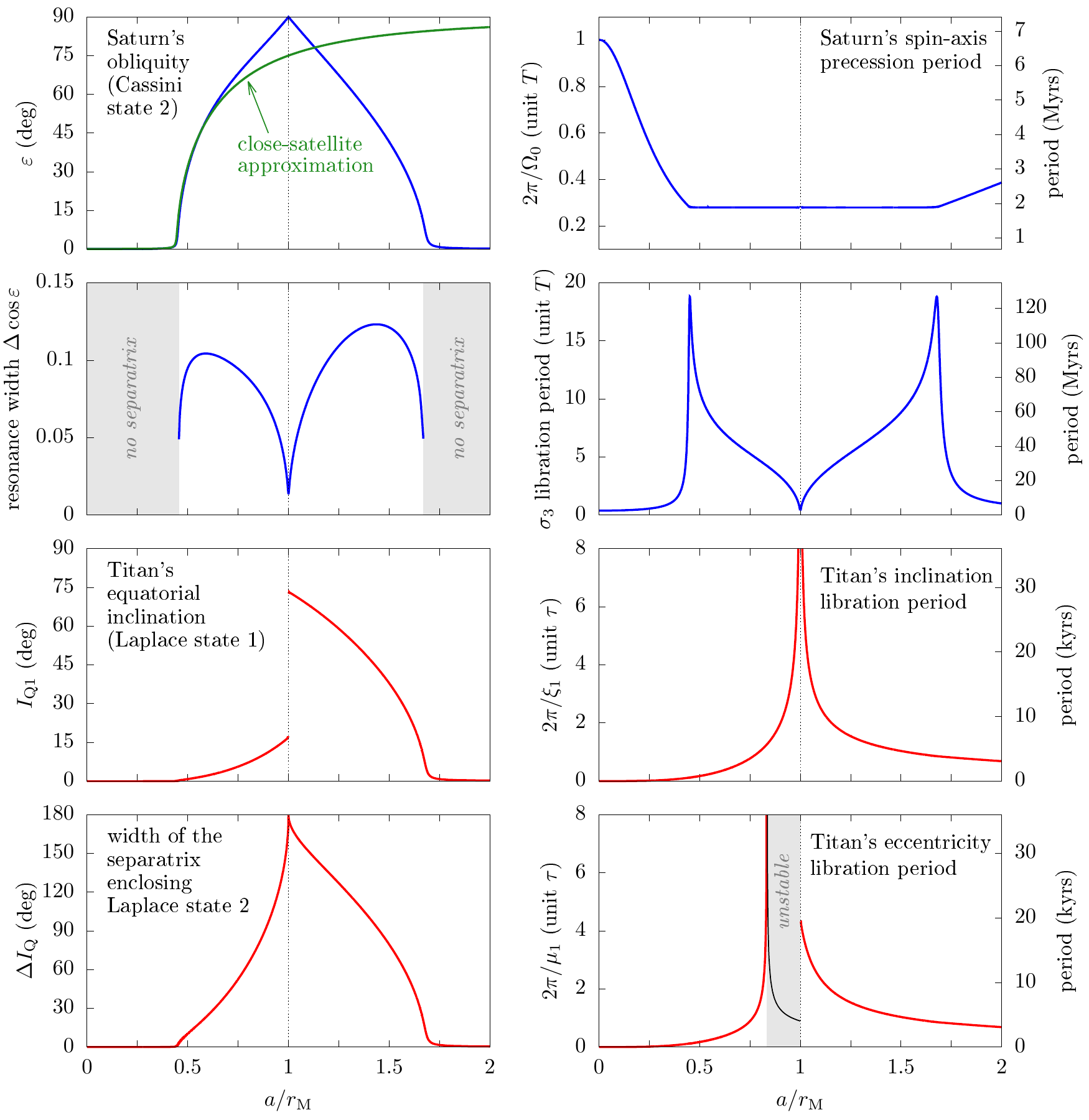}
   \caption{Evolution of Saturn's spin axis and Titan's orbit while following the centre of the secular spin-orbit resonance with $s_8$. We use blue for Saturn's spin axis and red for Titan's orbit. Each panel is described directly on the graph when the legend is not self-explanatory. In the right column, time is shown both in normalised units (left vertical axis) and in physical units (right vertical axis). The conversion factors are given in Table~\ref{tab:param}. All curves are obtained through the analytical formulas described in the text, assuming that Saturn closely oscillates near Cassini state 2 and Titan closely oscillates near Laplace state 1. In the top left panel, the green curve shows the location of the resonance centre according to the formula used by \cite{Saillenfest-etal_2021} valid for a close satellite. In the bottom right panel, the black curve in the unstable region corresponds to the period needed for the eccentricity to be multiplied by $\exp(2\pi)\approx 535$.}
   \label{fig:STEqevol}
\end{figure*}

Assuming that Saturn follows the centre of the resonance with $s_8$ (Cassini state 2) and Titan remains in its Laplace plane as it migrates (Laplace state 1), all the properties of the system can be monitored through the analytical formulas given in Sects.~\ref{sec:orb} and \ref{sec:plspin}. The general behaviour of the system is presented in Fig.~\ref{fig:STEqevol}. On the top left panel, we see that the close-satellite approximation used in previous articles underestimates the obliquity increase of Saturn, even though it remains valid up to an obliquity of about $60^\circ$. As detailed in Sect.~\ref{sec:orb}, we note that Titan's inclination is very different according to whether it reaches the singular point at $a=r_\mathrm{M}$ from above or from below. For an outward migration, Titan reaches the singularity with a quite moderate inclination of about $15^\circ$. Yet, since the width of the separatrix enclosing Laplace state 2 tends to $180^\circ$ at $a=r_\mathrm{M}$, this gives an idea of the large inclination variations expected if Titan deviates from the exact equilibrium point, for instance because of a coupling with the eccentricity becoming unstable (see Sect.~\ref{ssec:eccstab}). The right column of Fig.~\ref{fig:STEqevol} gives an idea of the large separation of timescales between the dynamics of Titan and of Saturn's spin axis. The two timescales become commensurate only in the vicinity of S$_1$, where the libration period of $\sigma_3$ tends to zero while Titan's libration periods tend to infinity because of the nearby instability. On the top right panel, we can appreciate how Saturn's spin-axis precession period is maintained to a constant value as it enters the resonance.

Titan is located today at a mean semi-major axis of $a_0\approx 0.49$~$r_\mathrm{M}$. If Titan goes on migrating outwards, it will eventually reach $a=r_\mathrm{M}$ at some time in the future. The simplified migration law for Titan provided by \cite{Lainey-etal_2020} is
\begin{equation}\label{eq:aTit}
a(t) = a_0\left(\frac{t}{t_0}\right)^b\,,
\end{equation}
where $t_0$ is Saturn's current age and $b$ is a real parameter. According to this formula, Titan will reach $a=r_\mathrm{M}$ after an interval of time from today equal to
\begin{equation}
\Delta t = t_0\left[\left(\frac{r_\mathrm{M}}{a_0}\right)^{1/b} - 1\right] \,.
\end{equation}
The astrometric measurements of \cite{Lainey-etal_2020} yield values of $b$ ranging in $[0.18,1.71]$, and their radio-science experiments yield values ranging in $[0.34,0.49]$. We deduce that if Titan goes on migrating as expected, it will reach $a=r_\mathrm{M}$ between about $2.4$ and $230$~Gyrs from now according to astrometric measurements, and in about $15$ to $33$~Gyrs according to radio-science experiments. Figures~\ref{fig:SaturnTitanRes} and \ref{fig:STEqevol} show that the system will first reach the unstable region E$_1$ when $a\approx 0.83$~$r_\mathrm{M}$, that is, between about $1.6$ and $80$~Gyrs from now according to astrometric measurements and in about $8.7$ to $17$~Gyrs according to radio-science experiments. These large values show that the system is unlikely to destabilise before the Sun leaves the main sequence. Yet, Saturn and Titan are not located exactly at their respective equilibrium points, but they oscillate around them with substantial amplitudes. As shown by \cite{Tamayo-etal_2013}, this can speed up the destabilisation process.

Because of the instabilities described above, Saturn and Titan are not expected to exactly follow Cassini state 2 and Laplace state 1, especially when they approach the singularity point S$_1$. For this reason, Fig.~\ref{fig:STEqevol} can only provide a qualitative picture of the evolution of the system. Because of the intricate and multi-timescale nature of the dynamics, building a self-consistent numerical model for the evolution of Saturn and Titan is challenging: the system involves the orbital dynamics of both Titan and Saturn, the spin-axis dynamics of Saturn torqued by the Sun and by Titan, and other planets of the Solar System should be included as well in some way to produce the multi-harmonic orbital precession of Saturn. Such a numerical model should also be fast enough to be usable for gigayear propagations (while Titan's orbital period today is only a few weeks). The design of such a model and the statistical exploration of the chaotic behaviour of Titan and Saturn are left for future works. Yet, a precise idea of the outcomes of the instability can already be obtained by mixing the two simplified models presented in Sects.~\ref{sec:orb} and \ref{sec:plspin}: on the one hand we explore numerically the dynamics of Titan when Saturn's spin-axis drifts inside the resonance, and on the other hand we explore the behaviour of Saturn while Titan migrates in the Laplace surface.

\subsection{Titan's orbit}\label{ssec:Titan}

Using the Hamiltonian function in Eq.~\eqref{eq:Hfirst}, our setting is similar to the migration simulations of \cite{Tremaine-etal_2009}, except that both the semi-major axis of Titan and the obliquity of Saturn evolve over time as slow-varying parameters. As a first approximation, their evolution law is provided by the top-left panel of Fig.~\ref{fig:STEqevol} (blue curve). Therefore, in our simulations we make the obliquity of Saturn and the semi-major axis of Titan vary simultaneously, the latter evolving according to the migration law given by Eq.~\eqref{eq:aTit}. We tried various values of $b$ in the full uncertainty range $[0.18,1.71]$, but the statistics of the simulations result to be absolutely independent of Titan's migration velocity. Indeed, the gigayear timescale of Titan's orbital expansion always remains extremely large as compared to the timescale of its secular dynamics ($\tau\approx 4501$~yrs; see Table~\ref{tab:param} and Fig.~\ref{fig:libT1}). Yet, when Titan reaches the unstable region, we do obtain several possible outcomes due to the intrinsic chaotic divergence of trajectories.

\begin{figure}
   \includegraphics[width=\columnwidth]{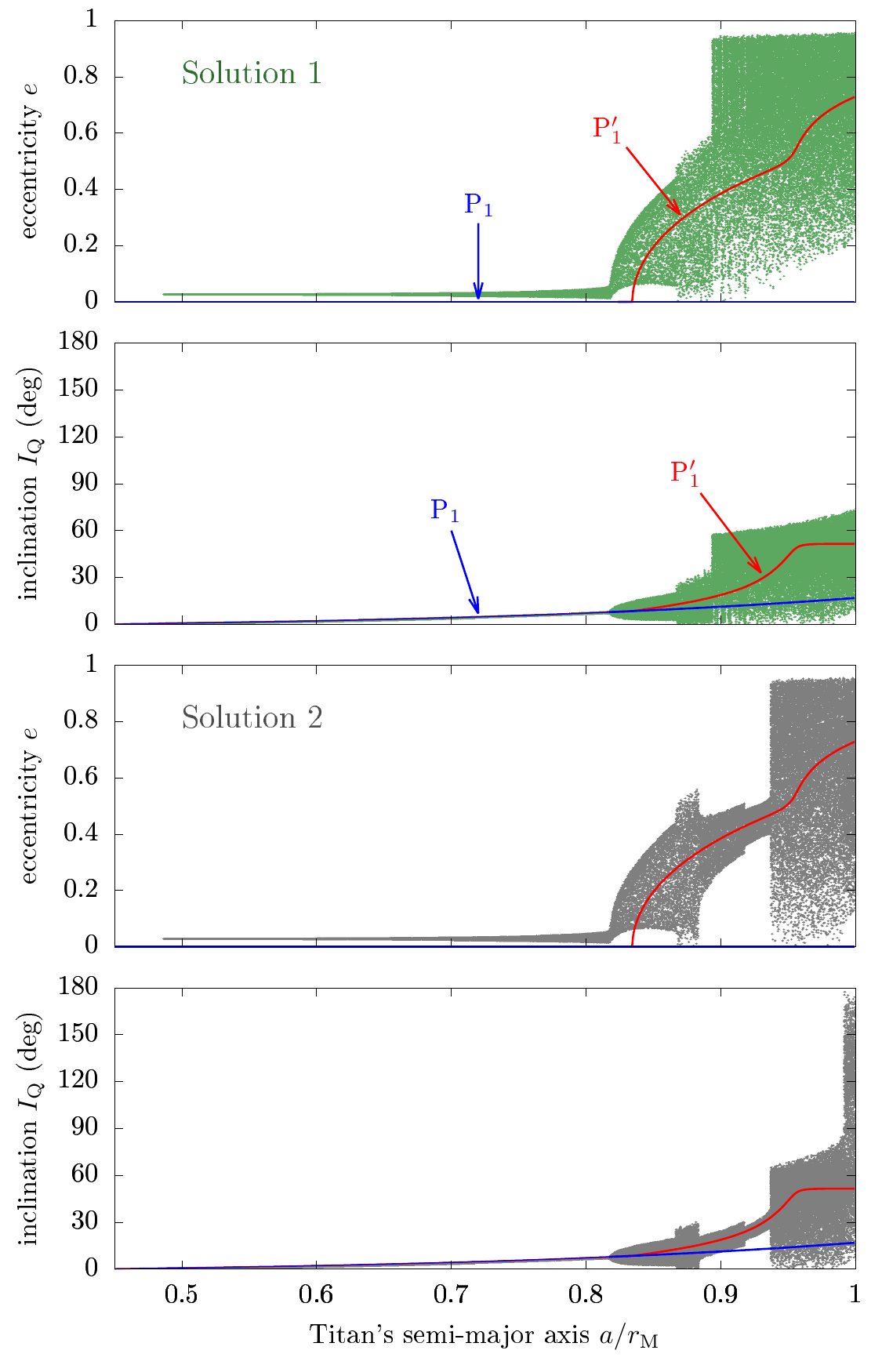}
   \caption{Long-term dynamics of Titan as it migrates away from Saturn. Saturn is assumed to remain locked in secular spin-orbit resonance with $s_8$ at all time; we take its obliquity evolution at the centre of the resonance (red curve in Fig.~\ref{fig:SaturnTitanRes}). Titan evolves according to the secular Hamiltonian function given in Eq.~\eqref{eq:Hfirst} in which its semi-major axis is a slowly varying parameter. The top and bottom panels give two outcomes of the chaotic transitions obtained by using a slightly different migration rate. In all panels, the blue curve shows the location of the circular Laplace state P$_1$, and the red curve shows the location of the eccentric Laplace state P$_1'$.}
   \label{fig:integTitan}
\end{figure}

Figure~\ref{fig:integTitan} shows two examples of simulations. As expected, Titan closely follows its local Laplace plane (blue curve), until it reaches the neighbourhood of the unstable region E$_1$. At this point, Titan's eccentricity increases, as the trajectory wanders in the vicinity of the stable eccentric equilibrium P$_1'$ (red curve). Then, P$_1'$ becomes unstable, as shown by the hatched region in Fig.~\ref{fig:P1primeStab}, and Titan's evolution becomes chaotic. This is where the two solutions in Fig.~\ref{fig:integTitan} diverge. In the case labelled `Solution~1', Titan jumps right away to a trajectory reaching very high eccentricity and inclination values. In the case labelled `Solution~2', Titan catches the eccentric equilibrium P$_1'$ when this equilibrium becomes stable again (see Fig.~\ref{fig:P1primeStab}), but it eventually goes back to the unstable zone where it reaches high eccentricity and inclination values. In this example, a new transition occurs just before the end of the simulation, and Titan's equatorial inclination starts oscillating between $0^\circ$ and $180^\circ$. In both cases presented in Fig.~\ref{fig:integTitan}, the ecliptic inclination (not shown) at the end of the integration also oscillates roughly between $0^\circ$ and $180^\circ$. Such extreme inclination variations are not surprising: in the circular case, Fig.~\ref{fig:phase3D} shows how the level curves of the Hamiltonian pass from a horizontal structure for $\varepsilon=0^\circ$ (panel c) to a vertical structure at S$_1$ (panel b), where the orbit can roll all the way around its nodal line. This is traduced by the extent of the separatrix that reaches $180^\circ$ at S$_1$ (see the bottom left panel of Fig.~\ref{fig:STEqevol}).

\begin{figure}
	\includegraphics[width=\columnwidth]{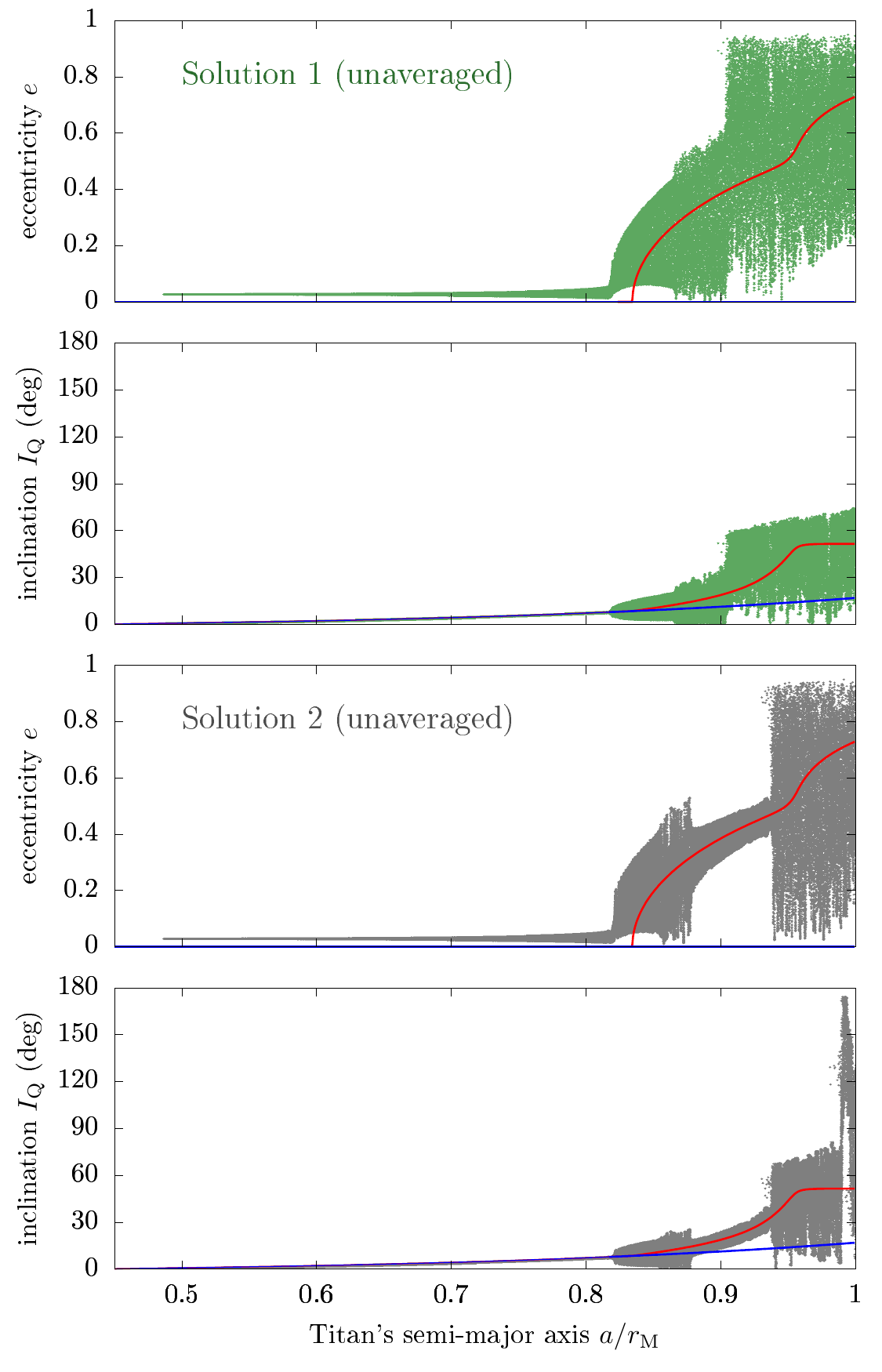}
	\caption{Same as Fig.~\ref{fig:integTitan}, except that the integration is performed using the unaveraged equations of the restricted three-body problem including Saturn's oblateness. The output is then digitally filtered to remove its short-period component. Due to the chaotic divergence of trajectories, we ran several simulations with different migration velocity and chose two of them that resemble the trajectories in Fig.~\ref{fig:integTitan}.}
	\label{fig:integTitan_osc}
\end{figure}

For comparison, we also performed direct unaveraged numerical integrations of the restricted three-body problem including Saturn's oblateness. In order to reproduce the drift in Titan's semi-major axis, we added in its equations of motion a small additional acceleration that depends on its velocity, and Saturn's obliquity is varied accordingly. No orbital or spin-axis precession motions for Saturn are included. In these simulations, Titan's migration is sped up by a factor of about $500$ as compared to Eq.~\eqref{eq:aTit}, which yields reasonable computation times (a few days or so). As explained above, this acceleration factor is justified by the extremely large separation between Titan's orbital timescale (thousands of years) and its migration timescale (billions of years). Figure~\ref{fig:integTitan_osc} shows two examples of such simulations, chosen for their similarity with Fig.~\ref{fig:integTitan}. They show that the secular model truncated at quadrupole order used throughout this article does capture the essence of the dynamics. In particular, the evection and `ivection' resonances reported by \cite{Speedie-etal_2020} to produce additional unstable regions are not found to play any role in Titan's future evolution.

Of course, when Titan's eccentricity increases and begins to oscillate widely, our model breaks down because the influence of Titan on Saturn's spin-axis precession is no longer given by Eq.~\eqref{eq:J2prime}. As a result, Saturn's obliquity should no longer follow the law given in Fig.~\ref{fig:SaturnTitanRes}. In order to explore the dynamics in the unstable region, Saturn's spin-axis motion should instead be integrated as well in a self-consistent way. Yet, Figs.~\ref{fig:integTitan} and \ref{fig:integTitan_osc} already give an idea of what can happen to Titan's orbit when the system reaches the unstable region E$_1$ and the vicinity of the singular point S$_1$. We see that its eccentricity and inclination can reach almost any value. In particular, Titan goes well below its Roche radius in Figs.~\ref{fig:integTitan} and \ref{fig:integTitan_osc}.

\subsection{Saturn's spin-axis}

Using the Hamiltonian function in Eq.~\eqref{eq:Hamrot}, our setting is similar to the migration simulations of \cite{Saillenfest-etal_2021}, except that both the semi-major axis and the inclination of Titan evolve over time as slowly varying parameters. The problem with this approach, and the reason why it has not been used in previous works, is that Titan's inclination (when it is not in the close-in or far-away regime) depends on Saturn's obliquity. Therefore, Saturn's effective precession constant $\alpha'$ depends on the obliquity in a complicated way (see Sect.~\ref{ssec:spinandsat}), and we lose the Hamiltonian structure described by Eq.~\eqref{eq:Hamrot}. 

Yet, except in the vicinity of the singular point S$_1$, the dependence of $\alpha'$ on the obliquity is weak. Therefore, at first level of approximation, the equations of motion obtained from the Hamiltonian in Eq.~\eqref{eq:Hamrot} are still valid, and the dependence of $\alpha'$ on the obliquity can be added afterwards in the equations of motion, in order to account for its long-term drift. Including Titan's Laplace plane inclination in $\alpha'$ means that for any obliquity variation of Saturn, Titan instantly moves at the new equilibrium configuration. As explained above, this approximation is justified by the large separation between the two timescales, but it fails near S$_1$, where the inclination variations of Titan as a function of obliquity are extremely sharp (and discontinuous exactly at S$_1$). But as shown in Sect.~\ref{ssec:Titan}, Titan is expected anyway to be destabilised before actually reaching S$_1$. Therefore, we stress that this model is not self-consistent; we use it here as a quick way to assess the relevance of our analytical predictions in Sect.~\ref{ssec:overview}, and to give a first qualitative picture of the different possible trajectories for Saturn. Apart from the obliquity dependence in $\alpha'$, our model is the same as that of \cite{Saillenfest-etal_2021}: the orbit of Saturn evolves according to the full series of \cite{Laskar_1990}, and Titan is made to migrate outwards according to the migration law in Eq.~\eqref{eq:aTit}. Since Saturn's polar moment of inertia and Titan's migration rate are not well known, we perform a large number of simulations with parameters $(b,\lambda)$ sampled in their uncertainty ranges. We refer to \cite{Saillenfest-etal_2021} for a discussion about all parameters and their uncertainties. We note that because of our rigid rotation model, $\lambda$ can be considered as an effective parameter that may slightly differ from what would be obtained using a refined model with differential rotation.

\begin{figure*}
	\centering
	\includegraphics[width=0.8\textwidth]{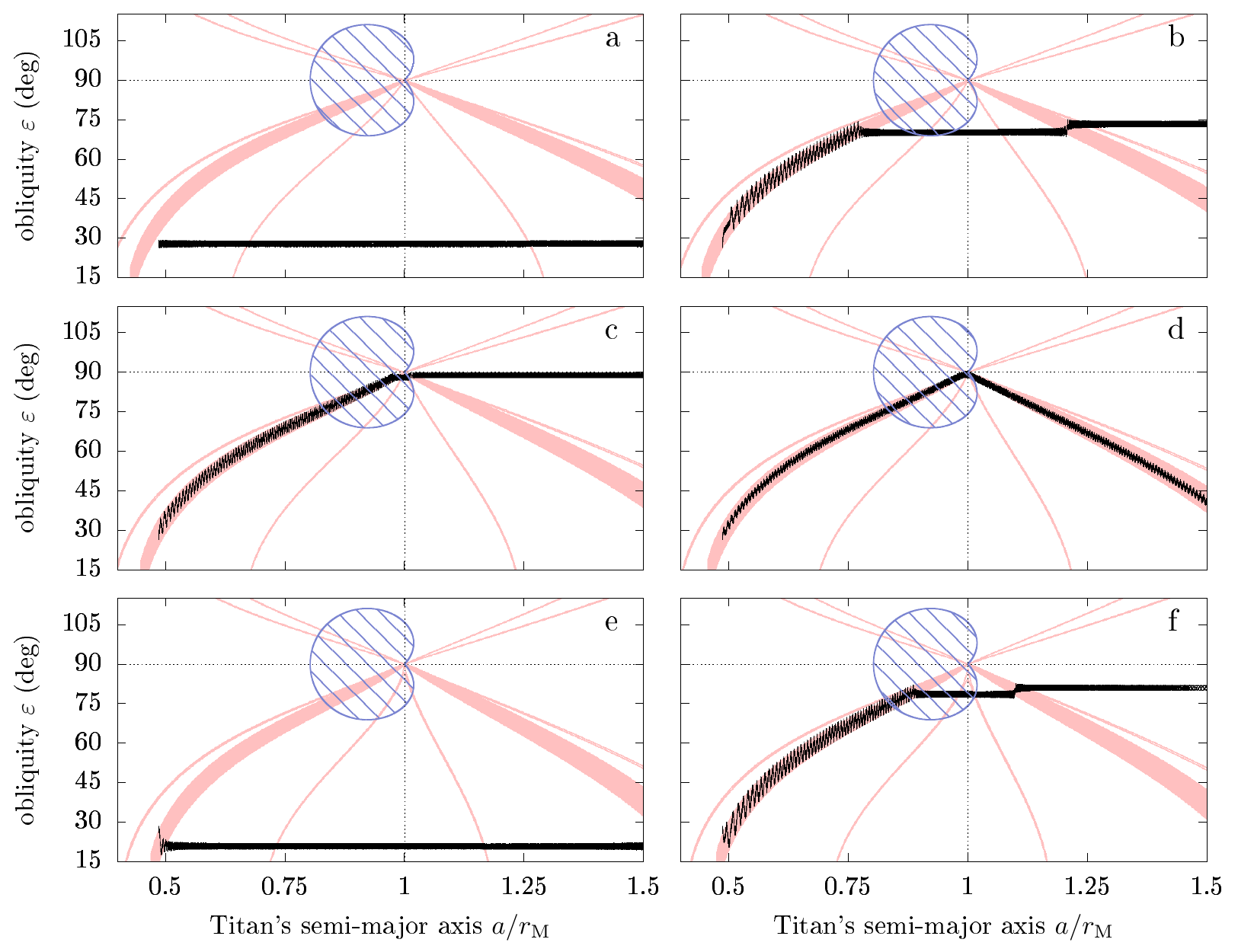}
	\caption{Numerical integration of Saturn's spin-axis dynamics for Titan remaining at all time in its circular Laplace equilibrium as it migrates away. Titan's migration parameter is set to $b=1$ (see Eq.~\ref{eq:aTit}) and each panel corresponds to a given value of Saturn's polar moment of inertia (see the red labels in Fig.~\ref{fig:obmax}). Trajectories are shown in black. The pink bands are Saturn's first-order secular spin-orbit resonances, and the blue hatched area is the region E$_1$ where the Titan's circular equilibrium is unstable (same as Fig.~\ref{fig:SaturnTitanRes}).}
	\label{fig:integSaturn}
\end{figure*}

Figure~\ref{fig:integSaturn} shows examples of trajectories for six different values of Saturn's normalised polar moment of inertia $\lambda$. Contrary to \cite{Saillenfest-etal_2021}, we do not represent the trajectories as a function of Saturn's precession constant because this constant is ill-defined near the singular point S$_1$ (see Sect.~\ref{ssec:spinandsat}). In order to compare Fig.~\ref{fig:integSaturn} with previous works, we stress that the initial point of the trajectory is the same on each panel; what differs here is the locations of the resonances, which are slightly shifted from one value of $\lambda$ to another. We recognise the different types of trajectories described by \cite{Saillenfest-etal_2021}:

In Panel~a, Saturn is currently out of the resonance and it goes farther away as Titan migrates. The two crossings of the $\nu_{51}$ resonance do not produce substantial obliquity variations.

In Panel~b, Saturn currently oscillates inside the resonance with a large amplitude. When the resonance width decreases, the adiabatic invariant cannot be conserved and Saturn escapes the resonance by crossing the separatrix. In this case, Saturn reaches a large obliquity, but Titan may remain stable anyway because it does not enter deep inside the unstable zone E$_1$ (hatched region). Eventually, Saturn crosses the $s_8$ resonance again when Titan passes in the far-satellite regime, producing an obliquity kick.

In Panels~c and d, Saturn currently oscillates closely around the resonance centre (with a minimum libration amplitude of $30^\circ$ or so; see \citealp{Ward-Hamilton_2004}). As shown in previous works, this configuration is the most likely in a dynamical point of view, regardless of the actual mechanism that is responsible for Saturn's resonance encounter with s$_8$ \citep{Hamilton-Ward_2004,Boue-etal_2009,Vokrouhlicky-Nesvorny_2015,Saillenfest-etal_2021}. In this case, we see that Saturn is able to get very close to the singular point S$_1$ before being destabilised, because the neighbouring resonances are thin and their overlap does not produce much chaos. After the chaotic transition, Saturn can either be ejected from the resonance with an obliquity of about $90^\circ$ or slightly more (Panel~c), or it can be recaptured at once when the resonance width increases again (Panel~d). However, as shown in Sect.~\ref{ssec:Titan}, Titan is expected to be completely destabilised before reaching S$_1$, so the evolution of Saturn's obliquity should remain frozen at some point as Titan is removed (collision or ejection). The questions of where this transition happens and what is the statistical outcome of the destabilisation would require a self-consistent numerical model; these questions are left for future works.

In Panels~e and f, Saturn did not reach yet the resonance with $s_8$ today. As discussed by \cite{Saillenfest-etal_2021}, this configuration would require a value of $\lambda$ that is slightly out of its expected range (namely $\lambda\gtrsim 0.241$ while we expect $\lambda\in[0.200,0.240]$). We include it here for completeness. As Saturn encounters the resonance, it can either cross it without being captured (Panel~e), in which case its obliquity suffers from a small kick, or it can be captured (Panel~f), in which case we end up with the same kind of evolution as in Panel~b.

\begin{figure*}
	\includegraphics[width=\textwidth]{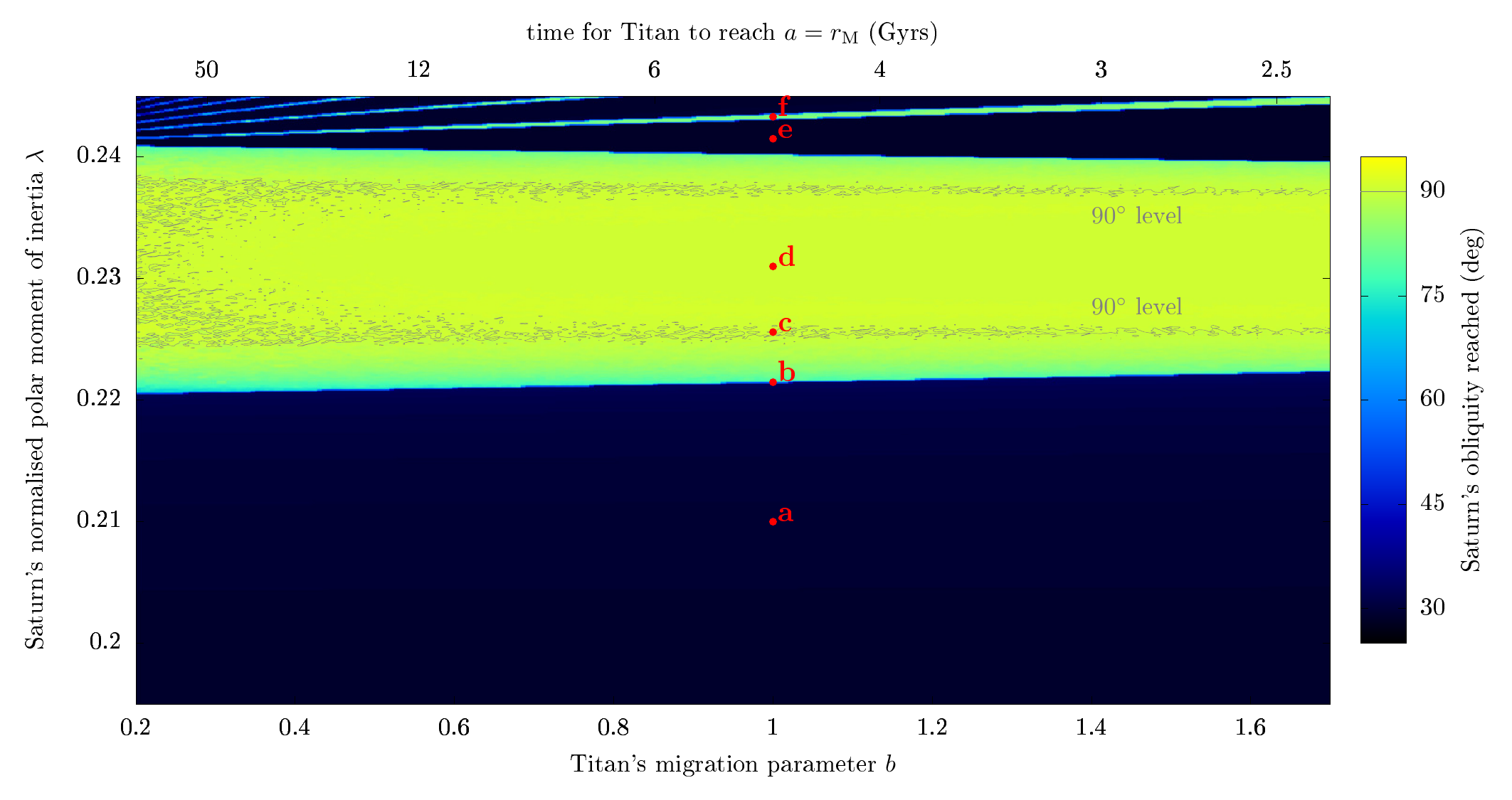}
	\caption{Maximum obliquity reached by Saturn for Titan migrating from its current location up to $a=r_\mathrm{M}$. Titan is assumed to remain at all time in its circular Laplace equilibrium. The integration time (top horizontal axis) depends on Titan's migration parameter $b$ (bottom horizontal axis). The noisy grey curve is the $90^\circ$ obliquity level. The red labels show the location of the six examples of trajectory presented in Fig.~\ref{fig:integSaturn}.}
	\label{fig:obmax}
\end{figure*}

As a summary of these numerical experiments, Fig.~\ref{fig:obmax} shows the maximum obliquity reached by Saturn for Titan migrating from its current location $a=a_0$ up to its midpoint radius $a=r_\mathrm{M}$. The figure shows the results obtained in a grid made of $250$ values of $b$ and $501$ values of $\lambda$. We stress that, contrary to previous works, the integration duration is not the same for each run, but depends on $b$ (see the top horizontal axis). The bottom dark-blue region in Fig.~\ref{fig:obmax} corresponds to the cases where Saturn is out of the resonance today and goes farther away as Titan migrates. The large coloured region corresponds to values of the parameters that put Saturn inside the resonance today. It is narrower for larger $b$ because large-amplitude librations are unstable if Titan's migration is too fast. The top region (which is out of the expected range for $\lambda$) corresponds to cases where Saturn did not reach yet the resonance today but will in the future, resulting in a capture (coloured stripe) or not (dark background). See Fig.~16 of \cite{Saillenfest-etal_2021} for a discussion about this striped pattern.

From Fig.~\ref{fig:obmax}, we conclude that Saturn gets extreme obliquities in a large region of the parameter space. If Saturn is inside the resonance today, then its obliquity reaches at least about $75^\circ$. This limit is robust in spite of our simplified model because it is reached before encountering the unstable region (see Fig.~\ref{fig:integSaturn}b). The preferred parameter range for Saturn in previous studies \citep{Boue-etal_2009,Vokrouhlicky-Nesvorny_2015,Saillenfest-etal_2021} is precisely the one producing the maximum obliquity increase (about $91^\circ$ in our simplified model); it is also the one producing the maximum destabilisation for Titan (see Sect.~\ref{ssec:Titan}). However, according to the exact migration rate of Titan, the system may not reach the instability by the end of the Sun's main sequence.

\section{Summary and discussions}\label{sec:ccl}
Titan is observed to migrate away from Saturn much faster than previously thought \citep{Lainey-etal_2020}. This migration is likely responsible for Saturn's current axis tilt of $26.7^\circ$ \citep{Saillenfest-etal_2021a}, in which case Saturn should still be trapped today in secular spin-orbit resonance with Neptune's nodal precession mode $s_8$. Since Titan goes on migrating today, Saturn's obliquity is expected to increase in the future \citep{Saillenfest-etal_2021}. In this article, we investigated the final outcome of this mechanism, and the behaviour of the system when Titan will cross its Laplace radius.

The intricate nature of the dynamics requires the evolution of both the satellite's orbit and the planet's spin axis to be studied. Building on the work of \cite{Tremaine-etal_2009}, we have shown that the three circular equilibria for the satellite (dubbed here Laplace states P$_1$, P$_2$, and P$_3$) are organised around two critical regions S$_1$ and S$_2$ in the parameter space, in which the pairs $(\mathrm{P}_1,\mathrm{P}_3)$ or $(\mathrm{P}_2,\mathrm{P}_3)$, respectively, are degenerate. In particular, $\mathrm{S}_1$ is defined as the point where the planet's obliquity is $\varepsilon=90^\circ$ and the satellite's semi-major axis is $a=2^{1/5}r_\mathrm{L}$ (where $r_\mathrm{L}$ is the Laplace radius defined by \citealp{Tremaine-etal_2009}).

We found that all three circular equilibria bifurcate to eccentric equilibrium configurations (noted $\mathrm{P}_1'$, $\mathrm{P}_2'$, $\mathrm{P}_3'$, and $\mathrm{P}_3''$) in some regions of the parameter space. The location of all eccentric equilibria can be expressed with explicit parametric representations. The critical regions $\mathrm{S}_1$ and $\mathrm{S}_2$ both have an eccentric continuation, in which the pairs $(\mathrm{P}_1',\mathrm{P}_3')$ or $(\mathrm{P}_2',\mathrm{P}_3'')$, respectively, are degenerate.

Regular satellites like Titan form in their classical Laplace plane (i.e. at equilibrium $\mathrm{P}_1$). As long as the equilibrium remains stable, they stay in its vicinity during their orbital migration, and their orbital inclination varies accordingly (see e.g. \citealp{Tremaine-etal_2009}). Using numerical integrations, we verified that this is indeed the case for Titan, even when taking into account its fast orbital expansion measured by \cite{Lainey-etal_2020}.

As shown in previous works, satellites increase the mean spin-axis precession rate $\Omega_0$ of their host planet, with a magnitude that depends on their orbital distance (see e.g. \citealp{Boue-Laskar_2006}). Consequently, if the planet is trapped in a secular spin-orbit resonance, any migration of its satellites is compensated by an obliquity change, so that $\Omega_0$ is maintained fixed at the resonant value (usually called Cassini state~2). In other words, the planet and its satellite follow a level curve of $\Omega_0$ in the plane $(a,\varepsilon)$. When the satellite's inclination is fully taken into account (i.e. without the usual close-in or far-away approximations), the singularity $\mathrm{S}_1$ in its orbital dynamics is transferred to the spin-axis precession rate of the planet. As a result, if the satellite is massive enough, most level curves of $\Omega_0$ converge towards the singularity $\mathrm{S}_1$. This means that if the satellite reaches the vicinity of its Laplace radius during its migration, the obliquity of its host planet is driven towards $\varepsilon=90^\circ$. If the resonance is large, the planet can even go beyond this limit. Simplified analytical formulas give the conditions required for a full $90^\circ$-tilt. These conditions are met for Titan and Saturn and the $s_8$ resonance, and confirmed using numerical simulations.

In the vicinity of $\mathrm{S}_1$, however, several kinds of instability are expected to happen. Firstly, $\mathrm{S}_1$ lies at the border of a region in the parameter space where the circular Laplace state $\mathrm{P}_1$ is unstable. As discussed in previous works \citep{Tremaine-etal_2009,Tamayo-etal_2013}, when a satellite crosses this border, it can transfer to the stable eccentric equilibrium P$_1'$, but this equilibrium soon becomes unstable, too. At this point, numerical integrations show that Titan's eccentricity and inclination completely destabilise, potentially allowing for the ejection of Titan or its collision on Saturn. In a hypothetical system in which the satellite would migrate inwards, the destabilisation is expected to be more violent, because in this case the system would smoothly reach the extreme vicinity of $\mathrm{S}_1$, where the instability is strongest, before being destabilised.

Secondly, all neighbouring secular spin-orbit resonances converge towards $\mathrm{S}_1$. The planet's spin axis is therefore expected to reach a chaotic region at some point, produced by resonance overlap. In the case of Saturn, the neighbouring resonances are very thin, and numerical experiments show that the chaotic region is restricted to a very small region near $\mathrm{S}_1$.

Thirdly, the widths of all secular spin-orbit resonances decrease in the vicinity of $\mathrm{S}_1$. As a result, if the libration amplitude of the planet inside the resonance is too large, it can be ejected before actually reaching $\varepsilon=90^\circ$. Using a simplified numerical model, we performed a preliminary survey of Saturn's behaviour and explored a large range for its poorly known moment of inertia. We found that Saturn can indeed be ejected from resonance with an obliquity $\varepsilon\gtrsim 75^\circ$, depending on its libration amplitude inside the resonance. Regardless of the mechanism responsible for Saturn's capture in secular spin-orbit resonance, previous studies show that Saturn is likely located deep inside the resonance today \citep{Boue-etal_2009,Vokrouhlicky-Nesvorny_2015,Saillenfest-etal_2021}. In this case, Saturn's obliquity increase is maximum, and it may reach $\varepsilon\approx 91^\circ$ in the future (provided that Titan is not ejected before; see above). Determining the statistical outcome of this double dynamical instability for Saturn and Titan is left for future works. It will require a self-consistent model coupling the orbital motions of Saturn and Titan, the spin-axis dynamics of Saturn torqued by the Sun and by Titan, and the multi-harmonic orbital precession of Saturn produced by the other planets of the Solar System.

Our results about Saturn and Titan rely on the assumption that Titan will go on migrating for gigayears in the future. If instead, Titan's migration rate strongly drops before reaching the unstable zone (i.e. if Titan is released out of the tidal resonance-locking mechanism of \citealp{Fuller-etal_2016}) the system would remain frozen, with roughly constant obliquity for Saturn and fixed orbit for Titan. However, to our knowledge, there is no evidence showing that Titan's migration would stop in the future (at least not before it becomes strongly unstable). The timescale required to tilt Saturn also plays an important role. According to the precise migration rate of Titan, Saturn and Titan should reach the instability region between a few gigayears and several tens of gigayears from now, provided that Titan goes on migrating as expected. Therefore, the evolution timescale may be too slow for the system to reach instability by the end of the Sun's main sequence. Yet, even if it does not reach instability in time, Saturn would anyway get very large obliquity values, as already described in Fig.~16 of \cite{Saillenfest-etal_2021}.

Moreover, the mechanism described here is generic: it only requires a secular spin-orbit resonance and a substantially massive migrating moon. The structure of the resonance and its convergence towards $\varepsilon=90^\circ$ near the moon's Laplace radius are the same for any (exo)planet considered, as well as the mechanism of destabilisation. In particular, tilting Uranus on a gigayear timescale by a former satellite now removed by the instability is a promising mechanism that has not been explored yet \citep{Boue-Laskar_2010,Rogoszinski-Hamilton_2020,Rogoszinski-Hamilton_2021}. This mechanism may also provide a dynamical explanation for the `super-puff' exoplanets thought to possess a massive face-on ring \citep{Akinsanmi-etal_2020,Piro-Vissapragada_2020}: the destroyed satellite would both provide the tilting mechanism (for the obliquity to be near $90^\circ$) and the ring material. More work is now needed to assess the feasibility of this scenario to these specific systems.

\begin{acknowledgements}
	We thank Ariane Courtot for her work that led us to a deeper understanding of the coupling between satellite and spin axis. M.~S. also thanks Aurore Mazur for her contribution to the study of Titan's destabilisation mechanism. We are grateful to the anonymous referee for having carefully read our manuscript and pointed out some ambiguity in terminology.
\end{acknowledgements}

\bibliographystyle{aa}
\bibliography{crazytitan}

\appendix

\section{Conversion formulas between the equator and ecliptic reference frames}\label{asec:CQ}

   The ecliptic reference frame is defined with the third axis perpendicular to the planet's orbit; we write $(I_\mathrm{C},\omega_\mathrm{C},\Omega_\mathrm{C})$ the Keplerian elements of the satellite measured in this frame, which are, respectively, the inclination, the argument of pericentre, and the longitude of the ascending node. The equator reference frame is defined with the third axis perpendicular to the planet's equator; we write $(I_\mathrm{Q},\omega_\mathrm{Q},\Omega_\mathrm{Q})$ the Keplerian elements of the satellite measured in this frame. Passing from one frame to the other corresponds to a rotation of $\pm\varepsilon$ around the mutual line of nodes of the two reference planes (i.e. the line joining both equinoxes), $\varepsilon$ being the obliquity of the planet. We obtain
   \begin{equation}
      \cos I_\mathrm{C} = \cos\varepsilon\cos I_\mathrm{Q} + \sin\varepsilon\sin I_\mathrm{Q}\cos\delta_\mathrm{Q} \,,
   \end{equation}
   and
   \begin{equation}
      \left\{
      \begin{aligned}
         \cos\omega_\mathrm{C}\sin I_\mathrm{C} &= \cos\varepsilon\cos\omega_\mathrm{Q}\sin I_\mathrm{Q} \\
         &+ \sin\varepsilon(\sin\omega_\mathrm{Q}\sin\delta_\mathrm{Q} - \cos\omega_\mathrm{Q}\cos I_\mathrm{Q}\cos\delta_\mathrm{Q}) \,,\\
         \sin\omega_\mathrm{C}\sin I_\mathrm{C}
         &= \cos\varepsilon\sin\omega_\mathrm{Q}\sin I_\mathrm{Q} \\
         &- \sin\varepsilon(\cos\omega_\mathrm{Q}\sin\delta_\mathrm{Q} + \sin\omega_\mathrm{Q}\cos I_\mathrm{Q}\cos\delta_\mathrm{Q}) \,,
      \end{aligned}
      \right.
   \end{equation}
   where $\delta_\mathrm{Q} \equiv \Omega_\mathrm{Q}-\Omega_\odot$ and $\Omega_\odot$ is the ascending node of the star measured along the equator of the planet. The first equation reverses as
   \begin{equation}
      \cos I_\mathrm{Q} = \cos\varepsilon\cos I_\mathrm{C} + \sin\varepsilon\sin I_\mathrm{C}\cos\delta_\mathrm{C} \,,
   \end{equation}
   where $\delta_\mathrm{C} \equiv \Omega_\mathrm{C}-\Omega_\mathrm{P}$ and $\Omega_\mathrm{P}$ is the ascending node of the planet's equator measured along the ecliptic. These three equations are all what we need to express the Hamiltonian function $\mathcal{H}$ in Eq.~\eqref{eq:Hfirst} in terms of the equator or ecliptic coordinates only. The expressions obtained are given in Eqs.~\eqref{eq:HpHs}, \eqref{eq:HpC} and \eqref{eq:HsQ}.

\section{Circular Laplace equilibria}\label{asec:circeq}

   In Sect.~\ref{sec:orb}, we describe the secular orbital dynamics of a massless satellite perturbed by the Sun and by the oblateness of its host planet at quadrupole order. We call the three kinds of circular equilibria the `Laplace states' $\mathrm{P}_1$, $\mathrm{P}_2$, and $\mathrm{P}_3$. Even though the equilibrium $\mathrm{P}_1$ is the one that matters most for regular satellites, all equilibria can play a role in the dynamics when the satellite becomes unstable and explores wide regions in the phase space (see Sect.~\ref{sec:sattit}). For this reason, we recall here the stability properties of all three circular equilibria.
   
   \subsection{Circular equilibrium $\mathrm{P}_1$}
   P$_1$ is stable to inclination variations (see Fig.~\ref{fig:phase}). For $\delta_\mathrm{Q}=0$, the equatorial inclination $I_{\mathrm{Q}1}$ of the satellite at $\mathrm{P}_1$ is given by Eq.~\eqref{eq:ILap13}. Inclination oscillations around the Laplace state $\mathrm{P}_1$ have frequency $\xi_1$ given in Eq.~\eqref{eq:nu1}. For a $\varepsilon=0^\circ$ or $180^\circ$, it simplifies into Eq.~\eqref{eq:nu10}.  Likewise, the value obtained for $\varepsilon=90^\circ$ is
   \begin{equation}
      \xi_1^2\Big|_{\varepsilon=90^\circ} =
      \left\{
      \begin{aligned}
         &-\frac{\kappa^2}{4}\left(\frac{r_\mathrm{M}^5}{a^5} - 1\right)\frac{r_\mathrm{M}^2}{a^2}
         &\quad\text{if}\quad a<r_\mathrm{M}\,,\\
         &-\frac{\kappa^2}{4}\left(1 - \frac{r_\mathrm{M}^5}{a^5}\right)\frac{a^3}{r_\mathrm{M}^3}
         &\quad\text{if}\quad a>r_\mathrm{M}\,,
      \end{aligned}
      \right.
   \end{equation}
   and we note that it tends to $0$ at the singular point S$_1$ (i.e. $a=r_\mathrm{M}$), since we retrieve the eigenfrequency $\xi_{13}^2=0$ along the degenerate stable circle $\mathscr{L}_{13}$. For a very close satellite, $\xi_1$ becomes independent of the obliquity:
   \begin{equation}\label{eq:nu1close}
      \xi_1^2\Big|_{a\rightarrow0} = -\frac{\kappa^2}{4}\frac{r_\mathrm{M}^7}{a^7} = - \left(\frac{3}{2}nJ_2\frac{R_\mathrm{eq}^2}{a^2}\right)^2 \,,
   \end{equation}
   where $n = \sqrt{\mu_\mathrm{P}/a^3}$ is the satellite's mean motion. This is the usual nodal precession frequency produced by the oblateness of the central body in absence of other perturber (see e.g.~\citealp{Murray-Dermott_1999}).
   
   P$_1$ is stable to eccentricity variations, except in the region $\mathrm{E}_1$ described in Sect.~\ref{ssec:eccstab}. Eccentricity oscillations around $\mathrm{P}_1$ have frequency $\mu_1$ given in Eq.~\eqref{eq:mu1}. For a zero-obliquity planet, Eq.~\eqref{eq:mu1} simplifies to
   \begin{equation}
      \mu_1^2\Big|_{\varepsilon=0} = -\frac{\kappa^2}{4}\left(\frac{r_\mathrm{M}^5}{a^5} + 1\right)^2\frac{a^3}{r_\mathrm{M}^3}\,,
   \end{equation}
   which is exactly equal to the inclination eigenfrequency $\xi_1^2$ given by Eq.~\eqref{eq:nu10}. Therefore, for a planet with a near-zero obliquity, the  oscillations of the inclination and eccentricity of a satellite around the classical Laplace state P$_1$ have the same frequencies. The value obtained for $\varepsilon=90^\circ$ is
   \begin{equation}\label{eq:mu190}
      \mu_1^2\Big|_{\varepsilon=90^\circ} =
      \left\{
      \begin{aligned}
         &-\frac{\kappa^2}{4}\left(\frac{r_\mathrm{M}^5}{a^5} - 3\right)\left(\frac{r_\mathrm{M}^5}{a^5} + 2\right)\frac{a^3}{r_\mathrm{M}^3}
         &\quad\text{if}\quad a<r_\mathrm{M}\,,\\
         &-\frac{\kappa^2}{16}\left(2 - \frac{r_\mathrm{M}^5}{a^5}\right)^2\frac{a^3}{r_\mathrm{M}^3}
         &\quad\text{if}\quad a>r_\mathrm{M}\,.
      \end{aligned}
      \right.
   \end{equation}
   We see that $\mu_1^2$ does not go to zero when $a\rightarrow r_\mathrm{M}$ and that it has two different limits; this means that going through the singularity S$_1$ produces a discontinuity of $\mu_1^2$. This could have been expected since P$_1$ and P$_3$ are inverted (i.e. there is a discontinuity in their location as well, jumping by $90^\circ$ in inclination; see Sect.~\ref{ssec:Lap}).
   
   For a very close satellite, $\mu_1$ becomes independent of the obliquity:
   \begin{equation}
      \mu_1^2\Big|_{a\rightarrow0} = -\frac{\kappa^2}{4}\frac{r_\mathrm{M}^7}{a^7} = - \left(\frac{3}{2}nJ_2\frac{R_\mathrm{eq}^2}{a^2}\right)^2 \,,
   \end{equation}
   which is the same value of $\xi_1^2$ in Eq.~\eqref{eq:nu1close}. This is the usual pericentre precession frequency produced by the oblateness of the central body in absence of other perturber (see e.g.~\citealp{Murray-Dermott_1999}).
   
   The eccentricity eigenfrequency $\mu_\mathrm{13}^2$ along the degenerate stable circle $\mathscr{L}_{13}$ (see Fig.~\ref{fig:phase3D}b) is given by
   \begin{equation}\label{eq:mu13}
      \mu_\mathrm{13}^2 = -\frac{\kappa^2}{16}\big(1+5\cos^2I_\mathrm{Q}\big)\big(1-5\cos^2I_\mathrm{Q}\big)\,.
   \end{equation}
   The value of $\mu_\mathrm{13}^2$ depends on the inclination of the satellite along the degenerate circle. We retrieve the discontinuous limits of Eq.~\eqref{eq:mu190} when $a\rightarrow r_\mathrm{M}$ by putting $I_\mathrm{Q}=0^\circ$ or $90^\circ$ in Eq.~\eqref{eq:mu13}. Along the degenerate circle, the satellite is stable only if $\cos^2I_\mathrm{Q}\leqslant 1/5$, that is, if $63^\circ\lesssim I_\mathrm{Q}\lesssim 117^\circ$. This corresponds to the region where the $J_2$-induced precession of $\omega_\mathrm{Q}$ is negative.
   
   \subsection{Circular equilibrium $\mathrm{P}_2$}
   P$_2$ is stable to inclination variations, and located at $\delta_\mathrm{Q}=\pm\pi/2$ and $I_\mathrm{Q}=90^\circ$ (see Fig.~\ref{fig:phase}). As described by \cite{Tremaine-etal_2009}, $\mathrm{P}_2$ corresponds to an orbit that is perpendicular to both the equator and the ecliptic planes. The frequency of small-amplitude inclination oscillations around P$_2$ is given by
   \begin{equation}\label{eq:xi2}
      \xi_2^2 = -\frac{\kappa^2}{4}\frac{r_\mathrm{M}^2}{a^2}\sin^2\varepsilon\,.
   \end{equation}
   As expected, the value of $\xi_2^2$ is always negative, and it tends to zero in the region S$_2$ of the parameter space (i.e. $\varepsilon=0^\circ$ or $180^\circ$), since we retrieve the eigenfrequency $\xi_{23}^2=0$ along the degenerate stable circle $\mathscr{L}_{23}$.
   
   The frequency of small-amplitude eccentricity oscillations around P$_2$ is given by
   \begin{equation}\label{eq:mu2}
      \mu_2^2 = -\frac{\kappa^2}{16}\left(\frac{r_\mathrm{M}^5}{a^5}+6\right)\left(\frac{r_\mathrm{M}^5}{a^5}-4\right)\frac{a^3}{r_\mathrm{M}^3} \,.
   \end{equation}
   We deduce that P$_2$ is stable to eccentricity variations only for close enough satellites that verify $a\leqslant r_3$, where we define
   \begin{equation}\label{eq:r3}
      r_3^5 = \frac{1}{4}r_\mathrm{M}^5\,.
   \end{equation}
   The region where P$_2$ is unstable to eccentricity variations is therefore
   \begin{equation}
      \mathrm{E}_2 = \Big\{a/r_\mathrm{M}>1/4^{1/5},\:\varepsilon\in[0^\circ,180^\circ]\Big\}\,.
   \end{equation}
   
   The eccentricity eigenfrequency $\mu_\mathrm{23}^2$ along the degenerate stable circle $\mathscr{L}_{23}$ (see Fig.~\ref{fig:phase3D}c) is also given by Eq.~\eqref{eq:mu2}. Hence, as for the Laplace state P$_2$, it is only stable for close-enough satellites.
   
   \subsection{Circular equilibrium $\mathrm{P}_3$}
   P$_3$ is unstable to inclination variations (see Fig.~\ref{fig:phase}). The squared eigenvalue $\xi_3^2$ of the linearised problem at P$_3$ is given by Eq.~\eqref{eq:nu1} where $I_{\mathrm{Q}1}$ must be replaced by $I_{\mathrm{Q}3}$, which is equal to $I_{\mathrm{Q}1}\pm 90^\circ$. As expected, $\xi_3^2>0$ whenever P$_3$ exists, and $\xi_3^2$ tends to $0$ at the singular regions S$_1$ and S$_2$, where we retrieve the eigenfrequencies $\xi_{13}^2=0$ and $\xi_{23}^2=0$ along the degenerate stable circles $\mathscr{L}_{13}$ and $\mathscr{L}_{23}$.
   
\begin{figure}
      \centering
      \includegraphics[width=0.9\columnwidth]{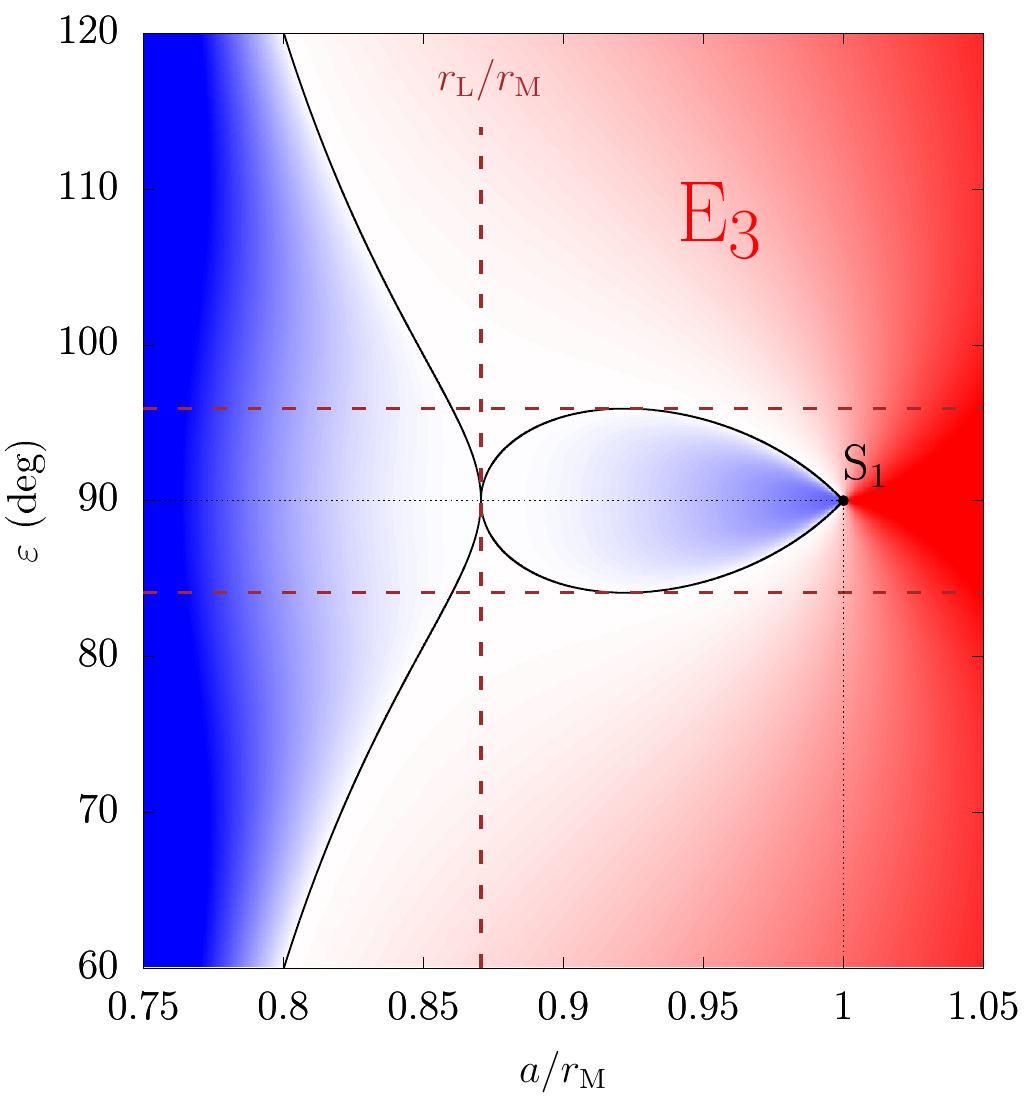}
      \caption{Region E$_3$ of the parameter space where the Laplace state P$_3$ is unstable to eccentricity variations. The colours and labels have the same meaning as in Fig.~\ref{fig:E1detail}, and the axes have the same scale. The border of E$_3$ is obtained using the closed-form expression in Eqs.~\eqref{eq:E3a} and~\eqref{eq:E1b}. The drop-like portion of the contour reaches its maximum width at $a^5/r_\mathrm{M}^5=2/3$ where the obliquity on the boundary is $\cos^2\varepsilon=(51-25\sqrt{3})/726$.}
      \label{fig:E3detail}
   \end{figure}

   Even though P$_3$ is unstable to inclination variations anywhere it exists in the parameter space, this is not the case for eccentricity variations \citep{Tremaine-etal_2009}. As illustrated in Figs.~\ref{fig:recap} and~\ref{fig:E3detail}, P$_3$ is stable to eccentricity variations for close-enough satellites and in a region resembling a water drop. The squared eigenvalue $\mu_3^2$ of the linearised problem at P$_3$ is given by Eq.~\eqref{eq:mu1} where $I_{\mathrm{Q}1}$ must be replaced by $I_{\mathrm{Q}3}=I_{\mathrm{Q}1}\pm 90^\circ$. We call E$_3$ the unstable region where $\mu_3^2>0$. Its border can be expressed piecewise as
   \begin{equation}\label{eq:E3a}
      \cos^2\varepsilon = \frac{-2u^2 + 13u + 38 - (6+u)\sqrt{-4u^2 + 12u + 41}}{2u}
   \end{equation}
   for $r_3\leqslant a\leqslant r_\mathrm{L}$, and Eq.~\eqref{eq:E1b} for $r_\mathrm{L}\leqslant a\leqslant r_\mathrm{M}$, recalling that $r_\mathrm{L}$ is the traditional `Laplace radius' defined in Eq.~\eqref{eq:rM}, and $r_3$ is given in Eq.~\eqref{eq:r3}.
   
\section{Eccentric Laplace equilibria}\label{asec:ecceq}

   The Hamiltonian function in Eq.~\eqref{eq:Hfirst} describes the secular orbital dynamics of a massless satellite perturbed by the star and by the oblateness of its host planet at quadrupole order. In Sect.~\ref{ssec:Lap}, the circular equilibrium configurations, called Laplace states P$_1$, P$_2$, and P$_3$ are described. Wherever they are defined, all three of them are stable to eccentricity variations, except in the regions E$_1$, E$_2$, and E$_3$, respectively, described in Sect.~\ref{ssec:eccstab} and illustrated in Fig.~\ref{fig:recap}.
   
   \cite{Tremaine-etal_2009} have shown that the system also admits eccentric equilibria, that is, configurations in which the satellite has a frozen eccentric orbit. Two of these configurations bifurcate away from P$_1$ and P$_2$ where these equilibria become unstable (i.e. at the borders of the E$_1$ and E$_2$ regions).
   
   In this section, we recall these results and go further in the analytical characterisation of the eccentric equilibria, allowing one to compute them more easily and in a form that is more directly usable in the context of our work. Furthermore, we show that the remaining eccentric equilibria described by \cite{Tremaine-etal_2009} bifurcate away from P$_3$ at the border of region E$_3$; therefore, all eccentric equilibria (that we call below P$_1'$, P$_2'$, P$_3'$, and P$_3''$) emerge from a bifurcation of a circular Laplace equilibrium.

   \subsection{Eccentric equilibrium P$_1'$}\label{ssec:P1prime}
   The eccentric equilibrium P$_1'$ corresponds to one of the configurations called `eccentric coplanar--coplanar Laplace equilibria' by \cite{Tremaine-etal_2009}. As such, the orbital angles of the satellite at P$_1'$ are $\omega_\mathrm{Q} = \pi/2\mod\pi$ and $\delta_\mathrm{Q} = 0$. As before, because of the symmetries of the secular problem, each eccentric equilibrium point has a twin obtained by the transformation $(\delta_\mathrm{Q},I_\mathrm{Q})\rightarrow(\pi+\delta_\mathrm{Q},\pi-I_\mathrm{Q})$ that corresponds to the same Laplace state with reversed orbital motion (see e.g. Fig.~\ref{fig:phase}). The equatorial inclination $I_{\mathrm{Q}1}'$ of the satellite at P$_1'$ is given in Eq.~\eqref{eq:IQ1p}, and its eccentricity $e$ is obtained by solving the equation
   \begin{equation}\label{eq:eqeP1p}
      \frac{r_\mathrm{M}^5}{a^5}(1 - 3\cos^2I_{\mathrm{Q}1}') = 2(1-e^2)^{5/2}\bigg(1 - 4\sin^2(\varepsilon - I_{\mathrm{Q}1}')\bigg)\,.
   \end{equation}
   By injecting the expression of $I_{\mathrm{Q}1}'$ into Eq.~\eqref{eq:eqeP1p}, we obtain an equation that only depends on $e$ and on the parameters of the problem, $a/r_\mathrm{M}$ and $\varepsilon$. If we solve it for $e\rightarrow 0$, we obtain again the boundary of the E$_1$ region defined at Eqs.~\eqref{eq:E1a} and \eqref{eq:E1b}; this shows that P$_1'$ indeed bifurcates away from P$_1$ where the latter becomes unstable.
   
\begin{figure*}
      \centering
      \includegraphics[width=0.85\textwidth]{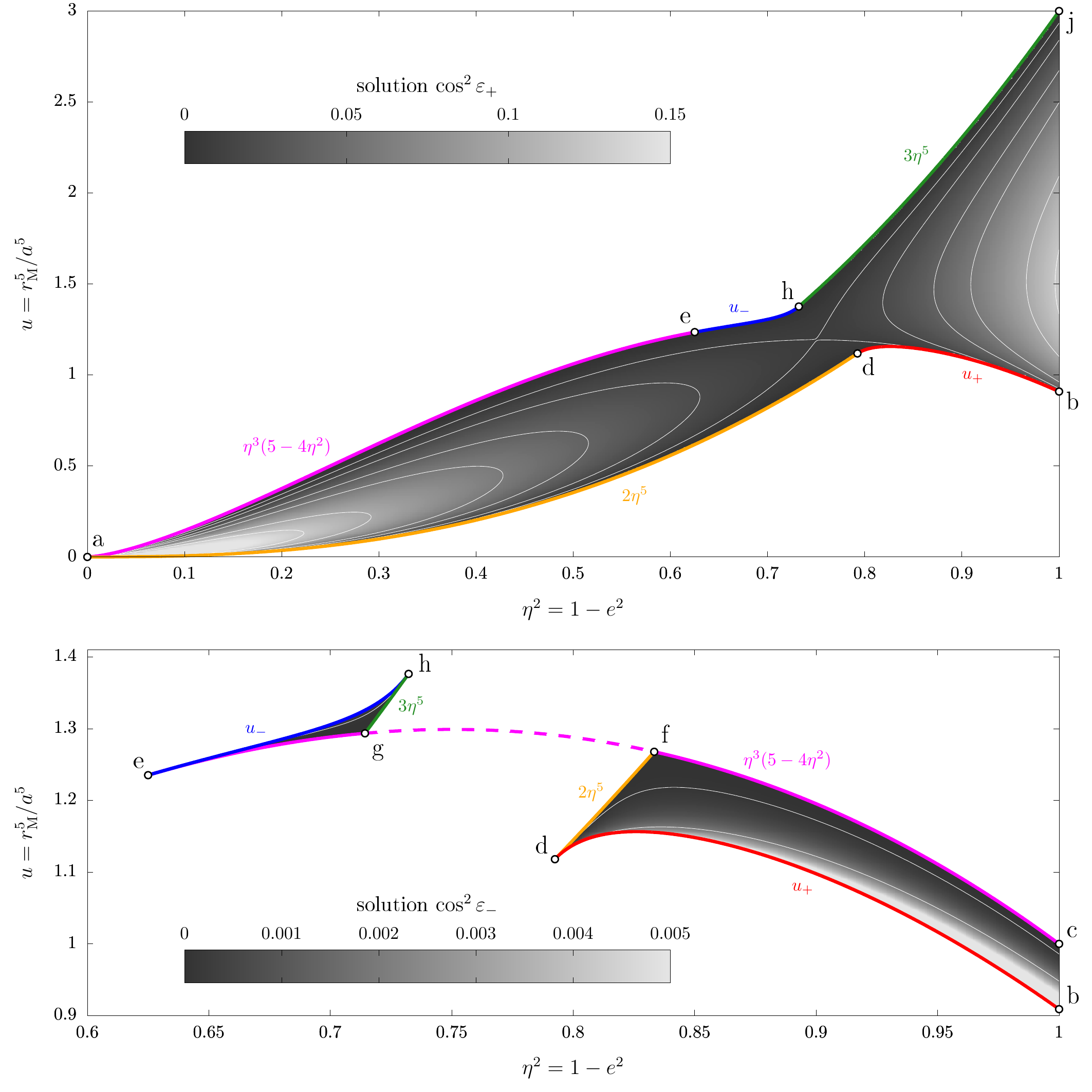}
      \caption{Location of the eccentric equilibrium P$_1'$ as a function of the parameters. The colour shades show the two closed-form obliquity solutions given by Eq.~\eqref{eq:cplusmoins}. Some levels are highlighted in white. The definition regions of the solutions are bounded by curves of the form $u = f(\eta)$, as labelled on the figures (coloured curves). Among those, $u_+$ and $u_-$ are given in Eq.~\eqref{eq:uplusmoins}. The junction points (black labels) are given in Table~\ref{tab:juncpt}.}
      \label{fig:cplusmoins}
   \end{figure*}

   When trying to solve Eq.~\eqref{eq:eqeP1p} for $\eta=\sqrt{1 - e^2}$ as a function of $u=r_\mathrm{M}^5/a^5$ and $\varepsilon$, we obtain an intractable polynomial of order $20$. However, there exists a straightforward analytical expression for $\varepsilon$ as a function of $u$ and $\eta$. Even though this is not exactly what we are looking for, this expression can be used to visualise the solutions in a graphical form and to set up efficient root-finding algorithms to reverse the expression for $\eta$. We define the obliquity solutions $\varepsilon_+$ and $\varepsilon_-$ as
   \begin{equation}\label{eq:cplusmoins}
      \cos^2\varepsilon_\pm = \frac{Z - 8X \pm Y}{2Z}\,,
   \end{equation}
   where
   \begin{equation}\label{eq:XYZ}
      \left\{
      \begin{aligned}
         X &= 24\eta^{12} - 90\eta^{10} + 75\eta^8 - 4u\eta^7 + 5u\eta^5 - u^2\eta^2 + 5u^2\,, \\
         Y &= (4\eta^5 - u)\sqrt{-8xu^2 - 8yu + 3z}\,, \\
         Z &= u\eta^3(4\eta^2 - 15)^2\,,
      \end{aligned}
      \right.
   \end{equation}
   and
   \begin{equation}\label{eq:xyz}
      \left\{
      \begin{aligned}
         x &= 8\eta^4 - 40\eta^2 + 25\,, \\
         y &= \eta^5(4\eta^2 - 5)^2\,, \\
         z &= \eta^6(4\eta^2 - 5)^2(32\eta^4 - 120\eta^2 + 75)\,.
      \end{aligned}
      \right.
   \end{equation}
   The values of $\cos^2\varepsilon_+$ and $\cos^2\varepsilon_-$ are shown in Fig.~\ref{fig:cplusmoins}. They give the location of the eccentric equilibrium P$_1'$ in the whole space of parameters. We see that solutions exist for any eccentricity value. However, they lie in restricted regions of the $(\eta,u)$ space, whose boundaries can be defined piecewise by analytical formulas of the form $u=f(\eta)$. These formulas are directly labelled on the figure, except $u_+$ and $u_-$ whose expressions are:
   \begin{equation}\label{eq:uplusmoins}
      \begin{aligned}
         u_\pm(\eta) &= \frac{-\eta^3(5 - 4\eta^2)}{4(8\eta^4 - 40\eta^2 + 25)}\Bigg[2\eta^2(5 - 4\eta^2) \\
         &\pm 5\sqrt{2(3 - 4\eta^2)(15 - 4\eta^2)(2\eta^4 - 8\eta^2 + 5)}\Bigg]\,.
      \end{aligned}
   \end{equation}
   Along the curves $u_+(\eta)$ and $u_-(\eta)$, the variable $Y$ cancels in Eq.~\eqref{eq:cplusmoins}, which means that $\cos^2\varepsilon_+$ and $\cos^2\varepsilon_-$ have the same value (double root). Along the remaining boundary curves in Fig.~\ref{fig:cplusmoins} (green, orange, and magenta curves), the solution $\cos\varepsilon$ represented is equal to zero.
   
   \begin{table}
      \caption{Junction points of the piecewise-defined boundaries in Figs.~\ref{fig:cplusmoins} and \ref{fig:cplusmoins_P3}.}
      \label{tab:juncpt}
      \vspace{-0.7cm}
      \begin{equation*}
         \begin{array}{rll}
            \hline
             \text{label} & \eta^2 & u \\
             \hline
             \hline
             \mathrm{a} & 0 & 0 \\
             \mathrm{b} & 1 & \left(2 + 5\sqrt{22}\right)/28 \\
             \mathrm{c} & 1 & 1 \\
             \mathrm{d} & \left(15 - 5\sqrt{3}\right)/8 & \frac{75}{32}\sqrt{\frac{5}{2}\left(33 - 19\sqrt{3}\right)} \\
             \mathrm{e} & 5/8 & 25\sqrt{10}/64 \\
             \mathrm{f} & 5/6 & 25\sqrt{30}/108 \\
             \mathrm{g} & 5/7 & 75\sqrt{35}/343 \\
             \mathrm{h} & \left(10-5\sqrt{2}\right)/4 & \frac{75}{16}\sqrt{5\left(58 - 41\sqrt{2}\right)} \\
             \mathrm{i} & 1 & 2 \\
             \mathrm{j} & 1 & 3 \\
             \hline
         \end{array}
      \end{equation*}
      \vspace{-0.3cm}
      \tablefoot{The variables are $\eta = \sqrt{1-e^2}$ and $u=r_\mathrm{M}^5/a^5$. The points are sorted by increasing value of $u$. The obliquity $\varepsilon$ is equal to $90^\circ$ at all points except at point b, where $\cos^2\varepsilon_\pm = (977-200\sqrt{22})/2002$.}
   \end{table}
   
\begin{figure*}
      \centering
      \includegraphics[width=0.495\textwidth]{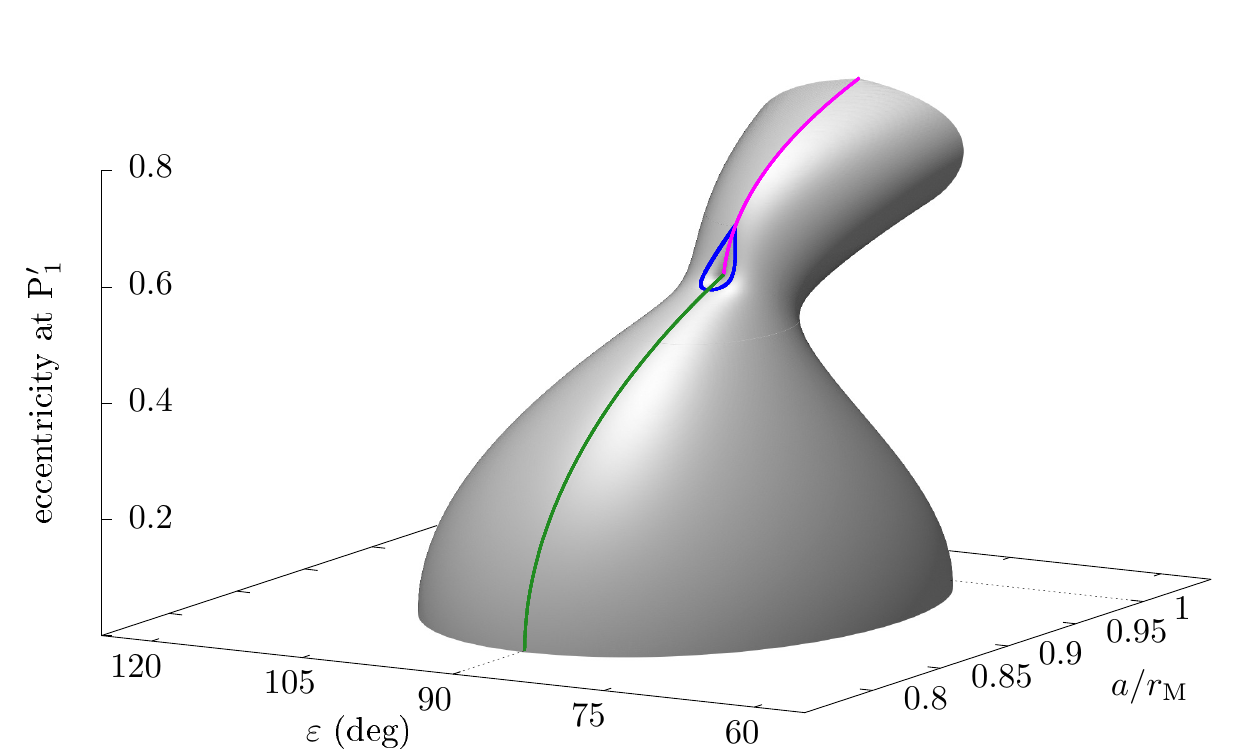}
      \includegraphics[width=0.495\textwidth]{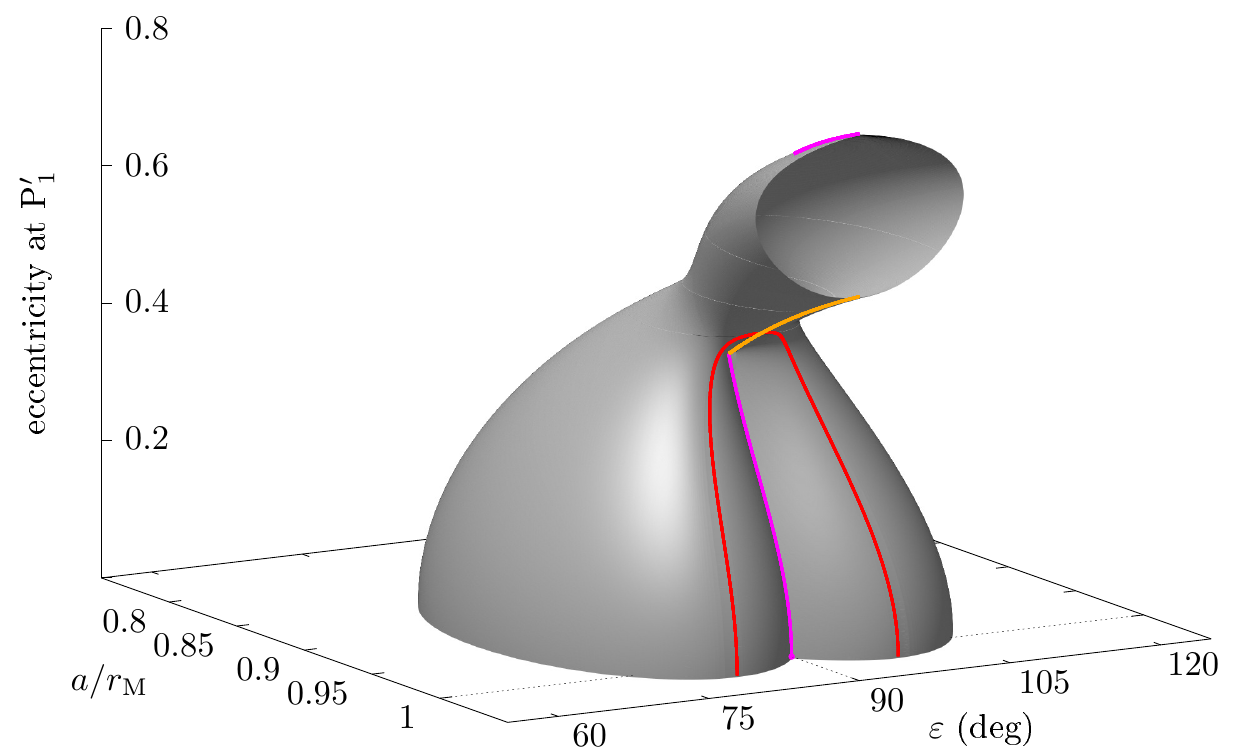}
      \caption{Eccentricity $e$ of the satellite at equilibrium P$_1'$ as a function of the parameters. The solutions lie on a three-dimensional surface shown here from two viewing angles. This figure is obtained by stitching together the patches represented in Fig.~\ref{fig:cplusmoins}. For better comparison, the dividing curves are drawn on the surface using the same colour code. The top tube-like portion of the surface has been cut; as shown in Fig.~\ref{fig:cplusmoins}, it extends up to $a\rightarrow\infty$ (i.e. $u\rightarrow 0$). The equilibrium P$_1'$ is singular along the magenta curve, called $\mathbb{S}_1$ in the text.}
      \label{fig:P1primewires}
   \end{figure*}

   Instead of projecting the solutions in the $(\eta,u)$ plane as in Fig.~\ref{fig:cplusmoins}, the closed-form expression in Eq.~\eqref{eq:cplusmoins} can be used to plot the eccentricity of the satellite at equilibrium P$_1'$ as a function of the parameters in a parametric form. We obtain a three-dimensional surface, as illustrated in Fig.~\ref{fig:P1primewires}. As expected, the bottom section of this surface (i.e. at $e=0$) coincides with the border of the region E$_1$ where the circular equilibrium P$_1$ is unstable (see Figs.~\ref{fig:recap} and \ref{fig:E1detail}). In the main text, Fig.~\ref{fig:P1prime} shows the value of the eccentricity at P$_1'$ as a colour scale, as well as the value of the inclination obtained using Eq.~\eqref{eq:IQ1p}. In order to draw this figure, the tube-like portion of the three-dimensional surface has been cut off.
   
   Along the three-dimensional curve $\mathbb{S}_1$ defined in Eq.~\eqref{eq:Ssing3D} and depicted by a magenta curve in Figs.~\ref{fig:cplusmoins} and \ref{fig:P1primewires}, the three-dimensional surface has a cusp. As discussed in the main text, the curve $\mathbb{S}_1$ is the eccentric continuation of the singular point S$_1$ described in Sect.~\ref{ssec:Lap}, where the equilibrium point P$_1$ is singular. We note that $\mathbb{S}_1$ pierces the three-dimensional surface in Fig.~\ref{fig:P1primewires} and reappears on the other side (this corresponds to the dotted portion in Fig.~\ref{fig:cplusmoins}).
   
\begin{figure*}
      \centering
      \includegraphics[width=0.495\textwidth]{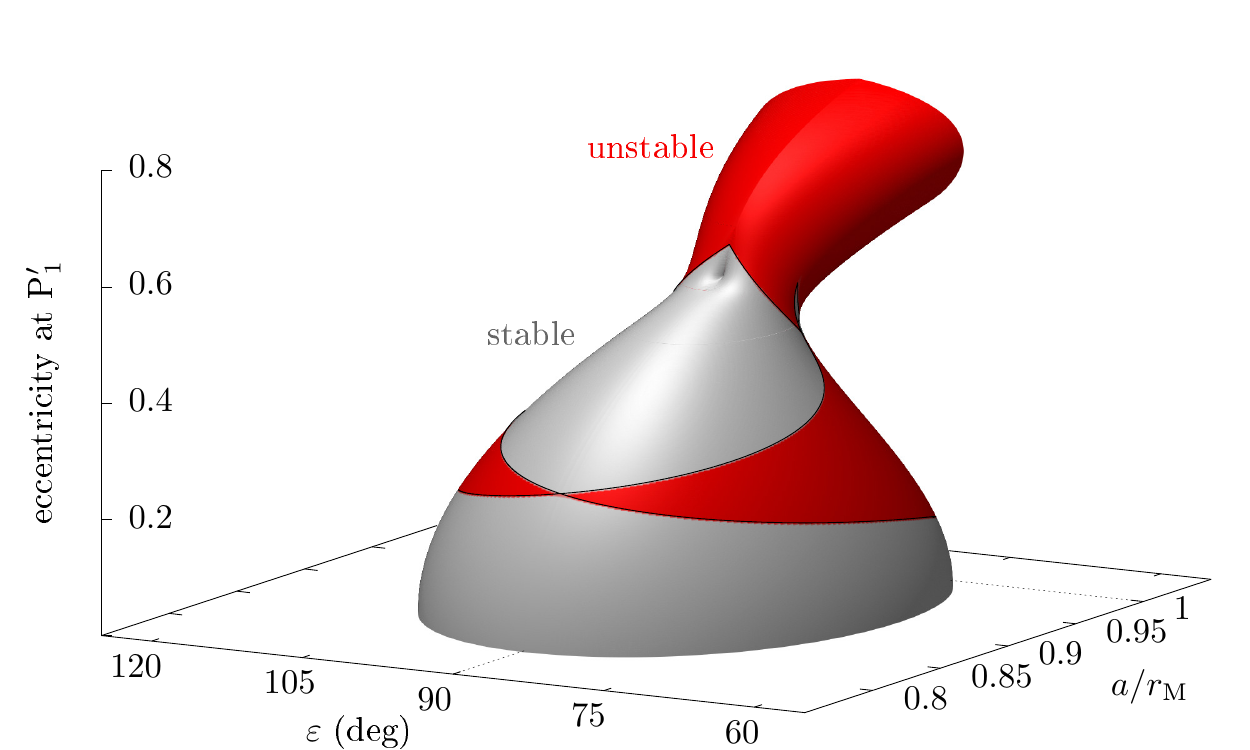}
      \includegraphics[width=0.495\textwidth]{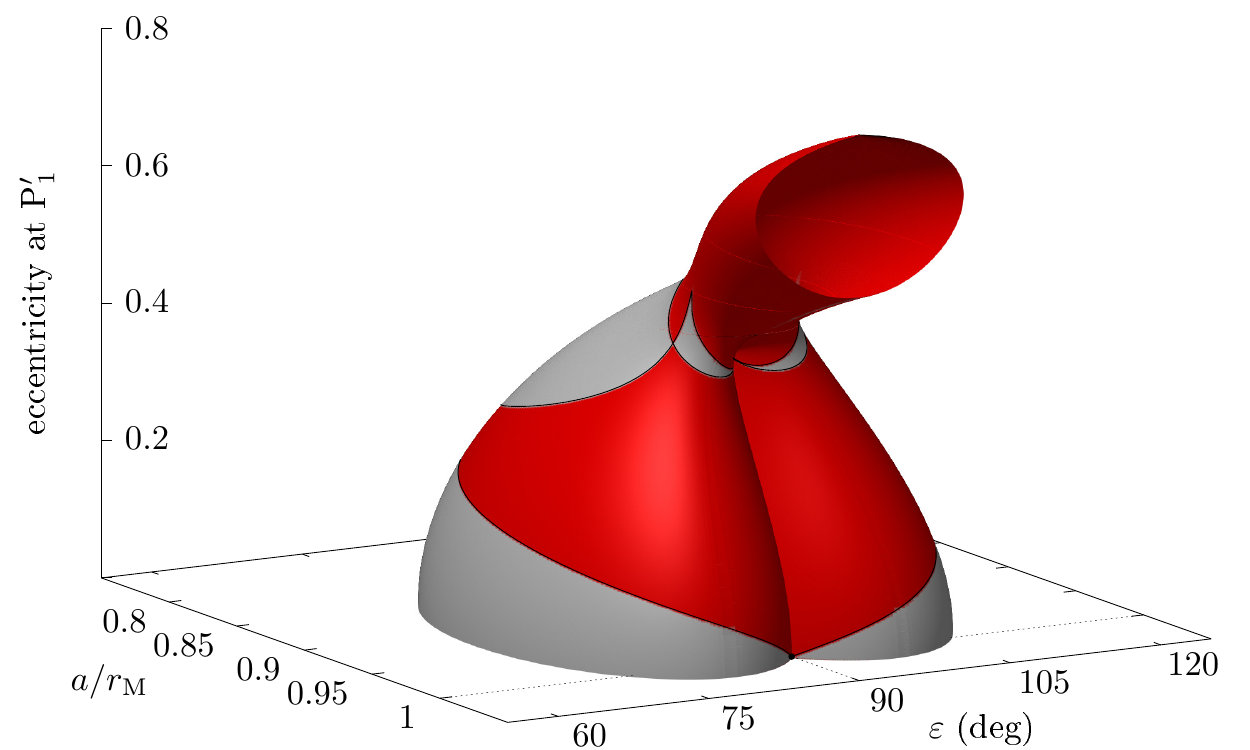}
      \caption{Stability of the equilibrium point P$_1'$ as a function of the parameters. The stable and unstable regions are painted in grey and red, respectively, and they are projected on the three-dimensional surface describing the eccentricity of the satellite. As in Fig.~\ref{fig:P1primewires}, the surface is seen from two viewing angles and the top tube-like portion has been cut for better readability.}
      \label{fig:P1prime3DStab}
   \end{figure*}

   We did not find a closed-form expression for the boundary dividing the regions of the parameter space where P$_1'$ is stable from the regions where it is unstable. However, the stability nature of P$_1'$ as a function of the parameters can be determined numerically: the equilibrium is stable wherever the linearised system has no eigenvalue with positive real part. Figure~\ref{fig:P1prime3DStab} shows the stable and unstable regions projected on the three-dimensional surface of P$_1'$. Sections of this surface can be seen in Fig.~6 of \cite{Tremaine-etal_2009}. In the stable regions, the oscillation frequencies of the satellite are illustrated in Fig.~\ref{fig:P1primeStab}.
   
\begin{figure}
      \centering
      \includegraphics[width=\columnwidth]{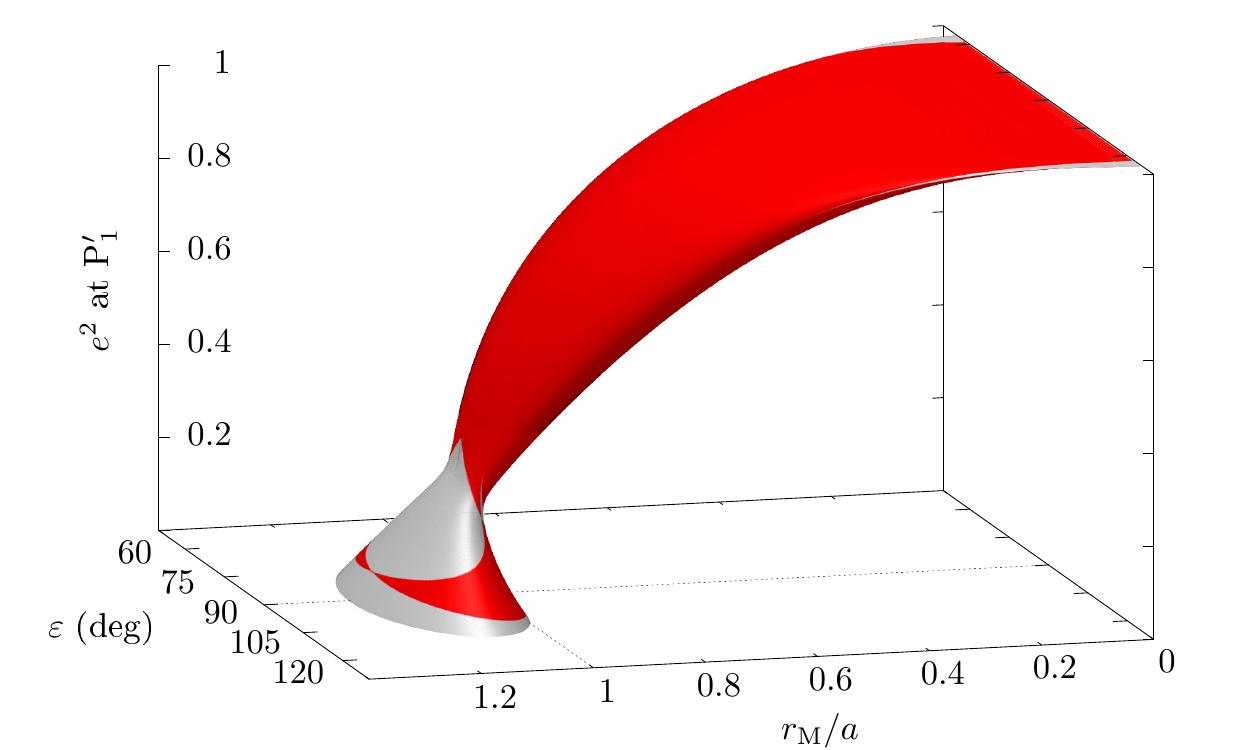}
      \caption{Same as Fig.~\ref{fig:P1prime3DStab} but replacing the coordinates $a/r_\mathrm{M}$ and $e$ by $r_\mathrm{M}/a$ and $e^2$, respectively. Contrary to previous figures, the top tube-like portion of the surface is seen in its totality.}
      \label{fig:P1prime3DStab2}
   \end{figure}

   For completeness, Fig.~\ref{fig:P1prime3DStab2} illustrates the stability nature of P$_1'$ in the space $(r_\mathrm{M}/a,\varepsilon,e^2)$, where the full three-dimensional shape can be visualised, including the region where $a\rightarrow\infty$. We see that there is a small additional region where P$_1'$ is stable, for a large semi-major axis and an eccentricity very close to $1$ (see the thin grey border on the right side of the figure). This thin stable region also appears in Fig.~5 of \cite{Tremaine-etal_2009} as the blue points lying at the tip of the middle triangle\footnote{Due to their similar properties, the equilibria P$_1'$ and P$_3'$ have been studied together by \cite{Tremaine-etal_2009}. Their Fig.~5 features therefore both P$_1'$ (for inclinations smaller than $90^\circ$) and P$_3'$ (for inclinations larger than $90^\circ$).}.
   
   \subsection{Eccentric equilibrium P$_2'$}
\begin{figure}
      \centering
      \includegraphics[width=\columnwidth]{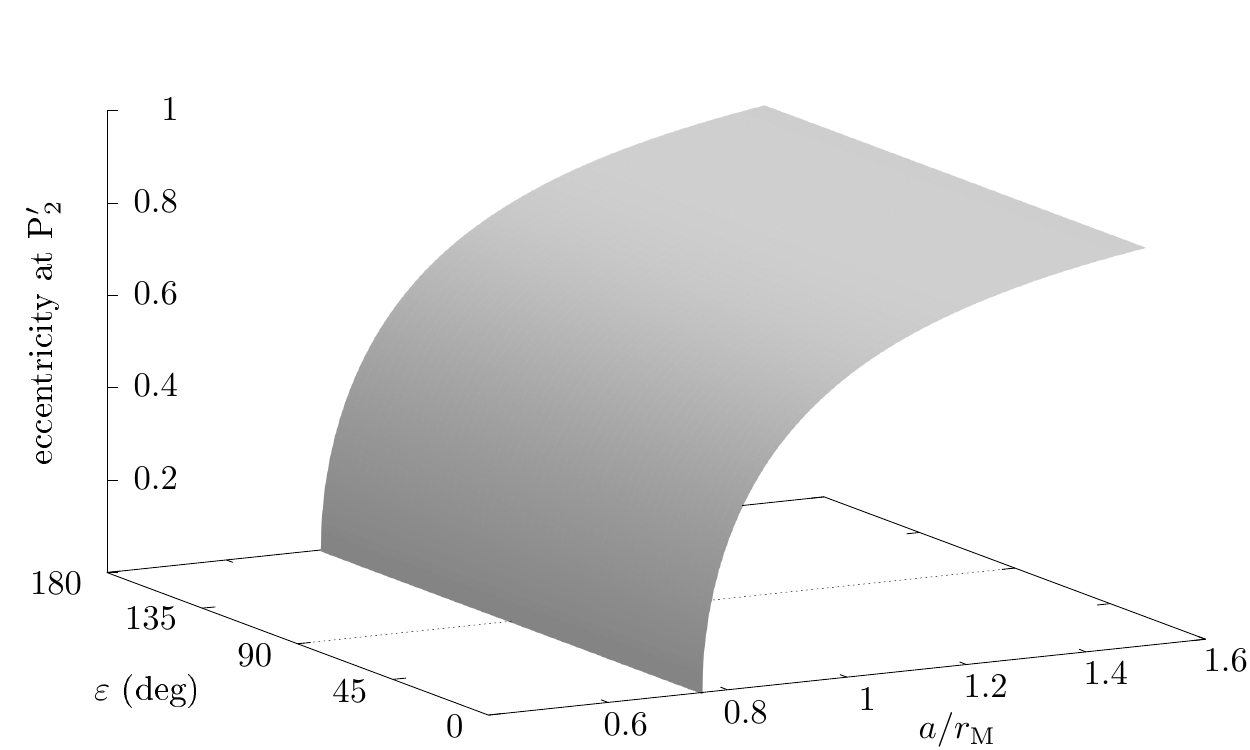}
      \caption{Eccentricity $e$ of the satellite at equilibrium P$_2'$ as a function of the parameters. The eccentricity is given by Eq.~\eqref{eq:e2prime}; it only depends on the semi-major axis $a/r_\mathrm{M}$ of the satellite. The equilibrium is stable wherever it exists.}
      \label{fig:P2prime3D}
   \end{figure}

   The eccentric equilibrium P$_2'$ corresponds to the configuration called `eccentric orthogonal--coplanar Laplace equilibrium' by \cite{Tremaine-etal_2009}. As such, the orbital inclination of the satellite at P$_2'$ is $I_\mathrm{Q} = I_\mathrm{C} = 90^\circ$, while its orbital angles are $\omega_\mathrm{C} = 0\mod\pi$ and $\delta_\mathrm{Q} = \pi/2$ (or $3\pi/2$ for its twin equilibrium with reversed orbital motion). The eccentricity of the satellite at P$_2'$ is given by the relation
   \begin{equation}
      u = 4\eta^5\,,
   \end{equation}
   where $u=r_\mathrm{M}^5/a^5$ and $\eta=\sqrt{1-e^2}$. If we solve it for $e\rightarrow 0$, we obtain again the critical distance $a=r_3$ defined in Eq.~\eqref{eq:r3}; this shows that P$_2'$ indeed bifurcates away from P$_2$ where the latter becomes unstable. Hence, P$_2'$ exists for $a\geqslant r_3$ and the eccentricity of the satellite at P$_2'$ is
   \begin{equation}\label{eq:e2prime}
      e_2' = \sqrt{1 - \left(\frac{r_\mathrm{M}/a}{4^{1/5}}\right)^2}\,.
   \end{equation}
   The bifurcation of P$_2'$ from P$_2$ can be visualised as a three-dimensional surface, illustrated in Fig.~\ref{fig:P2prime3D}.
   
   As shown by \cite{Tremaine-etal_2009}, P$_2'$ is stable wherever it exists. The eigenfrequencies of the linearised system are given by
   \begin{equation}
      \xi_2'^2 = -\kappa^2\left(\frac{a}{r_\mathrm{M}}\right)^3(1-e_2'^2)\sin^2\varepsilon \,,
   \end{equation}
   and
   \begin{equation}\label{eq:mu2prime}
      \mu_2'^2 = -\frac{25}{2}\kappa^2\left(\frac{a}{r_\mathrm{M}}\right)^3e_2'^2 \,.
   \end{equation}
   The eccentricity and inclination of the satellite are coupled in the vicinity of P$_2'$, so they their oscillation spectra both contain the two frequencies $\xi_2'$ and $\mu_2'$. At the transition radius $a=r_3$, we note that $\xi_2'$ and $\mu_2'$ are equal to the oscillation frequencies $\xi_2$ and $\mu_2$ around P$_2$ given in Eqs.~\eqref{eq:xi2} and \eqref{eq:mu2}, confirming that the eccentric equilibrium P$_2'$ bifurcates away from the circular equilibrium P$_2$.
   
   \subsection{Eccentric equilibria P$_3'$ and P$_3''$}
   As explained in Sect.~\ref{sec:orb}, the circular equilibrium P$_3$ is unstable to eccentricity variations in the region E$_3$ of the parameter space. In its illustrations in Figs.~\ref{fig:recap} and \ref{fig:E3detail}, we see that the border of E$_3$ can be divided into two components: a small drop-like boundary for $a>r_\mathrm{L}$ (whose expression is given by Eq.~\ref{eq:E1b}), and a V-shaped boundary for $a<r_\mathrm{L}$ spanning all values of obliquity $\varepsilon$ from $0$ to $180^\circ$ (and whose expression is given by Eq.~\ref{eq:E3a}). Below, we show that along these two boundaries, the circular equilibrium P$_3$ bifurcates into two distinct eccentric equilibria. We call them P$_3'$ and P$_3''$, respectively, for the eccentric equilibria bifurcating away from the drop-like boundary and from the V-shaped boundary.
   
   \paragraph{Eccentric equilibrium P$_3'$:} The eccentric equilibrium P$_3'$ corresponds to one of the configurations called `eccentric coplanar--coplanar Laplace equilibria' by \cite{Tremaine-etal_2009}. As such, the orbital angles of the satellite at P$_3'$ are $\omega_\mathrm{Q} = \pi/2\mod\pi$ and $\delta_\mathrm{Q} = 0$ (or $\delta_\mathrm{Q}=\pi$ for the twin equilibrium with reversed orbital motion). The equatorial inclination of the satellite at P$_3'$ can be written as
   \begin{equation}\label{eq:IQ3p}
      I_{\mathrm{Q}3}' = \frac{\pi}{2} + \frac{1}{2}\mathrm{atan2}\big[\sin(2\varepsilon),v+\cos(2\varepsilon)\big]\,,
   \end{equation}
   where $v$ depends on the eccentricity at equilibrium, as given in Eq.~\eqref{eq:uprime}. The inclination $I_{\mathrm{Q}3}'$ in Eq.~\eqref{eq:IQ3p} has the same form as $I_{\mathrm{Q}3}$ in Eq.~\eqref{eq:ILap13}, but where $u=r_\mathrm{M}^5/a^5$ is replaced by $v$. The two definitions coincide for $e=0$.
   
   The eccentricity $e$ of the satellite at P$_3'$ is obtained by solving the equation \eqref{eq:eqeP1p}, but where $I_{\mathrm{Q}1}'$ must be replaced by $I_{\mathrm{Q}3}'$. If we solve it for $e\rightarrow 0$, we obtain again the drop-like boundary of the E$_3$ region; this shows that P$_3'$ indeed bifurcates away from P$_3$ along this boundary. The resolution of the equation in the general case is very similar to what is presented in Appendix~\ref{ssec:P1prime} for P$_1'$: we obtain a closed-form analytical expression for $\varepsilon$ as a function of $u$ and $\eta$, which is again given by Eqs.~\eqref{eq:cplusmoins}, \eqref{eq:XYZ}, and \eqref{eq:xyz}, but with a differing domain of definition.
   
\begin{figure*}
      \centering
      \includegraphics[width=0.85\textwidth]{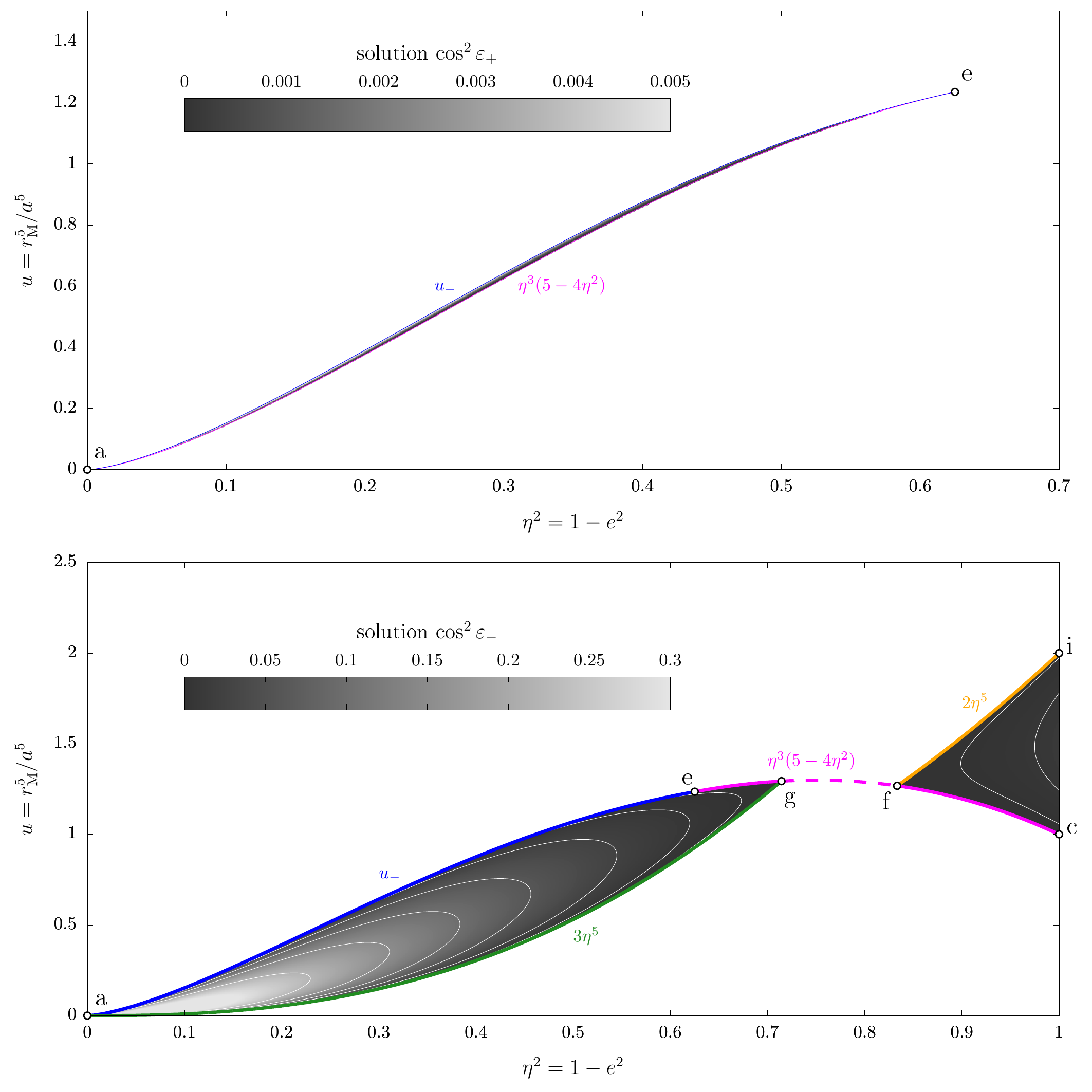}
      \caption{Same as Fig.~\ref{fig:cplusmoins}, but showing the location of the eccentric equilibrium P$_3'$ as a function of the parameters. On the top panel, the definition region of $\cos^2\varepsilon_+$ is very narrow, comprised between the curves $u=\eta^3(5-4\eta^2)$ and $u=u_-(\eta)$.}
      \label{fig:cplusmoins_P3}
   \end{figure*}

   The values of $\cos^2\varepsilon_+$ and $\cos^2\varepsilon_-$ for the eccentric equilibrium P$_3'$ are shown in Fig.~\ref{fig:cplusmoins_P3}. They give its location in the whole space of parameters. Contrary to P$_1'$, we see that solutions do not exist for all eccentricity values: there is a gap between $\eta_\mathrm{g}$ and $\eta_\mathrm{f}$ where P$_3'$ does not exist (see the corresponding values of eccentricity in Table~\ref{tab:juncpt}).
   
\begin{figure*}
      \centering
      \includegraphics[width=0.495\textwidth]{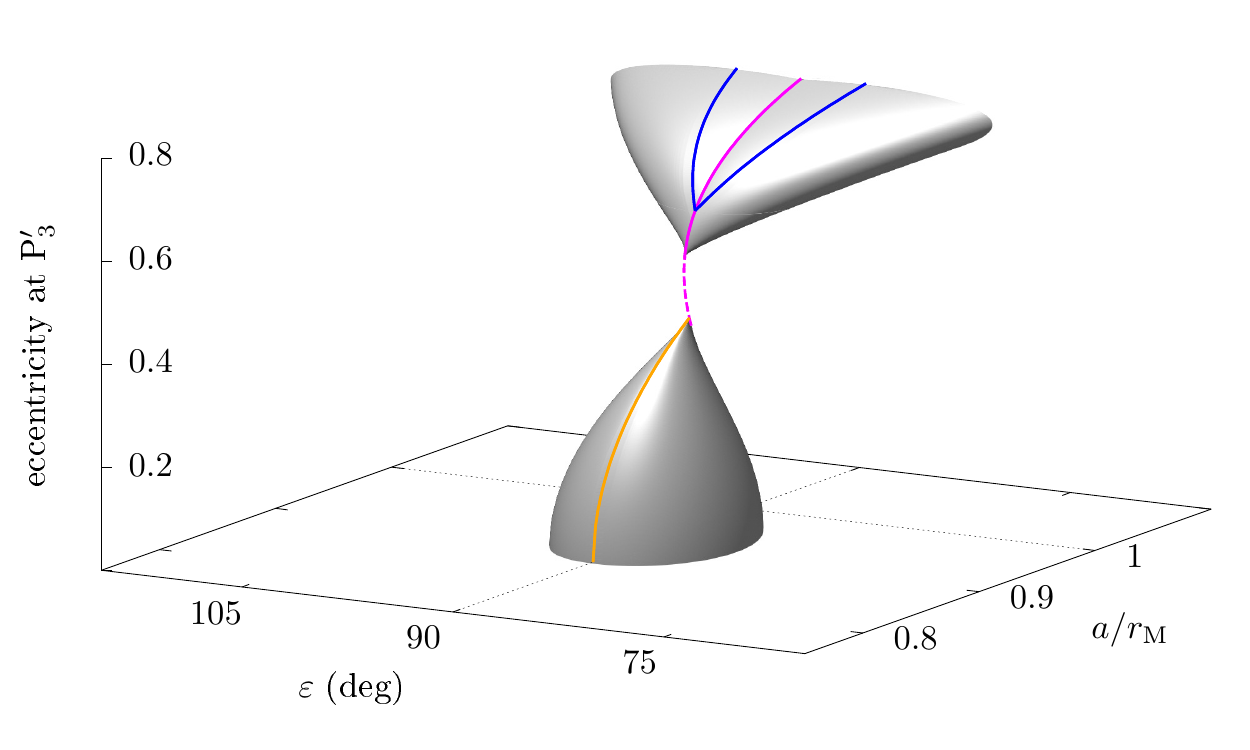}
      \includegraphics[width=0.495\textwidth]{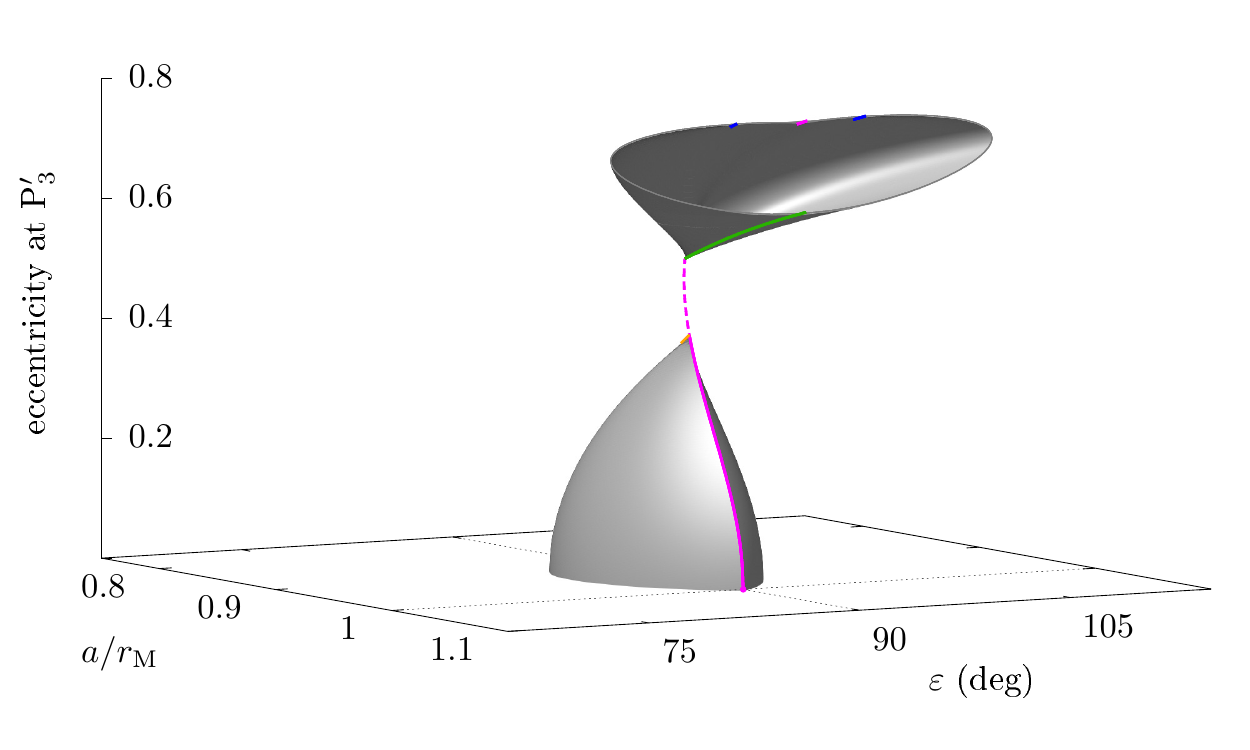}
      \caption{Eccentricity $e$ of the satellite at equilibrium P$_3'$ as a function of the parameters. This figure is obtained by stitching together the patches represented in Fig.~\ref{fig:cplusmoins_P3}. The top tube-like portion of the surface has been cut; as shown in Fig.~\ref{fig:cplusmoins_P3}, it extends up to $a\rightarrow\infty$ (i.e. $u\rightarrow 0$). The equilibrium P$_3'$ is singular along the magenta curve, called $\mathbb{S}_1$ in the text.}
      \label{fig:P3primewires}
   \end{figure*}

   The expression in Eq.~\eqref{eq:cplusmoins} can be used to plot the eccentricity of the satellite at equilibrium P$_3'$ in a parametric form. We obtain the three-dimensional surface illustrated in Fig.~\ref{fig:P3primewires}. As expected, the bottom section of this surface (i.e. at $e=0$) coincides with the drop-like boundary of the region E$_3$ where the circular equilibrium P$_3$ is unstable to eccentricity variations (see Figs.~\ref{fig:recap} and \ref{fig:E3detail}). As mentioned in Appendix~\ref{ssec:P1prime}, the eccentric equilibrium P$_3'$ is singular along the three-dimensional curve $\mathbb{S}_1$ given in Eq.~\eqref{eq:Ssing3D}, where $I_{\mathrm{Q}3}'$ is undefined. It corresponds to a degenerate equilibrium which results from the merging of P$_1'$ and P$_3'$. Indeed, we note that $\mathbb{S}_1$ is a contact region between the two three-dimensional surfaces shown in Figs.~\ref{fig:P1primewires} and \ref{fig:P3primewires} (see the magenta curve).
   
\begin{figure*}
      \centering
      \includegraphics[width=0.495\textwidth]{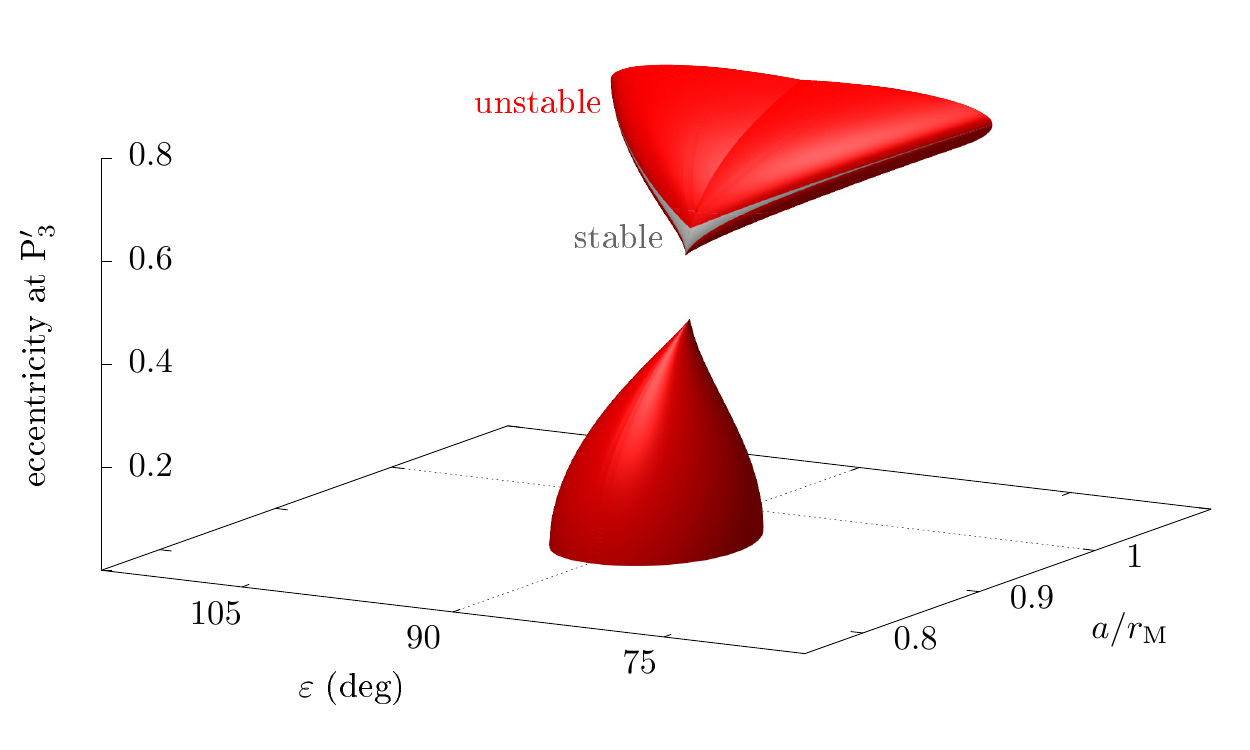}
      \includegraphics[width=0.495\textwidth]{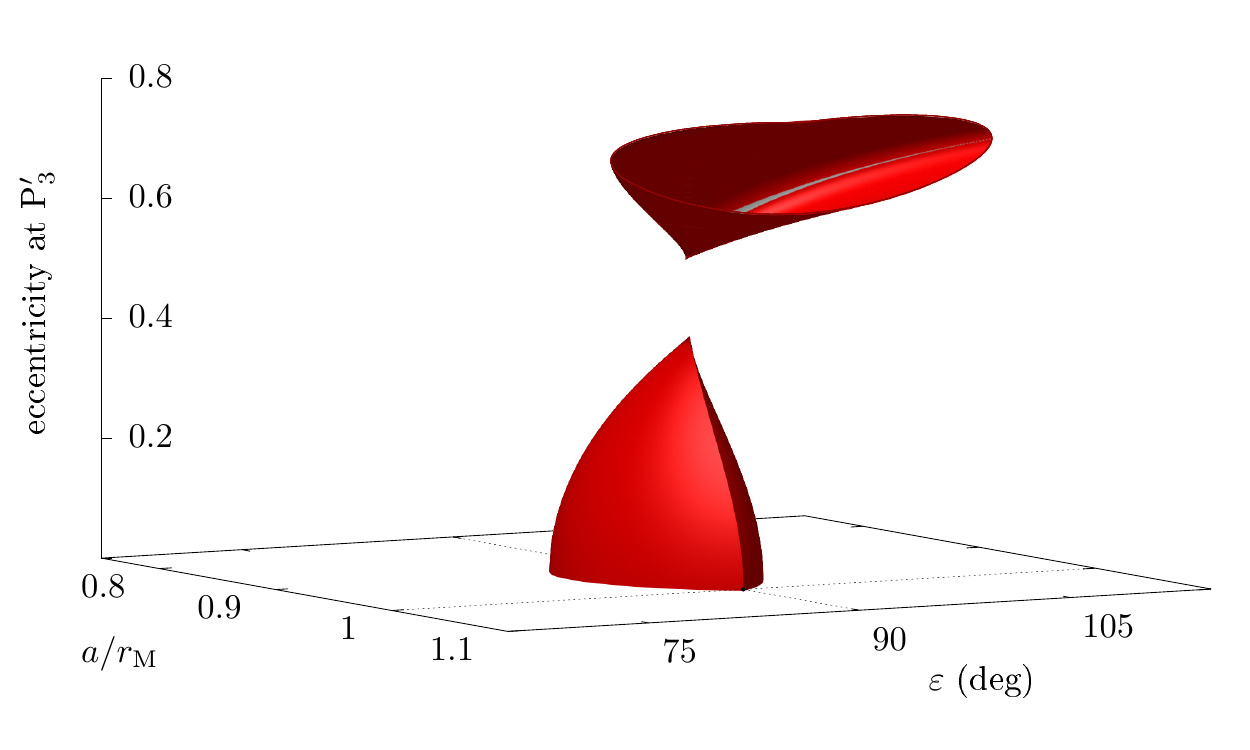}
      \caption{Stability of the equilibrium point P$_3'$ as a function of the parameters. The stable and unstable regions are painted in grey and red, respectively, and they are projected on the three-dimensional surface describing the eccentricity of the satellite. As in Fig.~\ref{fig:P3primewires}, the surface is seen from two viewing angles and the top tube-like portion has been cut for better readability.}
      \label{fig:P3prime3DStab}
   \end{figure*}

\begin{figure}
      \centering
      \includegraphics[width=\columnwidth]{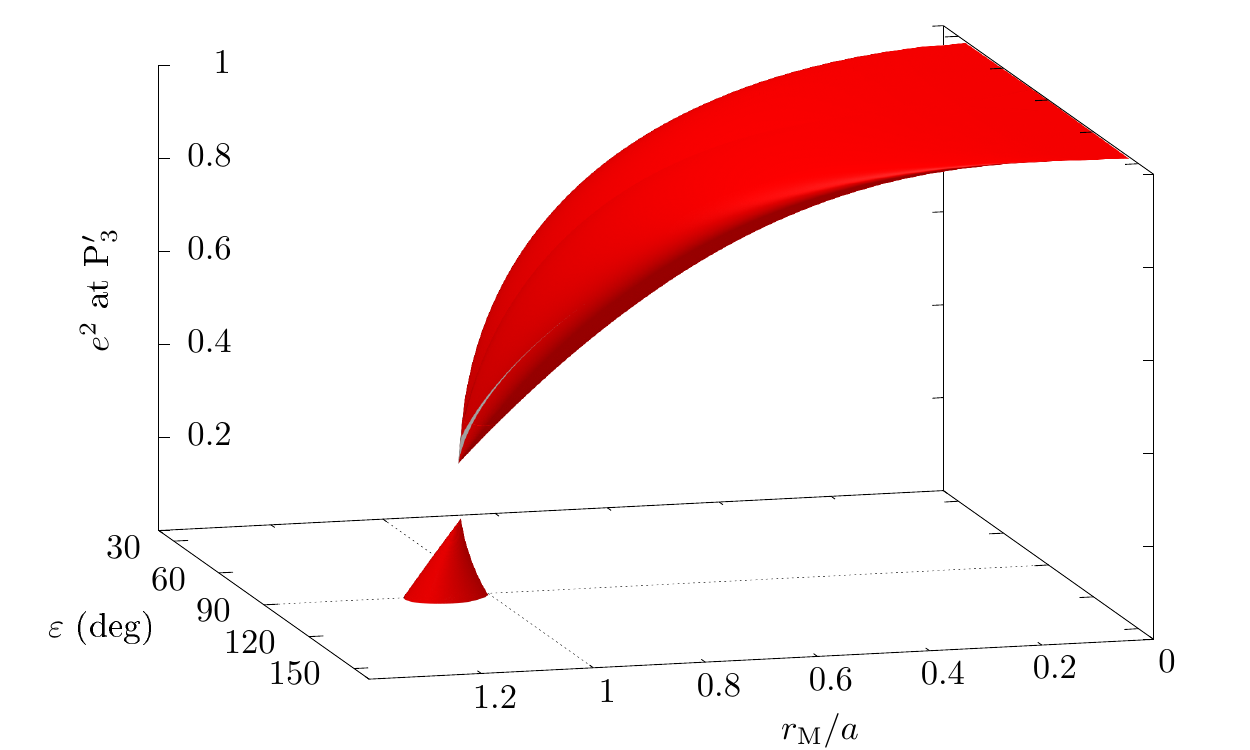}
      \caption{Same as Fig.~\ref{fig:P3prime3DStab} but replacing the coordinates $a/r_\mathrm{M}$ and $e$ by $r_\mathrm{M}/a$ and $e^2$, respectively. Contrary to previous figures, the top tube-like portion of the surface is seen in its totality.}
      \label{fig:P3prime3DStab2}
   \end{figure}

   Figure~\ref{fig:P3prime3DStab} highlights the regions of the parameter space where P$_3'$ is stable, projected on the three-dimensional surface. Interestingly, a small stable region exists on the top portion of the three-dimensional surface. This stable region is visible in Fig.~5 by \cite{Tremaine-etal_2009} as the blue points in the uppermost triangle. Therefore, even though the circular equilibrium P$_3$ described in Sect.~\ref{sec:orb} is unstable to inclination variations in the whole space of parameters, this is not everywhere the case for its eccentric counterpart P$_3'$. Yet, Fig.~\ref{fig:P3prime3DStab} shows that this stable region is disconnected from the low-eccentricity portion of the three-dimensional surface, which means that a regular satellite starting with a roughly circular orbit cannot reach it by a smooth change of parameters occurring after its formation. For completeness, Fig.~\ref{fig:P3prime3DStab2} illustrates the stability nature of P$_3'$ in the space $(r_\mathrm{M}/a,\varepsilon,e^2)$, where the full three-dimensional shape can be visualised, including the region where $a\rightarrow\infty$. No additional stable region can be seen.
   
   \paragraph{Eccentric equilibrium P$_3''$:} The eccentric equilibrium P$_3''$ corresponds to the configuration called `eccentric coplanar--orthogonal Laplace equilibrium' by \cite{Tremaine-etal_2009}. As such, the orbital angles of the satellite at P$_3''$ are $\omega_\mathrm{Q} = 0\mod\pi$ and $\delta_\mathrm{Q} = 0$ (or $\delta_\mathrm{Q}=\pi$ for the twin equilibrium with reversed orbital motion).  The equatorial inclination of the satellite at P$_3''$ can be written as
   \begin{equation}\label{eq:IQ3pp}
      I_{\mathrm{Q}3}'' = \frac{\pi}{2} + \frac{1}{2}\mathrm{atan2}\big[\sin(2\varepsilon),w+\cos(2\varepsilon)\big]\,,
   \end{equation}
   where we define
   \begin{equation}\label{eq:useconde}
      w = \frac{r_\mathrm{M}^5}{a^5}\frac{1}{(1-e^2)^{5/2}}\,,
   \end{equation}
   in which $e$ is the satellite's eccentricity at equilibrium. The inclination $I_{\mathrm{Q}3}''$ in Eq.~\eqref{eq:IQ3pp} has the same form as $I_{\mathrm{Q}3}$ in Eq.~\eqref{eq:ILap13}, but where $u=r_\mathrm{M}^5/a^5$ is replaced by $w$. The two definitions coincide for $e=0$.
   
\begin{figure}
      \centering
      \includegraphics[width=\columnwidth]{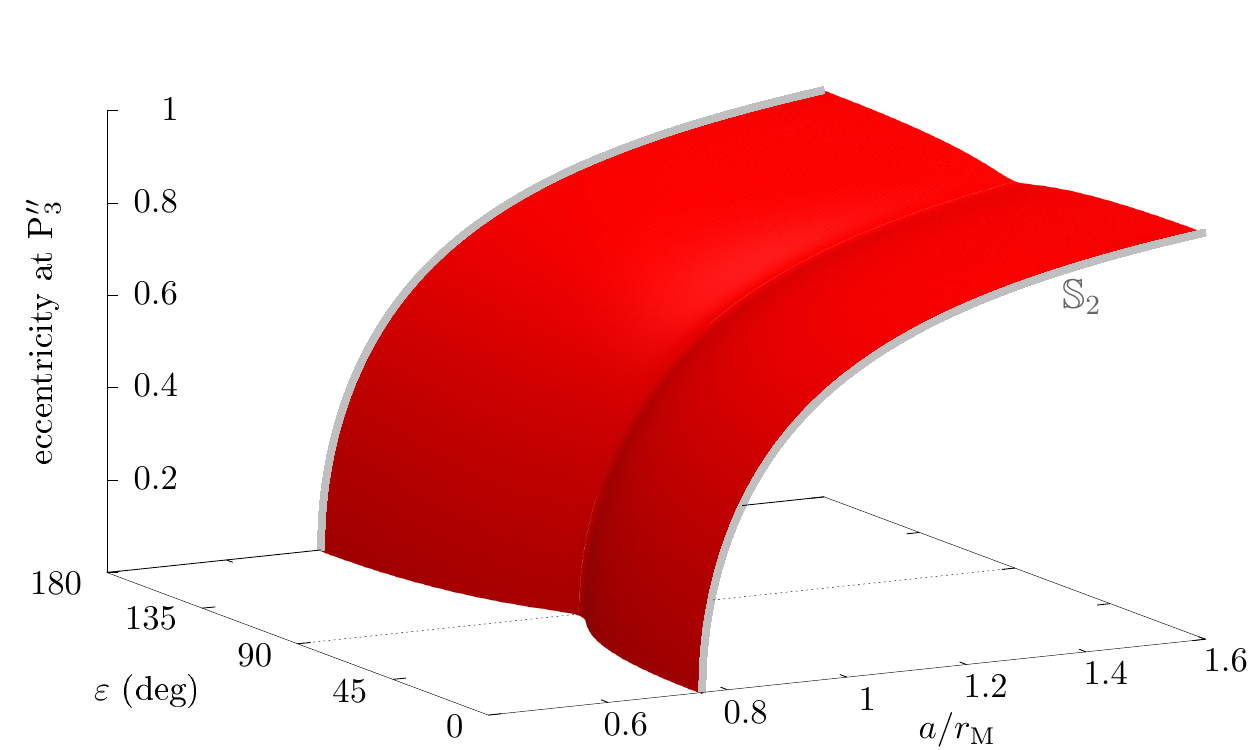}
      \caption{Eccentricity $e$ of the satellite at equilibrium P$_3''$ as a function of the parameters. The colour gives its stability: grey for stable and red for unstable. P$_3''$ is unstable in the whole parameter space except along $\mathbb{S}_2$, which results from the merging of P$_3''$ and P$_2'$ into a degenerate equilibrium.}
      \label{fig:P3second3D}
   \end{figure}

   The eccentricity $e$ of the satellite at P$_3''$ is given by the implicit equation
   \begin{equation}\label{eq:eP3pp}
      \cos^2\varepsilon = \frac{-2w^2 + 13w + 38 - (6+w)\sqrt{-4w^2 + 12w + 41}}{2w}\,,
   \end{equation}
   that must be inverted to obtain $w$ (and thus $e$) as a function of $\varepsilon$. We recognise the V-shaped boundary of the region E$_3$ (see Eq.~\ref{eq:E3a}), but where $u$ is replaced by $w$. This shows that P$_3''$ indeed bifurcates away from P$_3$ along this boundary. Through the inversion of Eq.~\eqref{eq:eP3pp}, the inclination $I_{\mathrm{Q}3}''$ is a function of $\varepsilon$ only and it does not depend on the orbital distance $a/r_\mathrm{M}$. The eccentric equilibrium P$_3''$ is unstable wherever it exists, except for $\varepsilon=0^\circ$ or $180^\circ$, where its unstable mode tends to zero. Indeed, the three-dimensional domain
   \begin{equation}
      \mathbb{S}_2 = \Big\{u=4\eta^5,\:\varepsilon=0^\circ\;\text{or}\;180^\circ\Big\}\,,
   \end{equation}
   is the eccentric continuation of the singular region S$_2$ described in Sect.~\ref{sec:orb}, where P$_2$ and P$_3$ merge with the separatrix. Similarly, $\mathbb{S}_2$ results from the merging of P$_2'$ and P$_3''$, which degenerate into an equilibrium region $\mathscr{L}_{23}'$. The bifurcation of P$_3''$ from P$_3$ and the two curves that define $\mathbb{S}_2$ can be visualised in Fig.~\ref{fig:P3second3D}. We see that $\mathbb{S}_2$ is the contact region between the two three-dimensional surfaces shown in Figs.~\ref{fig:P2prime3D} and \ref{fig:P3second3D}.
   
   Inside the region $\mathbb{S}_2$, the degenerate equilibrium $\mathscr{L}_{23}'$ is defined by an eccentricity $e$ given by Eq.~\eqref{eq:e2prime}, an inclination $I_\mathrm{Q}=90^\circ$, an argument of pericentre $\omega_\mathrm{Q}=0\mod\pi$, and any value for the longitude of node $\delta_\mathrm{Q}$. In the vicinity of $\mathscr{L}_{23}'$, the eigenfrequencies of the system are given by $\xi_{23}'=0$, and by Eq.~\eqref{eq:mu2prime} for $\mu_{23}'^2$.
   
\section{Self-consistent formula for the spin-axis precession rate}\label{asec:BL06}

   In Eq.~\eqref{eq:psidotsimp}, we give a simplified expression for the spin-axis precession rate of an oblate planet affected by a satellite. Taking advantage of the simplicity of this expression, many properties of the precession rate are given by closed-form analytical formulas, such as the location and magnitude of the maximum rate, or the shape of its level curves as a function of the parameters (see Sect.~\ref{ssec:spinandsat}). However, as stressed in the main text, Eq.~\eqref{eq:psidotsimp} is not self-consistent. The satellite is considered massless to compute its equilibrium inclination (Laplace state P$_1$), but massive when it comes to its influence on the planet's spin axis. Then, the spin axis is taken as fixed when studying the satellite's dynamics, but moving when taking into account secular spin-orbit resonances. Both these approximations are expected to be accurate for satellites with a small enough mass, because in this case the satellite's orbit evolves on a much shorter timescale than the planet's spin axis, and only the very long-term component of the satellite's dynamics (mean secular motion) affects the planet's obliquity. Yet, the validity range of our approximations are not known a priori, and since we explore a large range of mass parameters (see Fig.~\ref{fig:precrate}), a comparison with self-consistent models would be useful.
   
   For this purpose, we use the model of \cite{Boue-Laskar_2006}, which describes the orbital dynamics of an oblate planet and of its (massive) satellite, together with the spin-axis dynamics of the planet, considering the fully coupled equations of motion at quadrupolar order. Since the satellite is nonetheless considered to be less massive than its host planet, high-order terms in $\delta$ are neglected, where $\delta=m/(m+M)$ with our notations. Even though the model of \cite{Boue-Laskar_2006} stands for any eccentricity of the satellite, it is averaged over the satellite's argument of pericentre, which means that it cannot be used when the satellite oscillates around one of the stable eccentric equilibria described in Sect.~\ref{ssec:eccLap}. Under the approximation that the total angular momentum of the system is contained in the orbital motion of the planet (which is almost exactly verified in practice), \cite{Boue-Laskar_2006} give an analytical expression for the time evolution of the satellite's orbital pole and the planet's spin axis (see their Eq.~133). Due to its high level of generality, this expression is quite complicated and cumbersome to use. However, it shows that if the satellite is placed on its circular Laplace plane, then the planet and its satellite rigidly precess as a whole at the frequency
   \begin{equation}\label{eq:precBL}
      \Omega = \frac{T+\sqrt{\Delta}}{2}\,,
   \end{equation}
   where
   \begin{equation}
      \left\{
      \begin{aligned}
         T &= -\frac{\mathfrak{a}x + \mathfrak{b}yz}{\gamma} - \frac{\mathfrak{c}z + \mathfrak{b}xy}{\alpha}\,, \\
         \Delta &= \left(\frac{\mathfrak{a}x + \mathfrak{b}yz}{\gamma} - \frac{\mathfrak{c}z + \mathfrak{b}xy}{\alpha}\right)^2 + 4\frac{\mathfrak{b}^2xy^2z}{\gamma\alpha} \,,
      \end{aligned}
      \right.
   \end{equation}
   in which
   \begin{equation}
      x = \cos\varepsilon\,,
      \quad
      y = \cos I_\mathrm{Q}\,,
      \quad
      z = \cos(\varepsilon- I_\mathrm{Q})\,,
   \end{equation}
   are constant, and the physical parameters of the system are contained in the coefficients
   \begin{equation}
      \alpha = \frac{Mm}{M+m}\sqrt{\mu_\mathrm{P}(1+m/M)a}\,,
      \quad
      \gamma = \lambda MR_\mathrm{eq}^2\omega\,,
   \end{equation}
   and
   \begin{equation}
      \begin{aligned}
         \mathfrak{a} &= \frac{3}{2}\frac{\mu_\odot MJ_2R_\mathrm{eq}^2}{a_\odot^3(1-e_\odot^2)^{3/2}}\,,
         \quad\quad
         \mathfrak{b} = \frac{3}{2}\frac{\mu_\mathrm{P}mJ_2R_\mathrm{eq}^2}{a^3}\,, \\
         \quad
         \mathfrak{c} &= \frac{3}{4}\frac{\mu_\odot\frac{mM}{M+m}a^2}{a_\odot^3(1-e_\odot^2)^{3/2}}\,.
      \end{aligned}
   \end{equation}
   In this self-consistent model, the equatorial inclination $I_\mathrm{Q}$ of the satellite at its circular Laplace equilibrium is obtained by cancelling the nutation amplitude of the solution (coefficient $\mathfrak{s}$ in Eq.~133 of \citealp{Boue-Laskar_2006}). This amounts to solving for $I_\mathrm{Q}$ the equation
   \begin{equation}
      \mathfrak{a}\sin(2\varepsilon) + \mathfrak{b}\sin(2I_\mathrm{Q}) + 2\gamma\Omega\sin\varepsilon = 0\,,
   \end{equation}
   where $\Omega$ depends itself on $I_\mathrm{Q}$ through Eq.~\eqref{eq:precBL}. For a given set of parameters, this equation can easily be solved numerically. For small masses, one can check that we retrieve very accurately the inclination given by the classical massless-case formula in Eq.~\eqref{eq:ILap}.
   
\begin{figure}
      \centering
      \includegraphics[width=\columnwidth]{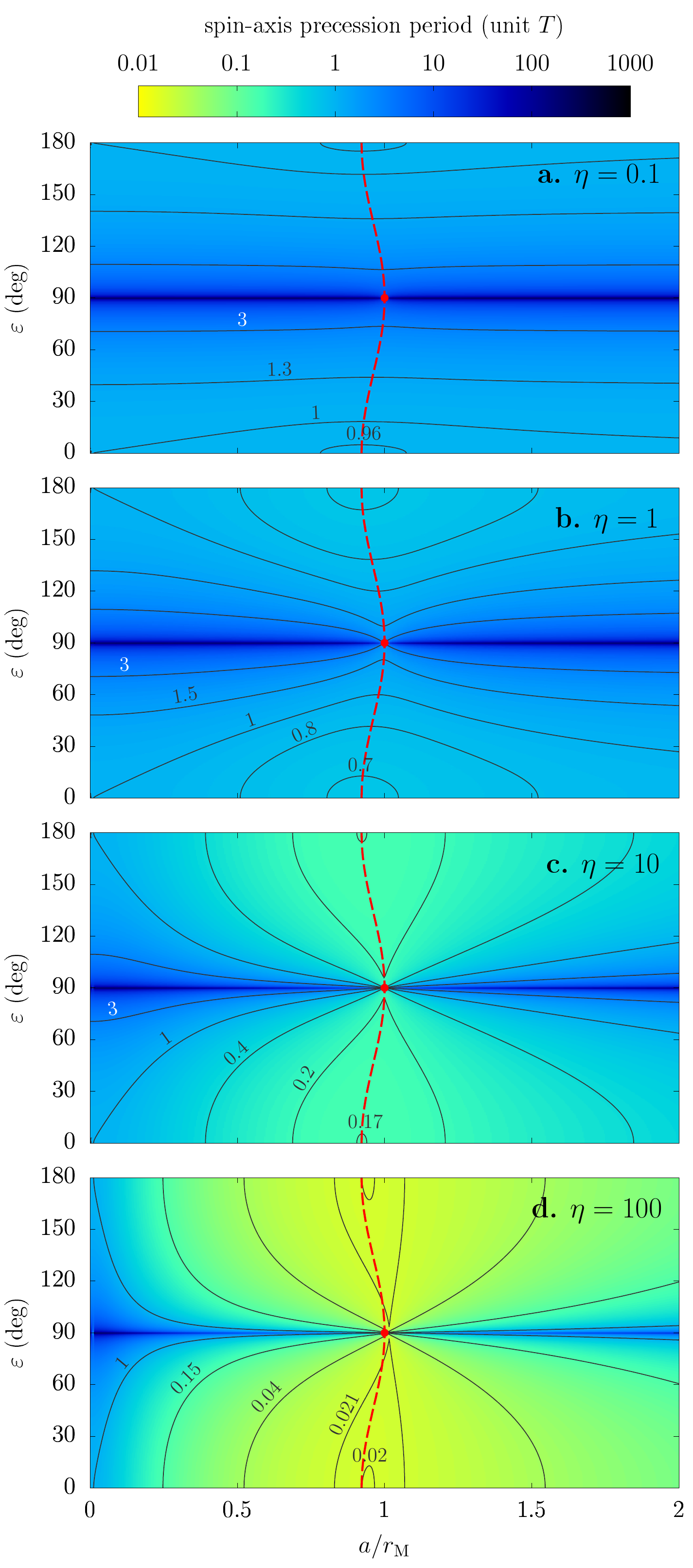}
      \caption{Same as Fig.~\ref{fig:precrate}, but using the self-consistent precession rate of \cite{Boue-Laskar_2006} given in Eq.~\eqref{eq:precBL}. The physical parameters are those of Titan and Saturn, except that the mass of Titan is adjusted to produce the same mass parameters $\eta$ as in Fig.~\ref{fig:precrate}. In order to ease the comparison, the ridge line and singular point (in red) are those of the simplified model, and the labelled level curves are the same as in Fig.~\ref{fig:precrate} (except the level $0.021$ in panel d which is added here).}
      \label{fig:precrateBL}
   \end{figure}

   Figure~\ref{fig:precrateBL} shows the precession period obtained for the parameters of Saturn and Titan. The mass of Titan is varied between each panel so as to produce the same mass parameter $\eta$ as in Fig.~\ref{fig:precrate}; all other physical parameters are left unchanged. Because physical parameters are deeply entangled in the self-consistent model of \cite{Boue-Laskar_2006}, the precession rate in Eq.~\eqref{eq:precBL} cannot be expressed in terms of a few macro-parameters, as done in Sect.~\ref{ssec:spinandsat}. As a result, Fig.~\ref{fig:precrateBL} is specific to the parameters of Titan and Saturn, whereas Fig.~\ref{fig:precrate} is generic.
   
   For small satellite masses, Figs.~\ref{fig:precrate} and \ref{fig:precrateBL} give undistinguishable results (panels a and b). For a mass of the order of Titan's (panel c), small differences in magnitude are noticeable near $a=r_\mathrm{M}$, as shown by the slight displacements of the level curves labelled $0.2$ and $0.17$. For a satellite about ten times as massive as Titan (panel d), the ridge line is slightly shifted right as compared to our simplified model, but its magnitude is still very similar to that of Fig.~\ref{fig:precrate}. For even larger masses (not shown), the ridge line is further shifted right, and the contribution of the satellite to $\lambda$ becomes non-negligible (see Eq.~\ref{eq:J2prime}). However, the overall structure of the dynamics remains the same, with a singular point at $\varepsilon=90^\circ$ towards which the level curves converge.
   
   From this comparison, we conclude that the simplified model presented in Sect.~\ref{ssec:spinandsat} provides a very good qualitative description of the dynamics, even though the exact magnitude of the precession rate can differ somewhat for large satellite-to-planet mass ratios. For the Earth-Moon system, whose mass ratio is as large as about $0.01$, a self-consistent precession model would be required to get precise quantitative results.
   
\end{document}